\documentclass[a4paper,11pt]{article}

\pdfoutput=1

\usepackage{a4wide}
\usepackage{amsthm}
\usepackage{amssymb}
\usepackage[usenames]{color}
\definecolor{DarkBlue}{rgb}{0,0,0.15}
\usepackage{amsmath}
\usepackage{amsfonts}
\usepackage{array}
\usepackage{enumitem}
\usepackage{subfigure}
\usepackage{supertabular}
\usepackage{captcont}
\usepackage[pdftex]{graphicx}
\usepackage{rotating}
\usepackage{setspace}
\usepackage{ifthen}
\usepackage{bussproofs}
\usepackage{xy,xypic}
\usepackage{index}

\ifnum\pdfoutput=1
  \usepackage[pdftex,colorlinks=true,pdfhighlight=/P,linkcolor=DarkBlue,urlcolor=DarkBlue,citecolor=DarkBlue,breaklinks=true,hypertexnames=false,hyperindex,plainpages=false,pdfpagelabels=true]{hyperref}
\else
  \usepackage[colorlinks=true,linkcolor=DarkBlue,urlcolor=DarkBlue,citecolor=DarkBlue,breaklinks=true]{hyperref}
\fi

\newindex{default}{idx}{ind}{Index}

\newtheorem{theorem}{Theorem}[section]
\newtheorem{lemma}[theorem]{Lemma}
\newtheorem{corollary}[theorem]{Corollary}

\newtheorem{remark}[theorem]{Remark}

\newtheorem{definition}[theorem]{Definition}
\newtheorem{notation}[theorem]{Notation}

\newcount\hh \newcount\mm
\hh=\time \divide\hh by 60
\mm=\hh \multiply\mm by 60 \mm=-\mm
\advance\mm by \time
\def\actTime{\number\hh:\ifnum\mm<10{}0\fi\number\mm} 

\usepackage{fancyhdr}
\def\N{{\mathbb N}}

\def\E{{\mathcal E}}
\def\P{{\mathcal P}}

\newcommand{\infer}[3][]{\ifthenelse{\equal{#1}{}}%
  {\ensuremath{\dfrac{#3}{#2}}}%
  {\ensuremath{\dfrac{#3}{#2}}\makebox[0pt][l]{\quad(\typeRule{#1})}}%
}

\newcommand{\reclist}[1]{\ensuremath{#1_{[\circlearrowright]}}}
\newcommand{\recfunc}[1]{\ensuremath{#1_{[\circlearrowright]}}}

\def\circL{{\circ_{\mathbf{L}}}}
\def\circM{{\circ_{\mathbf{M}}}}

\def\dom{\mathop{\mathrm{dom}}\nolimits}
\def\channel(#1,#2){\ensuremath{#1\!\!\models\!=\!\!\!\rotatebox[origin=c]{180}{$\models$}#2}}
\def\isdef{\stackrel{\text{\rm\tiny{}def}}{=}}
\def\linearize{\mathop{\mathrm{linearize}}\nolimits}
\def\listOf#1{\widetilde{#1}}
\def\set{\mathop{\mathrm{set}}\nolimits}
\def\Tr{\mathop{\mathrm{Tr\ }}\nolimits}
\def\val{\mathop{\mathrm{val}}\nolimits}
\def\value{\ensuremath{\mathbf v}}

\newcommand{\ket}[1]{\ensuremath{\vert#1\rangle}}

\newenvironment{operSem}[2][]{\subsubsection{#1}
#2

\renewcommand{\baselinestretch}{1}
\begin{center}
$\begin{array}{lrll}
\hline \\}{\\
\hline
\end{array}$
\end{center}}

\newenvironment{operSem*}{
\renewcommand{\baselinestretch}{1}
\begin{center}
$\begin{array}{lrll}
\hline \\}{\\
\hline
\end{array}$
\end{center}}

\newcommand{\intVal}[3]{\langle\!\langle #1, #2 \rangle\!\rangle_{\type{#3}}}
\newcommand{\opSemVerb}[4][]{\typesetSemRule{#2}\raisebox{\normalbaselineskip}[0pt]{\hypertarget{semRule:#2}{}}\hfill&#3 &\longrightarrow_{#1} &#4}
\newcommand{\opSem}[4][]{\opSemVerb[#1]{#2}{[gs\,\vert\,(lms, vp, \si{#3}\ ts)]}{#4}}
\newcommand{\semRule}[1]{\hyperlink{semRule:#1}{\typesetSemRule{#1}}}
\newcommand{\typesetSemRule}[1]{\textsc{#1}}

\def\synOr{\ |\ }
\newcounter{synRule}
\makeatletter
\newcommand\newSynRule[2][]{{%
  \addtocounter{synRule}{1}
#2
}}
\makeatother 

\newcommand{\baseterm}[1]{\ifmmode\text{\bf #1}\else{\bf #1}\fi}

\newcommand{\methodName}[2]{{{\it #1}(#2)}}
\newcommand{\callMethod}[2]{\methodName{#1}{#2}}
\newcommand{\reservedWord}[1]{{\bf #1{}} }

\newcommand{\type}[1]{{\sf #1}}
\newcommand{\indentedCode}[2]{\par\noindent#1 \{ \\%
\noindent\hspace*{3em}%
\hbox{%
  \addtolength{\textwidth}{-3em}%
  \parbox{\textwidth}{\noindent \strut\relax#2}%
} \\ \}}
\newcommand{\method}[4]{\hbadness=10000\indentedCode{\type{#1} \methodName{#2}{#3}}{#4}\\*[\parskip]}

\newcommand{\IfNoPar}[2]{\IF (#1) \{ \\
  \hspace*{3em}\hbox{\addtolength{\textwidth}{-3em}\parbox{\textwidth}{\noindent{}#2}} \\ \}
}
\newcommand{\IfElseNoPar}[3]{\IF (#1) \{ \\
  \hspace*{3em}\hbox{\addtolength{\textwidth}{-3em}\parbox{\textwidth}{\noindent{}#2}} \\ \} \ELSE \{ \\
  \hspace*{3em}\hbox{\addtolength{\textwidth}{-3em}\parbox{\textwidth}{\noindent{}#3}} \\ \}
}
\newcommand{\IfElseNoBlock}[3]{\par\noindent\IF (#1)
  \{ \\
  \hspace*{3em}\hbox{\addtolength{\textwidth}{-3em}\parbox{\textwidth}{\noindent{}#2}} \\ \} \ELSE #3}
\newcommand{\IfElseNoBlockNoPar}[3]{\IF (#1)
  \{ \\
  \hspace*{3em}\hbox{\addtolength{\textwidth}{-3em}\parbox{\textwidth}{\noindent{}#2}} \\ \} \ELSE #3}

\newcommand{\aliasfor}		{\reservedWord{aliasfor}}
\newcommand{\ELSE}			{\reservedWord{else}}
\newcommand{\fork}			{\reservedWord{fork}}
\newcommand{\IF}			{\reservedWord{if}}
\newcommand{\Measure}		{\reservedWord{measure}}
\newcommand{\new}			{\reservedWord{new}}
\newcommand{\recv}			{\reservedWord{recv}}
\newcommand{\return}		{\reservedWord{return}}
\newcommand{\send}			{\reservedWord{send}}
\newcommand{\withends}		{\reservedWord{withends}}
\newcommand{\while}			{\reservedWord{while}}

\newcommand{\var}[2]{\type{#2} $#1$}

\newcommand{\varChWithEnds}[2]	{\var{#1}{channel[#2]} \withends [$#1_0, #1_1$]}
\newcommand{\varChE}[2]	{\var{#1}{channelEnd[#2]}}
\newcommand{\varI}[1]	{\var{#1}{int}}
\newcommand{\varQ}[1]	{\var{#1}{qbit}}

\newcommand{\vardecl}[1]{\noindent{}#1 \\}

\newcommand{\displayRTEuninitVarUsage}{\ensuremath{\mathbf{UV}}}
\newcommand{\displayRTEqVarOverlap}{\ensuremath{\mathbf{OQV}}}
\newcommand{\displayRTEassignIncompatibleQuant}{\ensuremath{\mathbf{ISQV}}}
\newcommand{\RTEuninitVarUsage}{\refConcept[rte:UV]{\displayRTEuninitVarUsage}}
\newcommand{\RTEqVarOverlap}{\refConcept[rte:OQV]{\displayRTEqVarOverlap}}
\newcommand{\RTEassignIncompatibleQuant}{\refConcept[rte:ISQV]{\displayRTEassignIncompatibleQuant}}

\newcommand{\dispnone}{{\sf none}}
\renewcommand{\none}{{\hyperlink{ref:none}{\dispnone}}}

\newcommand{\si}[1]{\underline{\smash[b]{#1}}}

\newcommand{\void}{\type{void}}

\newcommand{\typesetTypeRule}[1]{\textsc{#1}}
\newcommand{\typeRule}[1]{\hyperlink{typeRule:#1}{\typesetTypeRule{#1}}}
\newcommand{\typeRuleDef}[2]{\typesetTypeRule{#1}\raisebox{\normalbaselineskip}[0pt]{\hypertarget{typeRule:#1}{}} & #2}

\newcounter{premise}
\makeatletter
\newcommand\newPremise[1]{{%
  \phantomsection%
  \addtocounter{premise}{1}
  \ifx#1\@empty\else\edef\@currentlabel{\arabic{premise}}\label{premise:#1}\fi%
  (\arabic{premise})%
}}
\newcommand\premise[1]{(\ref{premise:#1})}
\makeatother 

\makeatletter
\newcommand{\concept}[2][]{%
  \ifthenelse{\equal{#1}{}}%
   {\@concept{#2}{#2}}%
   {\@concept{#1}{#2}}%
}
\newcommand{\@concept}[2]{%
   \raisebox{\normalbaselineskip}[0pt]{\hypertarget{#1}}{}%
   \relax\ifmmode{#2\index{#1@$#1$|textbf}}\else{{\em #2}\index{#1|textbf}}\fi%
}

\makeatother
\newcommand{\refConcept}[2][]{%
  \ifthenelse{\equal{#1}{}}%
   {\hyperlink{#2}{#2}}%
   {\hyperlink{#1}{#2}}%
 }

\title{\bf Operational Semantics \\ and Type Soundness \\ of Quantum Programming Language LanQ}

\author{%
  Hynek Mlna\v{r}\'{\i}k\thanks{This work has been supported by the grants No. 201/04/1153 and MSM0021622419.}\\
  Faculty of Informatics, Masaryk University \\ Brno, Czech Republic\\
  \texttt{hmlnarik@mail.muni.cz}
}

\begin{document}

\maketitle

\begin{abstract}
\noindent
We present an imperative quantum programming language LanQ which was designed to support combination of quantum and classical programming and basic process operations -- process creation and interprocess communication. The language can thus be used for implementing both classical and quantum algorithms and protocols. Its syntax is similar to that of C language what makes it easy to learn for existing programmers. In this paper, we present operational semantics of the language and a proof of type soundness of the noncommunicating part of the language. We provide an example run of a quantum random number generator.
\end{abstract}

\newpage

\tableofcontents

\newpage

\section{Introduction}

Quantum computing is a young branch of computer science. Its power lies in employing quantum phenomena in computation. These laws are different to those that rule classical world: Quantum systems can be entangled. Quantum evolution is reversible. One can compute exponentially many values in one step.

Quantum phenomena were successfully used for speeding up a solution of computationally hard problems like computing discrete logarithm or factorisation of integers \cite{Shor94}. Another successful application of quantum phenomena in computing, namely in cryptography, is secure quantum key generation \cite{BB84,E91,B92}. Quantum key generation overcomes the classical in the fact that its security relies on the laws of nature, while classical key generation techniques rely on computational hardness of solving some problems. A nice example of quantum phenomena usage is a teleportation of an unknown quantum state \cite{Ben_93}.

For the formal description of quantum algorithms and protocols, several quantum programming languages and process algebras have already been developed. Some of them support handling quantum data only, however most of them allow combining of quantum and classical computations. Obtaining classical data from quantum systems is done by {\em measurement} which is probabilistic by its nature. This implies that quantum formalisms must be able to to handle probabilistic computation.

Existing formalisms are usually based on existing classical programming languages and process algebras. From imperative languages, we should mention \"{O}mer's QCL (Quantum Computation Language, \cite{Oem00}) whose syntax is based on that of C language; Betteli, Calarco and Serafini's Q language built as an extension of C++ basic classes \cite{BetCalSer01}. However, semantics of these imperative languages is not defined formally. Zuliani's qGCL (quantum Guarded Command Language, \cite{Zul01}) based on pGCL (probabilistic Guarded Command Language) has denotational semantics defined but does not support recursion.

Many of developed languages are functional because of relatively straightforward definition of its operational semantics. Van Tonder developed a quantum $\lambda$-calculus \cite{Ton03}; quantum $\lambda$-calculus was also developed by Selinger and Valiron \cite{SelVal05}; Selinger proposed functional static-typed quantum flow-chart programming language QFC and its text form QPL \cite{Sel04b}. Another functional programming language QML was developed by Altenkirch and Grattage \cite{AltGra04} and refined into nQML in \cite{LamGinPap06}.

Quantum process algebras differ to classical ones in the way they handle quantum systems. The main issue solved here is that they must guarantee that any quantum system is accessible by only one process at one time (because of the no-cloning theorem \cite{WooZur82}). The quantum process algebras QPAlg by Lalire and Jorrand \cite{JorLal04,JorLal04b,Lal05} and CQP by Gay and Nagarajan \cite{GayNag04,GayNag05,GayNag06} can describe both classical and quantum interaction and evolution of processes. QPAlg was inspired by CCS, originally using nontyped channels for interprocess communication. Recently \cite{Lal06}, Lalire has added support for fixpoint operator and typed channels to QPAlg.

The presented language LanQ is an imperative quantum programming language. It allows combination of quantum and classical computations to be expressed. Moreover, it has features of quantum process algebras -- it supports new process creation and interprocess communication. Its syntax is similar to the syntax of C language. In the present paper, we define its syntax, operational semantics, and prove type soundness of the noncommunicating part of the language.

The paper is structured as follows: we start with an example of an program written in LanQ in Section \ref{sec:infIntro}. We then formally define its concrete (Section \ref{sec:conSynt}) and internal syntax (Section \ref{sec:abSynt}). Then basic concepts used later in the paper are defined in Section \ref{sec:defs}, followed by typing system in Section \ref{sec:typing-rules}. In Section \ref{sec:opSem}, we define the operational semantics of the language and prove its type soundness in Section \ref{sec:type-soundness}. An example of a simple program execution can be found in Appendix \ref{sec:example}.

\section{Informal introduction}
\label{sec:infIntro}

We begin our description of LanQ by an example implementation of a well-known multiparty quantum protocol -- teleportation \cite{Ben_93}. Teleportation can be written as the program shown in Figure \ref{fig:teleportation}.

{\renewcommand{\baselinestretch}{1}
\begin{figure}[h]
\small
\hfill\begin{minipage}[t]{0.45\textwidth}
\method{void}{main}{}{%
	\vardecl{\varQ{\psi_A, \psi_B}; \\ 
		$\psi_{EPR}$ \aliasfor [$\psi_A, \psi_B$]; \\
		\varChWithEnds{c}{int};
	} \\
	$\psi_{EPR}$ = \callMethod{createEPR}{}; \\
	$c$ = \new \type{channel[int]}(); \\
	\fork \callMethod{bert}{$c_0$, $\psi_B$}; \\
	
	\noindent\callMethod{angela}{$c_1$, $\psi_A$};
}

\method{void}{angela}{\varChE{c_0}{int}, \varQ{ats}}{%
	\vardecl{\varI{r}; \\
	\varQ{\phi};
	} \\
	\noindent$\phi$ = \callMethod{doSomething}{}; \\
	$r$ = \Measure({\it BellBasis}, $\phi$, $ats$); \\
	\send($c_0$, $r$);
}
\end{minipage}
\hfill\vrule\hfill %
\begin{minipage}[t]{0.43\textwidth}
\method{int}{bert}{\varChE{c_1}{int}, \varQ{stto}}{%
	\vardecl{\varI{i};} \\
	$i$ = \recv($c_1$);
	\IfElseNoBlock{$i$ == 0}{ op$B_0$($stto$); }
	{\IfElseNoBlockNoPar{$i$ == 1}{ op$B_1$($stto$); }
	{\IfElseNoPar{$i$ == 2}{ op$B_2$($stto$); }
	{op$B_3$($stto$); }}} \\
	\callMethod{doSomethingElse}{$stto$};\\
	\return $i$;
}
\end{minipage}
\caption{Teleportation implemented in LanQ}
\label{fig:teleportation}
\end{figure}}

We now briefly describe the program. In LanQ, a program is a set of methods. Three methods, {\bf main}, {\bf angela} and {\bf bert}, are defined. The control is passed to a method called \methodName{main}{} when the program is run. This method takes no parameters and it returns no value what can be seen from the word \type{void} in front of the method name. The method \methodName{angela}{} has to be invoked with two parameters -- one end of a channel that can be used to transmit values of type \type{int}, and one qubit ({\it ie.} a quantum bit). It also returns no value. The method \methodName{bert}{} takes a channel end and a qubit, and returns a value of type \type{int}.

The method \methodName{main}{} declares variables $\psi_A, \psi_B, \psi_{EPR}, c, c_0$ and $c_1$ used in the method body in its first three lines: The type of variables $\psi_A,\psi_B$ is \type{qbit}. Variable $\psi_{EPR}$ is declared to be an alias for a two-qubit compound system $\psi_A \otimes \psi_B$. Channel $c$ capable of transmitting integers is declared on the next line. The ends of the channel are named $c_0$ and $c_1$.

On next lines, the method {\bf main} invokes method \methodName{createEPR}{} which creates an EPR-pair, and stores the returned reference to the created pair into the variable $\psi_{EPR}$. After that, a new channel is allocated and assigned to the variable $c$. This also modifies the channel end variables $c_0$ and $c_1$. The next command makes the running process split into two. One of the processes continues its run and invokes the method \methodName{angela}{}. The second process starts its run from the method \methodName{bert}{}.

The method \methodName{angela}{} receives one channel end and one qubit as arguments. After declaring variables $r$ and $\phi$, it assigns a result of invocation of a method \methodName{doSomething}{} to $\phi$. Then it measures qubits $\phi$ and $ats$ using the Bell basis, assigns the result of the measurement to the variable $r$ and sends it over the channel end $c_0$.

The method \methodName{bert}{} receives one channel end and one qubit as arguments. After declaring a variable $i$, it receives an integer value from the channel end $c_1$ and assigns it to the variable $i$. Depending on the received value, it applies one of the operators $opB_0$, $opB_1$, $opB_2$ and $opB_3$ on qubit $stto$. Then, it invokes a method \methodName{doSomethingElse}{} and passes the variable $stto$ as an argument to this method. Finally, it returns the value of the variable $i$ to the caller.

\section{Concrete syntax}
\label{sec:conSynt}

In this section, we introduce concrete syntax of LanQ programs. This syntax is used to write programs by a programmer. Semantics is defined using internal syntax which is described later (see Section \ref{sec:abSynt}).

The syntax is shown in Figure \ref{fig:conSynt}. Reserved words of the language are written in \textbf{bold} and the identifier names are in \textsc{capitals}. Grammar is given in nondeterministic extended Backus-Naur form (EBNF). The root of grammar is the nonterminal $program$.

For the sake of clarity, the concrete grammar nonterminals names are long and descriptive to indicate their meaning. We describe meaning of the most important nonterminals here: $program$ (words derived from this nonterminal represent LanQ \concept[program]{programs}), $code$ (\concept[csyntax-statement]{statements}), $pExpr$ (\concept[csyntax-promotable expression]{promotable expressions}, {\it ie.} expressions that can act as statements), $methodCall$ (\concept[csyntax-method call]{method calls}), $methodParams$ (\concept[csyntax-mparams]{method parameters}), $assignment$ (\concept[]{assignments}), $measurement$ (\concept[]{measurements}), $expr$ (\concept[]{expressions}), $indivExpr$ (\concept[]{individual expressions}, {\it ie.} expressions not containing any operators), $op$ (\concept[]{operators}), $method$ (\concept[method]{}\concept[method declaration]{method declarations}), $block$ (\concept[]{blocks} of code), $seq$ (\concept[]{block-forming statements}, {\it ie.} statements that can be used in blocks), and $varDeclaration$ (\concept[]{variable declarations}). The other nonterminals are auxiliary and their meaning is obvious.

\begin{definition}
Let $m$ be a \refConcept{method declaration}. We call the part of $m$ which was derived from nonterminal $methodHeader$ a \concept{method header}, and the part of $m$ which was derived from nonterminal $block$ a \concept{method body}.
\end{definition}

In the following example, a method named $mName$ is declared. The parts of the method declaration are annotated on the right side.

\begin{center}
\begin{figure}[h!]
\begin{tabular}{ll}
$\type{T}\ mName(\type{T_1}\ a_1,\dots,\type{T_n}\ a_n)	\qquad\qquad \left.\right\}$ & method header \\ %
$\left.\!\begin{array}{l}
\{ \\
\text{\qquad ... statements ...} \\
\}
\end{array}%
\hspace*{2.5cm}\right\}$ & method body
\end{tabular}
\caption{Declaration of a method named $mName$}
\end{figure}
\end{center}

{\renewcommand{\baselinestretch}{1}
\begin{figure}[ht!]
\label{fig:conSynt}
\noindent{\bf Code}\nopagebreak

$\begin{array}{lcl}
program & ::= &	\newSynRule[conSynt:program]{method+} \\
code & ::= & \newSynRule[conSynt:code:skip]{\baseterm{;}} \synOr
	\newSynRule[conSynt:code:pExpr]{pExpr \baseterm{;}} \synOr 
	\newSynRule[conSynt:code:fork]{fork} \synOr
	\newSynRule[conSynt:code:send]{send} \synOr
	\newSynRule[conSynt:code:ret]{return} \synOr
	\newSynRule[conSynt:code:blk]{block} \synOr
	\newSynRule[conSynt:code:if]{if} \synOr 
	\newSynRule[conSynt:code:while]{while} \\
pExpr & ::= & \newSynRule[conSynt:pExpr:asn]{assignment} \synOr
	\newSynRule[conSynt:pExpr:mc]{methodCall} \synOr
	\newSynRule[conSynt:pExpr:recv]{recv} \synOr
	\newSynRule[conSynt:pExpr:meas]{measurement} \synOr \\
& &  \newSynRule[conSynt:pExpr:new]{\baseterm{new}\ nonDupType \baseterm{()}} \\
methodCall & ::= &
	\newSynRule[conSynt:mc]{\text{\textsc{methodName}}\ \baseterm{(}\ (methodParams)?\ \baseterm{)}} \\
methodParams & ::= &
	\newSynRule[conSynt:mcParams]{expr\ (\baseterm{,}\ expr)*} \\
assignment & ::= &
	\newSynRule[conSynt:asn]{\text{\textsc{variableName}}\ \baseterm{=}\ expr} \\
measurement & ::= &
	\newSynRule[conSynt:meas]{\baseterm{measure}\ \baseterm{(}\ \text{\textsc{basisName}}\ (\baseterm{,}\ \text{\textsc{variableName}})+\ \baseterm{)}} \\
expr & ::= &
	\newSynRule[conSynt:expr]{indivExpr\ (op\ expr)?} \\
indivExpr & ::= &
	\newSynRule{\text{\textsc{const}}} \synOr \newSynRule{\text{\textsc{variableName}}} \synOr \newSynRule{\baseterm(\ expr\ \baseterm)} \synOr \newSynRule{pExpr} \\
op & ::= &
	\newSynRule[conSynt:op:plus]{\baseterm{+}} \synOr \newSynRule[conSynt:op:minus]{\baseterm{--}} \synOr \newSynRule[conSynt:op:otimes]{\mathbf{\otimes}} \synOr \dots  \\[0.7\normalbaselineskip]
\end{array}$

\vfill\noindent{\bf Block structure}\nopagebreak

$\begin{array}{lcl}
method & ::= & \newSynRule[conSynt:method]{methodHeader\ block} \\
block & ::= & \newSynRule[conSynt:block]{\baseterm{\{}\ (seq)?\ \baseterm{\}}} \\
seq & ::= & \newSynRule[conSynt:seq:varDecl]{varDeclaration\ (seq)?} \synOr \newSynRule[conSynt:seq:code]{code\ (seq)?} \\ 
methodHeader & ::= & \newSynRule[conSynt:methodHeader]{type\ \text{\textsc{methodName}}\ \baseterm{(}\ methodDeclParamList?\ \baseterm{)}} \\
methodDeclParamList & ::= & \newSynRule[conSynt:methodDeclParamList]{methodDeclParam (\baseterm{,}\ methodDeclParam)*} \\
methodDeclParam & ::= & \newSynRule[conSynt:methodDeclParam]{nonVoidType\ \text{\textsc{paramName}}} \\
varDeclaration & ::= &
	\newSynRule[conSynt:varDecl:nvt]{nonVoidType\ \text{\textsc{variableName}}(\baseterm{,}\ \text{\textsc{variableName}})*\ \baseterm{;}} \synOr \\
& &	\newSynRule[conSynt:varDecl:cht]{channelType\ \text{\textsc{variableName}}\ \baseterm{withends}} \\
& &	\quad \baseterm{[}\ \text{\textsc{variableName}}\ \baseterm{,}\ \text{\textsc{variableName}}\ \baseterm{]}\ \baseterm{;} \synOr \\
& &	\newSynRule[conSynt:varDecl:alf]{\text{\textsc{variableName}}\ \baseterm{aliasfor}} \\
& &	\quad \baseterm{[}\ \text{\textsc{variableName}}\ (\baseterm{,}\ \text{\textsc{variableName}})*\ \baseterm{]}\ \baseterm{;} \\[0.7\normalbaselineskip]
\end{array}$

\vfill
\noindent{\bf Program flow}\nopagebreak

$\begin{array}{lcl}
fork & ::= &
	\newSynRule[conSynt:fork]{\baseterm{fork}\ methodCall\ \baseterm{;}}\\
return & ::= &
	\newSynRule[conSynt:return]{\baseterm{return}\ (expr)?\ \baseterm{;}}\\[0.7\normalbaselineskip]
\end{array}$

\vfill\noindent{\bf Conditionals and loops}\nopagebreak

$\begin{array}{lcl}
if & ::= &
	\newSynRule[conSynt:if]{\baseterm{if}\ \baseterm{(}\ expr\ \baseterm{)}\ code\ (\baseterm{else}\ code)?} \\

while & ::= &
	\newSynRule[conSynt:while]{\baseterm{while}\ \baseterm{(}\ expr\ \baseterm{)}\ code} \\[0.7\normalbaselineskip]

\end{array}$

\vfill\noindent{\bf Communication}\nopagebreak

$\begin{array}{lcl}
recv & ::= &
	\newSynRule[conSynt:recv]{{\bf recv}\ \baseterm{(}\ expr\ \baseterm{)}} \\
send & ::= &
	\newSynRule[conSynt:send]{\baseterm{send}\ \baseterm{(}\ expr\ \baseterm{,}\ expr\ \baseterm{)}\ \baseterm{;}} \\[0.7\normalbaselineskip]
\end{array}$

\vfill\noindent{\bf Types}\nopagebreak

$\begin{array}{lcl}
type & ::= & \newSynRule{\baseterm{void}} \synOr \newSynRule[conSynt:type:nvt]{nonVoidType} \\
nonVoidType & ::= & \newSynRule[conSynt:nvt:dt]{dupType} \synOr \newSynRule[conSynt:nvt:ndt]{nonDupType} \\
dupType & ::= & \newSynRule{\baseterm{int}} \synOr \newSynRule{\baseterm{bool}} \synOr \dots \\
nonDupType & ::= & \newSynRule[conSynt:ndt:che]{\baseterm{channelEnd} \baseterm{[} nonVoidType \baseterm{]}} \synOr \newSynRule[conSynt:ndt:cht]{channelType} \synOr \newSynRule[conSynt:ndt:qt]{qType} \\
channelType & ::= &	\newSynRule[conSynt:cht]{\baseterm{channel} \baseterm{[} nonVoidType \baseterm{]}} \\
qType & ::= & \newSynRule[conSynt:qt:qbt]{qBasicType (\otimes qType)?} \\
qBasicType & ::= & \newSynRule{\baseterm{qbit}} \synOr \newSynRule{\baseterm{qtrit}} \synOr \dots \\[0.7\normalbaselineskip]
\end{array}$
\caption{Concrete syntax}
\end{figure}
}

\hypertarget{intSynt}{}
\section{Internal syntax}
\label{sec:abSynt}

In this section, we define the internal syntax of LanQ.

Using the concrete syntax, a LanQ program is written as a set of method declarations. This notation does not allow direct execution of the program. To define operational semantics, we need a representation for the program execution -- a syntax that allows us to evaluate a program by means of rewriting of program terms. The rewriting rules are presented later in Section \ref{sec:opSem} where the operational semantics is defined.

The internal syntax is defined in Figure \ref{fig:abSynt}. The syntax is similar to the concrete one while not containing declarative parts of the concrete syntax and being abbreviated. In the internal syntax, we define the following basic syntactic entities: \concept[syntax!number]{numbers} ($N$), \concept[syntax!lists]{lists} ($L$), \concept[syntax!recursive list]{recursive lists} ($RL$), \concept[syntax!reference]{references} ($R$), \concept[syntax!constant]{constants} ($C$), \concept[syntax!identifier]{identifiers} ($I$), \concept[syntax!type]{types} ($T$), and \concept[syntax!value]{internal values} ($\value$). 

\concept[syntax!promotable expression]{Promotable expressions} ($PE$) are expressions that can act as statements when postfixed by semicolon. \concept[syntax!expression]{Expressions} ($E$) can evaluate to an internal value. The syntactic classes of \concept[syntax!variable declaration]{variable declarations} ($VD$) and \concept[syntax!statement]{statements} ($S$) can together create a block. Therefore they are together called \concept[syntax!block-forming elementary statement]{block-forming elementary statements} ($Be$). A \concept[syntax!block-forming statement]{block-forming statement} ($B$) is built from zero or more such \refConcept[syntax!block-forming elementary statement]{block-forming elementary statements}.

\begin{remark}\label{note:gramAbbr}For the sake of clarity, we use the following notation in the rule body. We denote by $\bar{S}$ an abbreviation of BNF rule body ``$(S)*$'', and by $\listOf{E}$ an abbreviation of ``$(E\ (\baseterm{,}\ E)*)?$''.
\end{remark}

{\renewcommand{\baselinestretch}{1}
\begin{figure}[ht!]
$\begin{array}{lcl}
  N & ::= & \newSynRule{0}\synOr 1\synOr\dots \\
  L & ::= & \newSynRule{[]}\synOr \newSynRule{[\listOf{N}]}\\
  RL & ::= & \newSynRule{L}\synOr \newSynRule{[\listOf{RL}]}\\
  R & ::= & \newSynRule{\none}\synOr\newSynRule{\baseterm{(}\mathbf{Classical}\baseterm{,} N\baseterm{)}}\synOr\newSynRule{\baseterm{(}\mathbf{Quantum}\baseterm{,} RL\baseterm{)}}\synOr\newSynRule{\baseterm{(}\mathbf{Channel}\baseterm{,} N\baseterm{)}}\synOr\\
   & & \newSynRule{\baseterm{(}\mathbf{ChannelEnd_0}\baseterm{,} N\baseterm{)}}\synOr\newSynRule{\baseterm{(}\mathbf{ChannelEnd_1}\baseterm{,} N\baseterm{)}}\synOr\newSynRule{\baseterm{(}\mathbf{GQuantum}\baseterm{,} L\baseterm{)}}\synOr\newSynRule{\baseterm{(}\mathbf{GChannel}\baseterm{,} N\baseterm{)}} \\
  C & ::= & \newSynRule{R} \synOr \newSynRule[abSynt:C:id]{\baseterm{true}}\synOr \baseterm{false} \synOr \newSynRule{\bot} \synOr \dots \\
  I & ::= & \newSynRule[abSynt:id:var]{x}\synOr y\synOr z\synOr \dots \synOr\newSynRule[abSynt:id:op]+\synOr-\synOr\dots\\
  T & ::= & \newSynRule[abSynt:type:void]{\type{void}} \synOr \newSynRule[abSynt:type:int]{\type{int}}\synOr \newSynRule[abSynt:type:qbit]{\type{qbit}} \synOr \newSynRule[abSynt:type:cht]{\type{channel[}T\type{]}}\synOr \newSynRule[abSynt:type:che]{\type{channelEnd[}T\type{]}} \synOr  \newSynRule[abSynt:type:tt]{T\otimes T} \synOr \dots \\
  \value & ::= & \newSynRule{\intVal{R}{C}{\makebox{$T$}}}\\
  \\
  PE & ::= & \newSynRule[abSynt:pExpr:new]{\baseterm{new}\ T\baseterm{()}} \synOr 
  \newSynRule[abSynt:pExpr:asn]{I = E} \synOr 
  \newSynRule[abSynt:pExpr:mc]{I\baseterm{(}\listOf{E}\baseterm{)}} \synOr 
  \newSynRule[abSynt:pExpr:meas]{\baseterm{measure}\baseterm{(}\listOf{E}\baseterm{)}} \synOr 
  \newSynRule[abSynt:pExpr:recv]{\baseterm{recv}\baseterm{(}E\baseterm{)}} \\
  E & ::= & \newSynRule[abSynt:E:id]{I} \synOr 
  \newSynRule[abSynt:E:value]{\value} \synOr 
  \newSynRule[abSynt:E:brk]{\baseterm{(}E\baseterm{)}} \synOr 
  \newSynRule[abSynt:E:PE]{PE} \\
  VD & ::= & \newSynRule[abSynt:varDecl:nvt]{T\ \listOf{I} \baseterm{;}} \synOr 
  \newSynRule[abSynt:varDecl:cht]{\type{channel[}T\type{]}\ I\ {\bf withends[}I\baseterm{,}I\baseterm{];}} \synOr 
  \newSynRule[abSynt:varDecl:alf]{I\ \baseterm{aliasfor}\ [\listOf{I}]\baseterm{;}} \\
  S & ::= & \newSynRule[abSynt:code:skip]{\baseterm{;}} \synOr 
  \newSynRule[abSynt:code:PE]{PE \baseterm{;}} \synOr 
  \newSynRule[abSynt:code:block]{\{B\}} \synOr 
  \newSynRule[abSynt:code:if]{{\bf if}\ \baseterm{(}E\baseterm{)}\ S\ {\bf else}\ S} \synOr 
  \newSynRule[abSynt:code:while]{{\bf while}\ \baseterm{(}E\baseterm{)}\ S} \synOr \\
& & \newSynRule[abSynt:return:void]{\baseterm{return;}} \synOr 
  \newSynRule[abSynt:return:expr]{\baseterm{return}\ E\baseterm{;}} \synOr 
  \newSynRule[abSynt:fork]{{\bf fork}\ I\baseterm{(}\listOf{E}\baseterm{)}\baseterm{;}} \synOr 
  \newSynRule[abSynt:send]{{\bf send}\baseterm{(}E\baseterm{,} E\baseterm{)}\baseterm{;}} \\
  Be & ::= & \newSynRule[abSynt:Be:varDecl]{VD} \synOr  \newSynRule[abSynt:Be:S]{S} \\
  B & ::= & \overline{Be} \\

 \end{array}$
\caption{Internal syntax}
\label{fig:abSynt}
\end{figure}

Configuration syntax specifies formal notation of process configuration which is described in Subsection \ref{opSem:configuration}.

If a statement or an expression contains a subexpression, this subexpression is evaluated separately and the evaluation result is substituted in place of the subexpression. For this reason, we introduce a concept of a \concept{hole} ($\bullet$) which stands for the awaited result of subexpression evaluation. We call a term not containing a hole a \concept[closed expression]{}\concept[closed statement]{}\concept{closed term}.

The terms containing a hole are defined by nonterminals $Sc$ and $Ec$ which represent partially evaluated statements and expressions, respectively, whose subexpression is being evaluated. In other words, they represent {\em evaluation contexts}. We also define syntactic entities for \refConcept[runtime error]{\concept[syntax!runtime error]{}\concept[RtErr]{runtime errors}} ($RTErr$) and \concept[StkEl]{}\concept[syntax!term stack element]{term stack elements} ($StkEl$).

\begin{figure}[ht!]
$\begin{array}{lcl}
  Ec & ::= & \newSynRule{I = \bullet} \synOr \newSynRule{I\baseterm{(}\listOf{\value}\baseterm{,}\bullet\baseterm{,}\listOf{E}\baseterm{)}}\synOr
    \newSynRule{{\bf measure}\baseterm{(}\listOf{\value}\baseterm{,} \bullet\baseterm{,} \listOf{E}\baseterm{)}} \synOr \newSynRule{\baseterm{recv}\baseterm{(}\bullet\baseterm{)}}\\
  Sc & ::= & \newSynRule{\bullet \baseterm{;}} \synOr\
    \newSynRule{{\bf if}\ \baseterm{(}\bullet\baseterm{)}\ S\ {\bf else}\ S} \synOr
    \newSynRule{{\bf fork}\ I\baseterm{(}\listOf{\value}\baseterm{,} \bullet\baseterm{,} \listOf{E}\baseterm{)}\baseterm{;}} \synOr 
    \newSynRule{{\bf send}\baseterm{(}\bullet\baseterm{,} E\baseterm{)}\baseterm{;}} \synOr 
    \newSynRule{\baseterm{return\ }\bullet\baseterm{;}} \\
  
  \concept{RTErr} & ::= & \newSynRule{\RTEuninitVarUsage}\synOr\newSynRule{\RTEqVarOverlap}\synOr\newSynRule{\RTEassignIncompatibleQuant} \\
\\
  StkEl & ::= & \newSynRule{B}\synOr{E}\synOr{Ec}\synOr{Sc}\synOr\newSynRule{RTErr} \synOr 
  \newSynRule{\circL} \synOr 
  \newSynRule{\circM}
   \\
  LMS & ::= & \text{local memory state as described in Subsection \ref{opSem:configuration}} \\
  VP & ::= & \text{variable properties as described in Subsection \ref{opSem:configuration}} \\ \\
  LS & ::= & \baseterm{(}LMS\baseterm{,} VP\baseterm{,} \overline{StkEl}\baseterm{)} \\
  P & ::= & LS \synOr LS \parallel P \\
  GS & ::= & \baseterm{(}\baseterm{(}DM\baseterm{,}L\baseterm{)}\baseterm{\baseterm{,}}[\listOf{\channel(I,I)}]\baseterm{)} \text{\quad where $DM$ is a density matrix} \\
  Sys & ::= & \baseterm{[}GS\,|\,P\baseterm{]} \\
  Univ & ::= & p \bullet Sys \synOr p \bullet Sys \boxplus Univ
 \end{array}$
 \caption{Configuration syntax}
\label{fig:abSynt:conf}
\end{figure}}

Before a method can be invoked to be run, we must transform its body to the internal representation. Fortunately, the method bodies derived using concrete syntax and internal syntax rules differ only in the following:
\begin{itemize}
 \item In internal representation, all \IF statements have \ELSE part, {\it ie.} a statement $\IF\ \baseterm{(}E\baseterm{)}\ P$ is rewritten to $\IF\ \baseterm{(}E\baseterm{)}\ P\ \ELSE\ \baseterm{;}$,
 \item In internal representation, all constants $C$ are represented by internal values $\intVal{\none}{C}{T}$ where \type{T} is the type of the constant $C$,
 \item In internal representation, all operators are written in the prefix notation and seen as method calls, {\it ie.} $E\ \odot\ F$ is converted to $\odot\baseterm{(}E\baseterm{,}F\baseterm{)}$.
\end{itemize}

There is obviously an algorithm which rewrites any method body derived using the concrete syntax to the internal representation.

An example of a block written using the concrete syntax and its internal representation is shown in Figure \ref{abSynt:fig:methodConversion}.

{\renewcommand{\baselinestretch}{1}
\begin{figure}[h!]
\subfigure[][]{\label{abSynt:fig:methodConversion:orig}\begin{minipage}[t]{0.47\textwidth}%
\indentedCode{}{%
	\vardecl{\varI{r}; \\
	\varQ{\phi};
	} \\
	\noindent$\phi$ = \callMethod{doSomething}{}; \\
	$r$ = \Measure({\it BellBasis}, $\phi$, $ats$); \\
	\send($c_0$, $r$);\\
	\IfNoPar{$r == 0$}{\Measure({\it StdBasis}, $\phi$);}
}
\end{minipage}}%
\ \hfill\vline\hfill\ %
\subfigure[][]{\label{abSynt:fig:methodConversion:converted}\begin{minipage}[t]{0.45\textwidth}%
\indentedCode{}{%
	\vardecl{\varI{r}; \\
	\varQ{\phi};
	} \\
	\noindent$\phi$ = \callMethod{doSomething}{}; \\
	$r$ = \Measure({\it BellBasis}, $\phi$, $ats$); \\
	\send($c_0$, $r$);\\
	\IfElseNoBlockNoPar{$==(r,0)$}{\Measure({\it StdBasis}, $\phi$);}{;}
}\\
\end{minipage}}
\caption{Block derived using concrete syntax (a) and the same block converted to internal syntax (b).}
\label{abSynt:fig:methodConversion}
\end{figure}}

\section{Typing}
\label{sec:typing}

\subsection{Typing rules}
\label{sec:typing-rules}

LanQ is a typed language. This feature enables us to detect errors arising from incorrect usage of methods, variables and constants at compile time.

First, we define the ground types used in LanQ: 
\begin{itemize}
 \item \concept[\void]{\type{void}} -- a type with only one value: $\bot$,%
   \footnote{This type is a unit type. In several other languages, it is called differently, {\it eg.} \type{unit} in OCaml or \type{()} in Haskell. We have decided to use the name \type{void} as usual in C-based languages because LanQ follows C language also in many other aspects.}
 \item \concept[type!int]{\type{int}} -- a type of integers,
 \item \concept[type!bool]{}\type{bool} -- a type of truth values {\bf true} and {\bf false},
 \item \concept[type!q$d$it]{\type{q}$d$\type{it}} -- a type of references to $d$-dimensional quantum systems. Types which are often used are given special names: instead of writing \type{q2it}, one can write \concept[type!qbit]{\type{qbit}}, and in place of \type{q3it} one can use \concept[type!qtrit]{\type{qtrit}}.
 \item \type{\refConcept{Ref}} -- a type of references, defined later in Section \ref{opSem:memModel},
 \item \type{RTErr} -- a type of runtime errors, and
 \item \type{MeasurementBasis} -- a type of measurement bases.
\end{itemize}
The ground type system can be indeed extended when needed.

If \type{T} is a type, then \type{channel[T]} is a type of references to a channel capable of transmitting values of type \type{T}, and \type{channelEnd[T]} is a type of references to ends of such a channel. Further, let $\type{S_1}, ..., \type{S_n}, \type{S}$ be types for $n \geq 0$. Then a \concept{method type} $\type{T}$ is defined to be the type $\type{S_1}, \type{S_2}, ..., \type{S_n} \rightarrow \type{S}$. We call types $\type{S_1},\dots,\type{S_n}$ \concept{argument types} and the type \type{S} a \concept{return type}.

\begin{definition}
We define a set \concept[Types]{}$\refConcept{Types}$ of types of classical values. We denote by $\val(\type{T})$ a set of values of type \type{T} and define a set \concept[Values]{}$\refConcept{Values}$ as a set of values of all types:
$$\refConcept{Values} = \bigcup_{\type{T} \in \refConcept{Types}} \val(\type{T}).$$
\end{definition}

After parsing a program, the method declarations are stored in a triplet $(M_T, M_H, M_B)$ of partial functions. We call this triplet a \concept{method typing context} where:
 \begin{itemize}
  \item $M_T(m)$ which returns the \refConcept{method type} for a \refConcept{method} $m$ (the method type is straightforwardly determined from the method header), 
  \item $M_H(m)$ returns the \refConcept{method header} for a \refConcept{method} $m$, and 

\begin{figure}[p]
$\begin{array}{m{0.2\linewidth}c}
\typeRuleDef{T-Program}{\infer{\vdash (M_T, M_H, M_B)}{\forall m \in \dom(M_T):\ (M_T, M_H, M_B) \vdash_T M_H(m) : M_T(m)}} \\

\quad \\

\typeRuleDef{T-Method}{\infer{(M_T, M_H, M_B) \vdash_T T\ m\baseterm{(}T_1\ a_1\baseterm{,}\dots\baseterm{,}T_n\ a_n\baseterm{)} : T_1,\dots,T_n\rightarrow T}{\begin{array}{l}
T = \void \vee \refConcept{RetOk}(M_B(m)), \\
(M_T, M_H, M_B);a_1:T_1,\dots,a_n:T_n,@retVal:T \vdash_T M_B(m) : \void
\end{array}}} \\
\end{array}$
\caption{Typing rules for program and method declaration}
\label{fig:typRules:prog}
\end{figure}

\begin{figure}[p]
$\begin{array}{m{0.20\linewidth}c}
\typeRuleDef{T-VarDecl}{\infer{M;\Gamma \vdash_T T\ I_0\baseterm{,}\dots\baseterm{,}I_n\baseterm{;} B : \type{T}}{M;\Gamma,I_0:T,\dots,I_n: T \vdash_T B : \type{T}}} \\[\normalbaselineskip]

\typeRuleDef{T-VarDeclChE}{\infer{M;\Gamma \vdash_T \type{channel}\baseterm{[}T\baseterm{]}\ I_0\ \text{\bf withends }\baseterm{[}I_1\baseterm{,} I_2\baseterm{];} B : \type{T}}{M;\Gamma,I_0:\type{channel[}T\type{]}, I_1:\type{channelEnd[}T\type{]}, I_2:\type{channelEnd[}T\type{]} \vdash_T B : \type{T}}} \\[\normalbaselineskip]

\typeRuleDef{T-VarDeclAlF}{\infer{M;\Gamma \vdash_T I_0\ \baseterm{ aliasfor [}I_1\baseterm{,}\dots\baseterm{,} I_n\baseterm{];} B : \type{T}}
{\begin{array}{rl}
\forall i\in \{1,\dots,n\}: M;\Gamma &\vdash_T I_i : \type{T_i} \text{ where } \type{T_i} \text{ is a quantum type}, \\
M;\Gamma,I_0:\type{\bigotimes_{i=1}^n \type{T_i}} &\vdash_T B : \type{T}
\end{array}}} \\[\normalbaselineskip]
\end{array}$
\caption{Typing rules for variable declarations}
\label{fig:typRules:varDecl}
\end{figure}

\begin{figure}[p]
$\begin{array}{m{0.23\linewidth}c}
\typeRuleDef{T-Skip}{\overline{M;\Gamma \vdash_T \baseterm{;} : \void}} \\[\normalbaselineskip]

\typeRuleDef{T-PromoExpr}\infer{M;\Gamma \vdash_T PE\baseterm{;} : \void}{M;\Gamma \vdash_T PE : \type{T}} \\[\normalbaselineskip]


\typeRuleDef{T-Block}{\infer{M;\Gamma \vdash_T \baseterm{\{} B \baseterm{\}} : \void}{M;\Gamma \vdash_T B : \void}} \\[\normalbaselineskip]

\typeRuleDef{T-BlockHead}{\infer{M;\Gamma \vdash_T Be_0\ Be_1\ \dots\ Be_n : \void}{M;\Gamma \vdash_T Be_0 : \type{void} \quad M;\Gamma \vdash_T Be_1\ \dots\ Be_n : \void}  \\
\multicolumn{2}{r}{\text{where }Be_0 \neq VD}} \\[\normalbaselineskip]

\typeRuleDef{T-If}{\infer{M;\Gamma \vdash_T \baseterm{if}\ \baseterm{(}E\baseterm{)}\ S_0\ \baseterm{else}\ S_1  : \void}{M;\Gamma \vdash_T E : \type{bool} \quad M;\Gamma \vdash_T S_0 : \void \quad M;\Gamma \vdash_T S_1 : \void}}\\[\normalbaselineskip]

\typeRuleDef{T-While}{\infer{M;\Gamma \vdash_T \baseterm{while}\ \baseterm{(}E\baseterm{)}\ S : \void}{M;\Gamma \vdash_T E : \type{bool} \quad M;\Gamma \vdash_T S : \void}} \\[\normalbaselineskip]


\typeRuleDef{T-ReturnVoid}{\overline{M;\Gamma, @retVal : \void \vdash_T \baseterm{return;}\ : \void}} \\[\normalbaselineskip]

\typeRuleDef{T-ReturnExpr}{\infer{M;\Gamma \vdash_T \baseterm{return}\ E\baseterm{;}\ : \void}{M;\Gamma, @retVal : \type{T} \vdash_T E : \type{T}}} \\[\normalbaselineskip]

\typeRuleDef{T-Fork}{\infer{M;\Gamma \vdash_T \baseterm{fork}\ I\baseterm{(}\listOf{E}\baseterm{)}\baseterm{;}\ : \void}{M;\Gamma \vdash_T I\baseterm{(}\listOf{E}\baseterm{)} : \type{T} \quad \text{$I$ is a classical method}}} \\[\normalbaselineskip]

\typeRuleDef{T-Send}{\infer{M;\Gamma \vdash_T \baseterm{send}\baseterm{(}E_0, E_1\baseterm{);}\ : \void}{M;\Gamma \vdash_T E_0 : \type{channelEnd[T]} \quad M;\Gamma \vdash_T E_1 : \type{T}}} \\[\normalbaselineskip]

\end{array}$
\caption{Typing rules for statements}
\label{fig:typRules:stat}
\end{figure}

\begin{figure}[h]
$\begin{array}{m{0.2\linewidth}c}
\typeRuleDef{T-Var}{\overline{M;\Gamma, I:\type{T} \vdash_T I : \type{T}}} \\[0.5\normalbaselineskip]

\typeRuleDef{T-Value}{\overline{M;\Gamma \vdash_T \intVal{R}{C}{T} : \type{T}}} \\[0.5\normalbaselineskip]

\typeRuleDef{T-Bracket}{\infer{M;\Gamma \vdash_T \baseterm{(}E\baseterm{)} : \type{T}}{M;\Gamma \vdash_T E : \type{T}}} \\[\normalbaselineskip]

\typeRuleDef{T-Alloc}{\infer{M;\Gamma \vdash_T \baseterm{new}\ \type{T}\baseterm{()} : \type{T}}{\text{\type{T} is either a quantum type ($\type{Q_d}$) or a channel type ($\type{channel[T]}$)}}} \\[\normalbaselineskip]

\typeRuleDef{T-Assign}{\infer{M;\Gamma \vdash_T I = E : \type{T}}{M;\Gamma, I:\type{T} \vdash_T E : \type{T}}} \\[\normalbaselineskip]

\typeRuleDef{T-MethodCall}{\infer{M;\Gamma \vdash_T I\baseterm{(}E_0\baseterm{,}\dots\baseterm{,} E_n\baseterm{)} : \type{T}}{M_T(I) = \type{S}_0,\dots,\type{S}_n \rightarrow \type{T}\text{\quad}
M;\Gamma \vdash_T E_0 : \type{S}_0\text{\quad\dots\quad}M;\Gamma \vdash_T E_n : \type{S}_n}} \\
\multicolumn{2}{r}{\text{where $M=(M_T, M_H, M_B)$}} \\[\normalbaselineskip]

\typeRuleDef{T-Measurement}{\infer{M;\Gamma \vdash_T \baseterm{measure}\baseterm{(}E_0\baseterm{,} E_1\baseterm{,}\dots\baseterm{,} E_n\baseterm{)} : \type{int}}{
\begin{array}{rl}
M;\Gamma &\vdash_T E_0 : \type{MeasurementBasis}, \\
\forall i\in \{1,\dots,n\}: M;\Gamma &\vdash_T E_i : \type{T_i} \text{ where } \type{T_i} \text{ is a quantum type}
\end{array}
}} \\[\normalbaselineskip]

\typeRuleDef{T-Recv}{\infer{M;\Gamma \vdash_T \baseterm{recv}\baseterm{(}E\baseterm{)} : \type{T}}{M;\Gamma \vdash_T E : \type{channelEnd[T]}}} \\[\normalbaselineskip]
\end{array}$
\caption{Typing rules for expressions}
\label{fig:typRules:expr}
\end{figure}


  \item $M_B(m)$ which returns the \refConcept{method body} represented using \hyperlink{intSynt}{internal syntax} for a \refConcept{method} $m$.
 \end{itemize}

We provide typechecking rules in Figures \ref{fig:typRules:prog}, \ref{fig:typRules:varDecl}, \ref{fig:typRules:stat} and \ref{fig:typRules:expr}.
These rules use a \concept{typing context} which is a pair $(M;\Gamma)$ consisting of:
\begin{itemize}
 \item $M$ is a \refConcept{method typing context},
 \item $\Gamma$ is a \concept{variable typing context} -- a partial mapping $\Gamma: \refConcept{Names} \rightharpoonup \refConcept{Types}$ that assigns a type to a variable name. We write a variable typing context $\Gamma$ as $\Gamma = a_1:\type{T_1},\dots,a_n:\type{T_n}$ meaning that the type $\type{T}_i$ is assigned to the variable $a_i$.
 
 The extension of a context $\Gamma$ by a variable $b$ of type $\type{T}_b$ is written as $\Gamma, b:\type{T}_b$. It is undefined if $\Gamma(b)$ is defined and $\Gamma(b) \neq \type{T}_b$. Otherwise it is defined as:
  $$(\Gamma, b:\type{T}_b)(x) = \begin{cases}
                               \type{T}_b & \text{if } x=b, \\
                               \Gamma(x)	& \text{otherwise.}
                              \end{cases}$$
 \end{itemize}

Let $P$ be a \refConcept{program} whose internal representation is stored in a \refConcept{method typing context} $(M_T, M_H, M_B)$. We call this program \concept[well-typed program]{well-typed} if the premise of the rule \typeRule{T-Program} is satisfied for this method typing context. This check is passed iff all the declared methods satisfy the typing rule \typeRule{T-Method}.

Typechecking of a method in the rule \typeRule{T-Method} is a little more complicated. The reason is that we require any method $m$ whose \refConcept{return type} is not \void{} to return a value of appropriate type in all possible control paths which the evaluation of this method can take. The return type is for the sake of typechecking stored in the formal variable $@retVal$.

This can be checked at compile time. We split this requirement into two: 
\begin{enumerate}
 \item[(1)] during evaluation, the method $m$ body always reaches a \return{}statement (or invokes a \refConcept{runtime error} or diverges), and
 \item[(2)] any value returned by a \return{} statement during evaluation of the method $m$ body is of appropriate type. This is checked by typing rules \typeRule{T-ReturnVoid} and \typeRule{T-ReturnExpr} in cooperation with \typeRule{T-Method}.
\end{enumerate}
For formal definition of the condition (1), we define a predicate $\refConcept{RetOk}$.

\begin{definition}
Let $B$ be a \refConcept[syntax!block-forming statement]{block-forming statement}. We define a predicate \concept[RetOk]{$RetOk$} as:

\begin{eqnarray*}
RetOk(B) & = & \begin{cases}
\mathbf{true} & \text{if $B = \baseterm{return}\ $E$\baseterm{;}$} \\
RetOk(S_t) \wedge RetOk(S_e) & \text{if $B = \IF\ \baseterm{(}E\baseterm{)}\ S_t\ \ELSE\ S_e$} \\
\bigvee_{Be_i} RetOk(Be_i) & \text{if $B = Be_0\ Be_1\ \dots\ Be_n$} \\
RetOk(B') & \text{if $B = \baseterm{\{}\ B'\ \baseterm{\}}$} \\
\mathbf{false} & \text{otherwise}
\end{cases}
\end{eqnarray*}
\end{definition}

This predicate does not handle the case $B = \text{\bf while ($E$) $S$}$ because evaluation of the condition $E$ is undecidable at compile-time. Hence even for the straightforwardly always-terminating case $B = \text{\bf while (true) return $1$;}$, $\refConcept{RetOk}(B)$ is not satisfied. Therefore the predicate is only approximate.

Later we prove a lemma stating that if the predicate $\refConcept{RetOk}$ is satisfied on $B$ then any control path of evaluation of $B$ reaches a \baseterm{return;} or \baseterm{return} $E$\baseterm{;} statement, or a \refConcept{runtime error}, or diverges (see Lemma \ref{lemma-eval-retok-to-return}). Thus, if $B$ is a method $m$ body and $RetOk(B)$ is satisfied, the evaluation of method can either reach some {\bf return} statement, lead to a \refConcept{runtime error}, or diverge.

\begin{definition}
We call a method $m$ \concept[well-typed method]{well-typed} if the premises of rule \typeRule{T-Method} are satisfied for this method.
\end{definition}

\begin{remark}
Note that if a method $m$ is \refConcept[well-typed method]{well-typed} and its \refConcept{return type} is not \void{} then its \refConcept[method body]{body} contains only \baseterm{return} $E$\baseterm{;} statements, {\it ie.} no \baseterm{return;} statements.
\end{remark}

The rest of the typing rules is usual: The rules for typechecking variable declarations in Figure \ref{fig:typRules:varDecl} check the \refConcept[syntax!block-forming statement]{block-forming statement} can be typechecked with the variable context extended with the newly declared variables.

We formally regard all statements to be of type \void{} what is seen in the typechecking rules in Figure \ref{fig:typRules:stat}. These rules are quite usual up to the rule \typeRule{T-Fork}. This rule requires that the method, which should be a new process run from, is classical. This is natural requirement as running a new process, which is by its nature a classical object from a quantum operator, would be a nonsense.

The typechecking rules for expressions shown in Figure \ref{fig:typRules:expr} are designed as usual.

\section{Basic concepts}
\label{sec:defs}

Before we continue with formal definition of the semantics, we must define several useful functions and structures. First we define notation used in the rest of the article.

\subsection{Notation}

\begin{notation}
Let $S$ be a set, $\bot \notin S$. Then $S_\bot = S \cup \{\bot\}$. We denote a set of natural numbers with zero $\N \cup \{0\}$ by $\N_0$.
\end{notation}

\begin{definition}
Let $S$ be a set. An \concept{$S$-list} $s = [s_1,\dots,s_n]$ is a list where $n\in \N_0$ and $s_1,\dots,s_n \in S$. Set of all finite $S$-lists $\{s \ | \ s \text{ is a finite } S\text{-list}\}$ is denoted by $S_{[]}$.
\end{definition}

\begin{definition}
Let $m,n \in \N_0$. Let $L = [l_1,\dots,l_n], K = [k_1,\dots,k_m]$ be lists. Then $|L|$ is a \concept[list length]{length of a list $L$}, $|L| = n$. Concatenation of lists $L$ and $K$ is defined as $L \cdot K = [l_{1},\dots,l_{n},k_{1},\dots,k_{m}]$. Set of list $L$ elements is defined as $\set(L) = \{l_1,\dots,l_n\}$.
\end{definition}

\subsection{Reference-related concepts}

We use the following specially formed lists for storing references to quantum systems.

\begin{definition}
\noindent Let $S$ be a set. We define a \concept{recursive $S$-list} recursively as:
\begin{itemize}
 \item Any $S$-list $[s_1,\dots,s_k]$ is a recursive $S$-list,
 \item A list $[e_1,\dots,e_m]$ is a recursive $S$-list for any $m \in \N_0$ if $e_1,\dots,e_m$ are recursive $S$-lists.
\end{itemize}
\end{definition}

For example, $[[[1,2,3],[2,3]],[1]]$ and $[]$ are recursive $\N$-lists.

Recursive $S$-lists are used for the representation of quantum system references in the following way:

A reference to a quantum system, be it compound or not, is specified by a recursive $\N$-list. Quantum systems are stored in indexed registers in the quantum memory, one quantum system per one register. The (unique) index is assigned to a quantum system when it is allocated. The reference to the system with index $n$ is a recursive $\N$-list $[n]$.

Let us have two quantum systems $\phi$ and $\psi$ whose indices are $1$ and $2$, respectively. The references to these quantum systems are $r_\phi=[1]$ and $r_\psi=[2]$, respectively. A reference to a compound system $\rho$ consisting of the two quantum systems $\phi$ and $\psi$ is then a recursive $\N$-list $r_\rho = [r_\phi, r_\psi] = [[1],[2]]$. Note that the structure of $\rho$, {\it ie.} that it consists of {\em two} systems, corresponds to the structure of the reference $r_\rho$ -- it is built up from {\em two} elements.

\begin{notation}
The set of all finite recursive $S$-lists $\{s \ | \ s \text{ is a finite recursive } S\text{-list}\}$ is denoted by $\reclist{S}$.
\end{notation}

Recursive $S$-lists allow us to nicely capture quantum system structure in the reference. However, when working with referred quantum systems, {\it eg.} applying some unitary operator, we do not want to bother with the structure -- we only need a list of indices of the affected quantum systems. To get such a {\em linearized} list out of the structured one, we define the following function:

\begin{definition}
For a recursive $S$-list $s \in \reclist{S}$, we define a function $\concept[linearize]{\linearize}: \reclist{S}\rightarrow S_{[]}$ which converts a recursive $S$-list into an $S$-list:
\begin{eqnarray*}
\hline\\
 \linearize(s) &\isdef &s\text{ if }s \in S_{[]} \\
 \linearize([e_1,\dots,e_m]) &\isdef &\linearize(e_1)\cdot{\dots}\cdot\linearize(e_m) \text{ where } e_1,\dots,e_m \in \reclist{S}\\
 \hline
\end{eqnarray*}
\end{definition}

If a reference $R$ contains a special element $\bot$ which denotes a reference to nonexistent quantum system, we consider the reference $R$ itself to refer to nonexistent quantum system. Hence linearization of such a reference returns $\bot$:

\begin{definition}
For a set $S_\bot$, we define a function $\concept[linearize_bot]{\linearize_\bot}: \reclist{(S_\bot)} \rightarrow (S_\bot)_{[]} \cup \{\bot\}$ as:
\begin{eqnarray*}
\hline\\
 \linearize_\bot(s) &\isdef &\begin{cases}
\linearize(s) = [s_1,\dots,s_n] & \text{if } s_i \neq \bot \text{ for all } 1 \leq i \leq n \\
\bot & \text{otherwise}
\end{cases}\\
 \hline
\end{eqnarray*}
\end{definition}

\begin{definition}
Let $S$ be a set. We define a function $\recfunc{\set}: \reclist{S} \rightarrow S$ for getting the set of recursive list items regardless of the structure as $\recfunc{\set}(l) \isdef \set(\linearize(l))$. We also define function $\recfunc{|-|}: \reclist{S} \rightarrow \N$ for getting length of a recursive list regardless of the structure as $\recfunc{|l|} \isdef |\linearize(l)|$.
\end{definition}

\subsection{Variable-related concepts}

We use partial functions to capture variable properties, {\it eg.} mapping a variable name to a place in memory where the variable value is stored, or a mapping to the variable type. We define several useful functions for handling these partial functions describing variables.

Adding a new variable to a set of known variables is represented by extending the domain of the appropriate partial function with the new variable name. We call this function an {\em update}. Sometimes we only want to update a variable property if the updated variable is already contained in the domain of the updated function, {\it eg.} change a memory reference referred by the variable. We call such an operation an {\em replacement}. In general, we define these functions in the following way:

\begin{definition}
Let $f: X \rightharpoonup Y$ be a partial function. For $x\in X, y\in Y$, we define \concept{replacement} $f[x \mapsto y]: X \rightharpoonup Y$ and \concept{update} $f[x \mapsto y]_+: X \rightharpoonup Y$ as:
$$f[x \mapsto y](z) = \begin{cases}
y & \text{if $x=z$ and $f(x)$ is defined,} \\
f(z) & \text{otherwise,}
\end{cases}$$
and 
$$f[x \mapsto y]_+(z) = \begin{cases}
y & \text{if $x=z$,} \\
f(z) & \text{otherwise.}
\end{cases}$$
\end{definition}

Note that the $f[x\mapsto y](x)$ is defined iff $f(x)$ is defined while $f[x\mapsto y]_+(x)$ is defined even if $f(x)$ is not defined.\footnote{The $+$ sign in function index means ``add the mapping from $x$ to $y$ even if it was not defined yet''.}

We need to store different variable properties: variable type, names of channel ends corresponding to given channel {\it etc.} Each property is represented by a separate partial mapping. Hence variable properties are described by a tuple of such partial functions.

\begin{definition}
A \concept{partial function tuple} is a tuple $f = (f_0,\dots,f_n)$ where $f_0,\dots,f_n$ are partial functions.
\end{definition}

We will need to capture variable scope during a method evaluation. For this reason, we define concepts of {\em \refConcept{list of partial function tuples}} and {\em \refConcept{list of lists of partial function tuples}}. Their usage is in more detail explained in Subsection \ref{subsec:varstack}. We also extend update and replacement functions to \refConcept[list of partial function tuples]{lists of partial function tuples} and \refConcept[list of lists of partial function tuples]{lists of lists of partial function tuples}.

\begin{definition}
We define \concept{list of partial function tuples} recursively as:

\begin{itemize}
 \item \concept[square]{}$\square$ is an (empty) list of partial function tuples,
 \item $[K\circ_L f]$ is a list of partial function tuples if $K$ is a list of partial function tuples and $f$ is a \refConcept{partial function tuple}. The symbol $\circ_L$ serves as an element separator only.\footnote{$L$ in $\circ_L$ stands for {\em local}.}
\end{itemize}
\end{definition}

\begin{definition}
For a \refConcept{list of partial function tuples} $K$, we define a \concept[replacement!list of partial function tuples]{replacement} $K[x \mapsto y]_i$ and an \concept[update;list of partial function tuples]{update} $K[x \mapsto y]_{+,i}$ of outermost $x$ in an $i$-th partial function of a \refConcept{list of partial function tuples} $K$ as:

\begin{eqnarray*}
\hline\\\relax
[K \circ_L (f_0,\dots,f_n)][x \mapsto y]_i &\isdef &\begin{cases}
 [K \circ_L (f_0,\dots,f_i[x \mapsto y],\dots,f_n)] & \text{if }f_i(x)\text{ is defined} \\
 [K[x\mapsto y]_i \circ_L (f_0,\dots,f_n)] & \text{otherwise}
\end{cases}\\\relax
\refConcept[square]{\square}[x\mapsto y]_i &\isdef &\refConcept[square]{\square} \\
\\\relax
[K \circ_L (f_0,\dots,f_n)][x \mapsto y]_{+,i} &\isdef &[K \circ_L (f_0,\dots,f_i[x \mapsto y]_+,\dots,f_n)]\\\relax
\refConcept[square]{\square}[x\mapsto y]_{+,i} &\isdef &\refConcept[square]{\square} \\
\hline
\end{eqnarray*}
\end{definition}

\begin{definition}
We define a \concept{list of lists of partial function tuples} recursively as:
\begin{itemize}
 \item \concept[blacksquare]{}$\blacksquare$ is an (empty) list of lists of partial function tuples,
 
 \item $[L_1\circ_G K]$ is a list of lists of partial function tuples if $L_1$ is a list of lists of partial function tuples and $K$ is a \refConcept{list of partial function tuples}. The symbol $\circ_G$ serves as an element separator only.\footnote{$G$ in $\circ_G$ stands for {\em global}.}
\end{itemize}
\end{definition}

\begin{definition}
For a \refConcept{list of lists of partial function tuples} $L$, a {\em replacement} $L[x \mapsto y]_i$ and an {\em update} $L[x \mapsto y]_{+,i}$ of mapping of $x$ in the outermost \refConcept{list of partial function tuples} is defined as:

\begin{eqnarray*}
\hline\\\relax
[L_1 \circ_G K][x \mapsto y]_i &\isdef & [L_1 \circ_G K[x\mapsto y]_i] \\\relax
\refConcept[blacksquare]{\blacksquare}[x\mapsto y]_i &\isdef &\refConcept[blacksquare]{\blacksquare} \\
\\\relax
[L_1 \circ_G K][x \mapsto y]_{+,i} &\isdef & [L_1 \circ_G K[x\mapsto y]_{+,i}] \\\relax
\refConcept[blacksquare]{\blacksquare}[x\mapsto y]_{+,i} &\isdef &\refConcept[blacksquare]{\blacksquare} \\
\hline
\end{eqnarray*}
\end{definition}

Last, we define a coalesce\footnote{The name {\em coalesce} is given because this function is similar to the COALESCE function defined in SQL-92 standard.} function $*$ of two partial functions $f,g: A\rightharpoonup B$. Coalesce $g*f$ is a partial function which for $(g*f)(x)$ results into $f(x)$ if $f(x)$ is defined, otherwise to $g(x)$:

\begin{definition}
Let $f,g: A\rightharpoonup B$ be partial functions. We define a \concept{coalesce of $g$ and $f$} $\concept{g*f}: A\rightharpoonup B$ as: $$(g*f)(x) = \begin{cases} f(x) &\text{if $f(x)$ is defined} \\ g(x) &\text{otherwise.} \end{cases}$$
\end{definition}

\begin{definition}
We define $\concept{Names}$ to be a set of all identifier names.
\end{definition}

\section{Operational semantics}
\label{sec:opSem}

In this section, we define the operational semantics of the LanQ programming language.

\subsection{Memory model}
\label{opSem:memModel}

In this subsection, we describe the memory model used in LanQ implementation.

\begin{figure}[ht]
 \centerline{\includegraphics[width=0.7\linewidth]{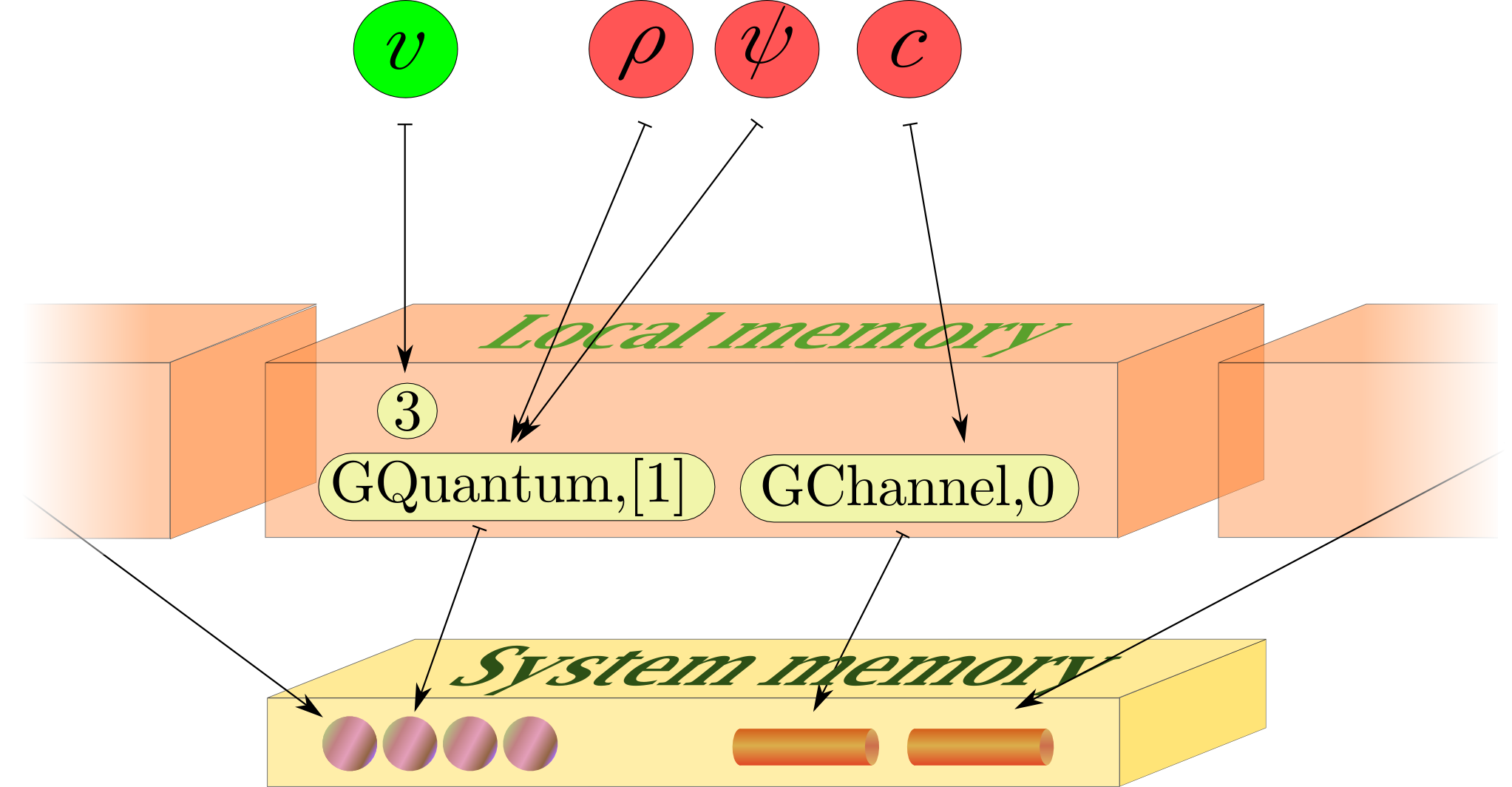}}
 \caption{Memory model of LanQ. Several processes are currently running, their memories are shown as red boxes. Variable names are shown in a circle, duplicable variables are shown in green, nonduplicable ones in red. The variables $v$ (integer), $\rho$ (quantum system), $\psi$ (quantum system) and $c$ (channel) belonging to one process refer to places in its local memory shown as yellow containers. The containers refer to values -- a value 3 in the case of variable $v$ container and references to the system memory in the rest of the cases.}
 \label{fig:memModel}
\end{figure}

Our model abstract machine uses a memory to store values. As we work both with duplicable and nonduplicable data, we have two types of memory: \concept[system memory]{system} and \concept[]{local}. All processes manage their own memory -- a \concept{local process memory} where duplicable values are stored. System manages the system memory where nonduplicable resources are stored. Processes cannot access the system memory directly, they work with resources by means of references to the system memory. This is transparent to the programmer. Memory model is depicted in Figure \ref{fig:memModel}.

This memory model allows us to assign a reference to the same nonduplicable resource to many different variables. It also allows simple definition of communication of resources -- the \refConcept{global reference} to system memory is unmapped from the sending process and moved to the receiver.

It is however also a possible source of \refConcept[runtime error]{runtime errors}. Consider a situation when a process sends away a qubit. Then the mapping from its \refConcept{local process memory} to the place \refConcept{system memory} where the qubit physically exists is removed but the mapping from variables to the \refConcept{local process memory} is preserved. If a process then tries to use the value of such a variable, a \refConcept{runtime error} signalling uninitialized variable usage occurs.%
\footnote{Note that the same behaviour would be exhibited if we unmapped the references directly from variables to local memory -- in that case we would have to deinitialize all the variables referring to the resource (this operation would however cost more time than the previously described one) and the issue with uninitialized variables remains.}

A memory reference specifies a position of a value in the memory. We distinguish two basic types of references:
\begin{itemize}
 \item \concept[local reference]{Local references} specify places in a \refConcept{local process memory} where variable values are stored. All variables used during evaluation are assigned local references. A special \refConcept{local reference} \hypertarget{ref:none}\dispnone{} refers to no value. We have four kinds of \refConcept[local reference]{local references}:
 
   \begin{itemize}
\item References to local classical value memory: $\concept{Ref_{Cl}} = \{\none\} \cup (\{Classical\} \times \N_\bot)$,
\item References to local quantum systems reference memory: \\$\concept{Ref_Q} = \{\none\} \cup (\{Quantum\} \times \reclist{(\N_\bot)})$,
\item References to local channel reference memory: $\concept{Ref_{Ch}} = \{\none\} \cup (\{Channel\} \times \N)$,
\item References to local channel end reference memory: \\$\concept{Ref_{ChE}} = \{\none\} \cup (\{ChannelEnd_0, ChannelEnd_1\} \times \N)$.
   \end{itemize}
 The set of local references is defined as: $$\concept{Ref_L} = \refConcept{Ref_{Cl}} \cup \refConcept{Ref_Q} \cup \refConcept{Ref_{Ch}} \cup \refConcept{Ref_{ChE}}.$$
 
 \item \concept[global reference]{Global references} refer to resources stored in \refConcept{system memory}.
  \begin{itemize}
\item References to system channel memory: $\concept{Ref_{GCh}} = \{\bot\} \cup (\{GChannel\} \times \N_\bot)$,
\item References to system quantum memory: $\concept{Ref_{GQ}} = \{\bot\} \cup (\{GQuantum\} \times\N_{[]})$.
   \end{itemize}
 The set of global references is defined as: $$\concept{Ref_G} = \refConcept{Ref_{GCh}} \cup \refConcept{Ref_{GQ}}.$$
\end{itemize}

A set of all references is defined as: $$\concept{Ref} = \refConcept{Ref_G} \cup \refConcept{Ref_L},$$ and a set of references to nonduplicable values is denoted by: $$\concept{Ref_{nd}} = \refConcept{Ref_Q} \cup \refConcept{Ref_{Ch}} \cup \refConcept{Ref_{ChE}}.$$

\begin{definition}
\concept[type!\type{Ref}]{}A \concept{memory reference} is an element from the set $\refConcept{Ref}$. We define \type{\refConcept{Ref}} to be the type of memory references.
\end{definition}

\begin{remark}
Note that a memory reference is a classical value, therefore $\type{\refConcept[type!\type{Ref}]{Ref}} \in \refConcept{Types}$.
\end{remark}

\subsection{Variable properties storage}
\label{subsec:varstack}

In this subsection, we informally introduce structure used for handling variable properties.

\refConcept[variable properties]{Variable properties} is a structure where properties of variables neccessary for correct handling the variables are stored while respecting their scope: The actually running method has access only to variables declared in this method, it cannot access any variable from the caller method. Moreover, the validity of a variable is limited to the block in which the variable is declared.

The properties of a variable are formally described later in Subsection \ref{opSem:configuration}. They are represented by partial functions which, given a variable name, return:
\begin{itemize}
 \item A reference to \refConcept{local process memory} where the value of the variable is stored,
 \item Variable names representing individual ends of the channel (if the given variable represents a channel),
 \item List of variable names representing quantum systems which are subsystems of the compound quantum system (if the given variable represents a compound quantum system),
 \item A type of the given variable.
\end{itemize}

Therefore we have a quadruple of partial functions that represent variable properties: $f = (f_{var},f_{ch},f_{qa},f_{type})$. Indeed, this quadruple is not enough for handling variable scope.

Respecting a variable scope is achieved by using \refConcept[list of partial function tuples]{lists of partial function tuples} and \refConcept[list of lists of partial function tuples]{lists of such lists}:
\begin{itemize}
\item {\it A variable can be accessed only from within the block where it was declared.} This is ensured by using a \refConcept{list of partial function tuples} (separated by $\circ_L$), where a new tuple is appended to the list when a block is started and removed when the block ends,

\item {\it Only variables from the currently running method are accessible to this method.} This is ensured by using a \refConcept{list of lists of partial function tuples} (separated by $\circ_G$) where a new \refConcept{list of partial function tuples} is appended when a method is invoked, and removed when a method finishes.
\end{itemize}

We show the manipulation with a \refConcept{variable properties} structure on an example. Consider the following method $a$:

\noindent$\begin{array}{ll}
\\
	& \type{int}\ a\baseterm(\type{int}\ c\baseterm)\\
\text{\footnotesize 1} & \baseterm\{  \\
\text{\footnotesize 2} & $\qquad$ \type{bool}\ b\baseterm;  \\
\text{\footnotesize 3} & $\qquad$ b = {\bf true}\baseterm;  \\
\text{\footnotesize 4} & $\qquad$ {\bf if}\ (b)\ \baseterm\{  \\
\text{\footnotesize 5} & $\qquad \qquad$ \type{int}\ i\baseterm; \\ 
	& $\qquad \qquad$ \vdots \\
\text{\footnotesize 6} & $\qquad$ \baseterm\} \\
\text{\footnotesize 7} & $\qquad$ \return\ 3+c\baseterm; \\
\text{\footnotesize 8} & \baseterm\}	 \\
\\
\end{array}$

We show the \refConcept{variable properties} construction as the individual lines of the method are executed. As the formal notation of the \refConcept{variable properties} is not well-readable, we also provide the reader with its graphical representation. The representation uses the following notation:

\begin{itemize}
 \item A \refConcept{variable properties} tuple $(f_{var},f_{ch},f_{qa},f_{type})$ is represented as: {\small $\begin{pmatrix}
f_{var} \\ \hline f_{ch} \\ \hline f_{qa} \\ \hline f_{type}
\end{pmatrix}$},
 \item A \refConcept[list of partial function tuples]{list of variable properties} $[K \circ_L f]$ is represented as: $\begin{array}{|c|}\hline f \\ K \\ \hline\end{array}$,
 \item A \refConcept[list of partial function tuples]{list of lists of variable properties} $[L_1 \circ_G K]$ is represented as: \raisebox{-15pt}{\includegraphics{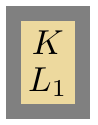}}.
\end{itemize}

\begin{figure}
\subfigure[Before execution of line 1]{\label{fig:vps-0}\includegraphics[scale=0.75]{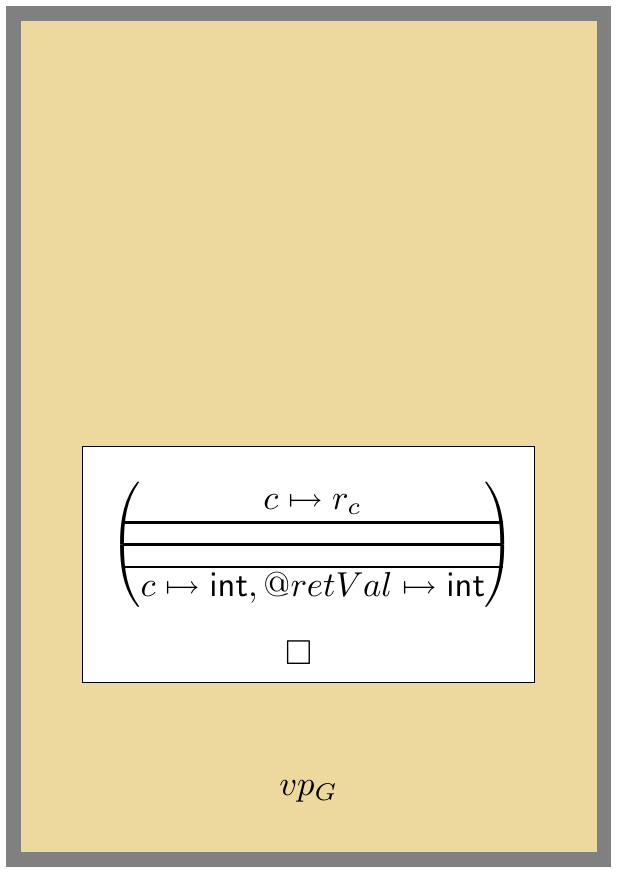}}
\hfill
\subfigure[After execution of line 1]{\label{fig:vps-1}\includegraphics[scale=0.75]{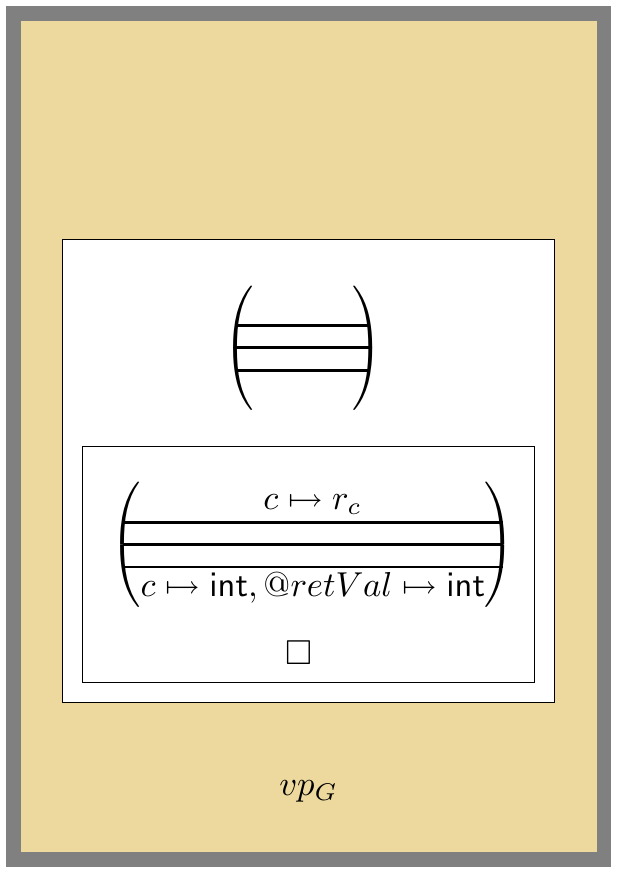}}
\hfill
\subfigure[After execution of line 2]{\label{fig:vps-2}\includegraphics[scale=0.75]{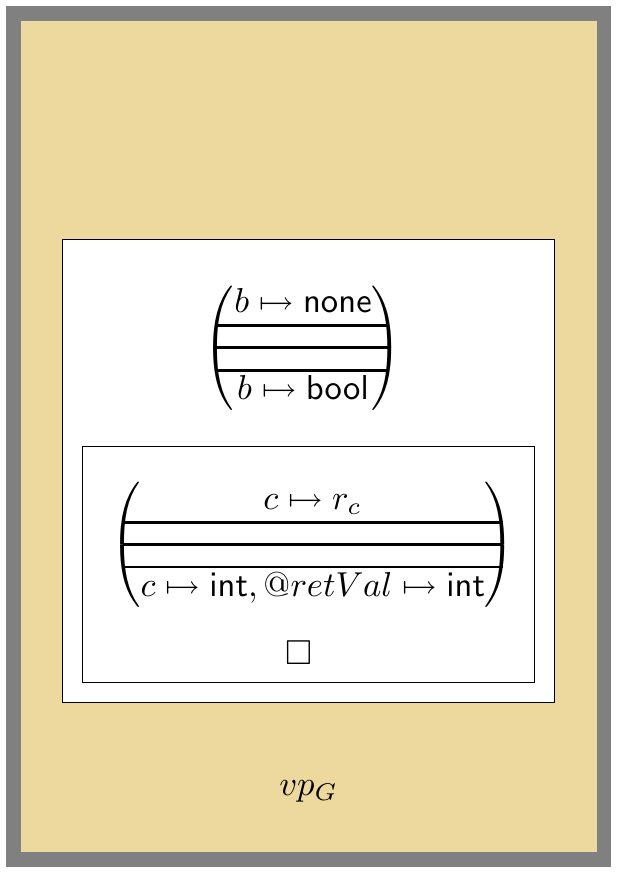}}

\subfigure[After execution of line 3]{\label{fig:vps-3}\includegraphics[scale=0.75]{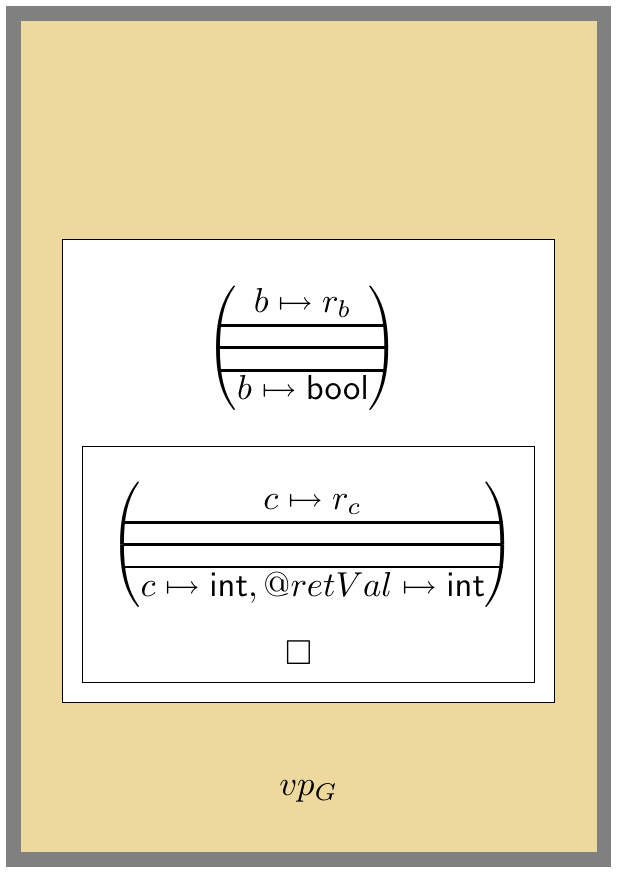}}
\hfill
\subfigure[After execution of line 4]{\label{fig:vps-4}\includegraphics[scale=0.75]{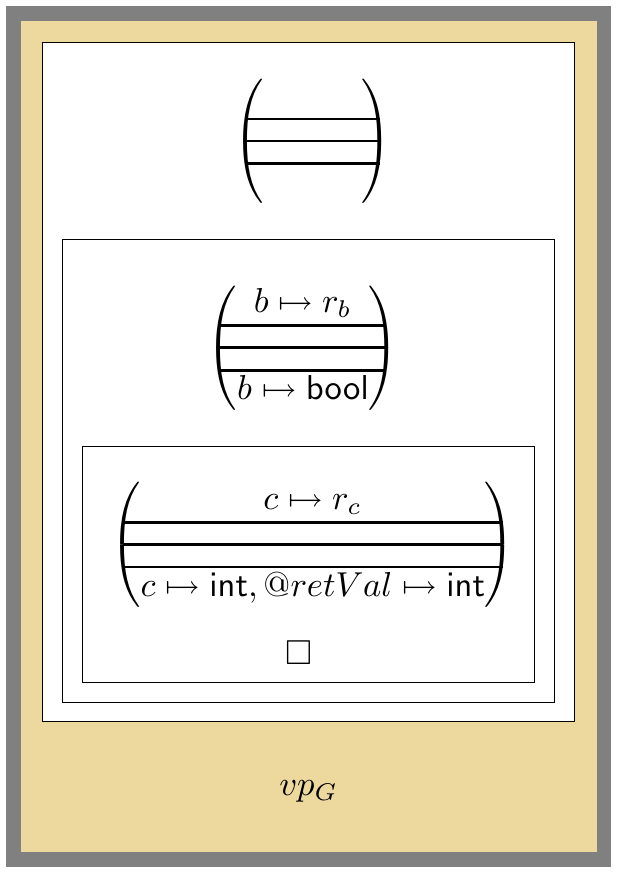}}
\hfill
\subfigure[After execution of line 5]{\label{fig:vps-5}\includegraphics[scale=0.75]{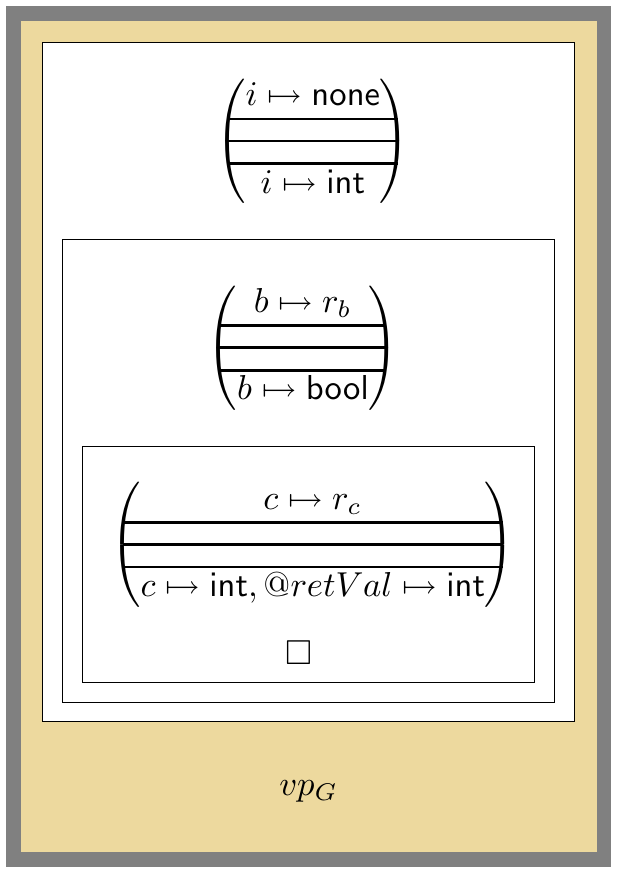}}

\subfigure[After execution of line 6]{\label{fig:vps-6}\includegraphics[scale=0.75]{varPropsPict/varPropsPict-fig5}}
\hfill
\subfigure[After execution of line 7]{\label{fig:vps-7}\includegraphics[scale=0.75]{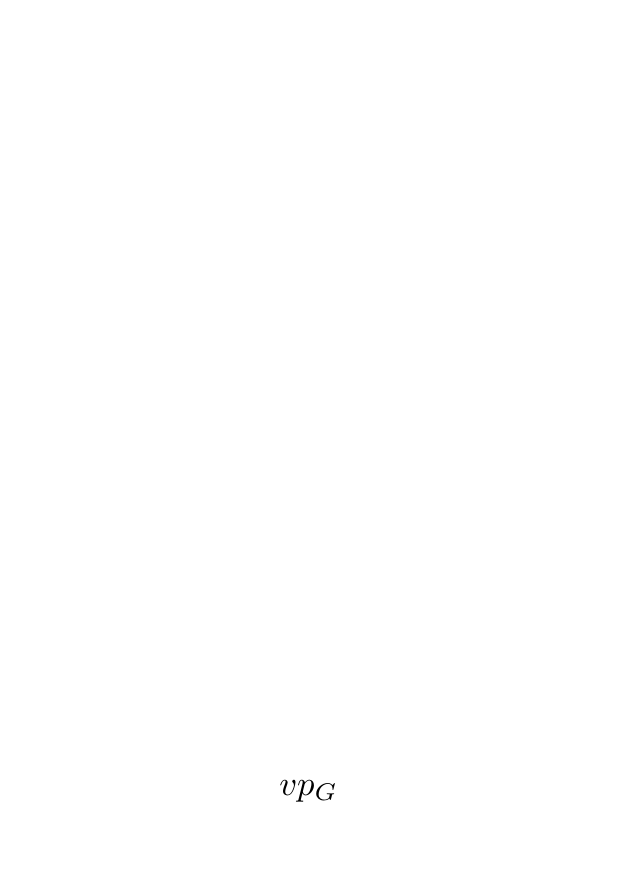}}
\hfill
\includegraphics[scale=0.75]{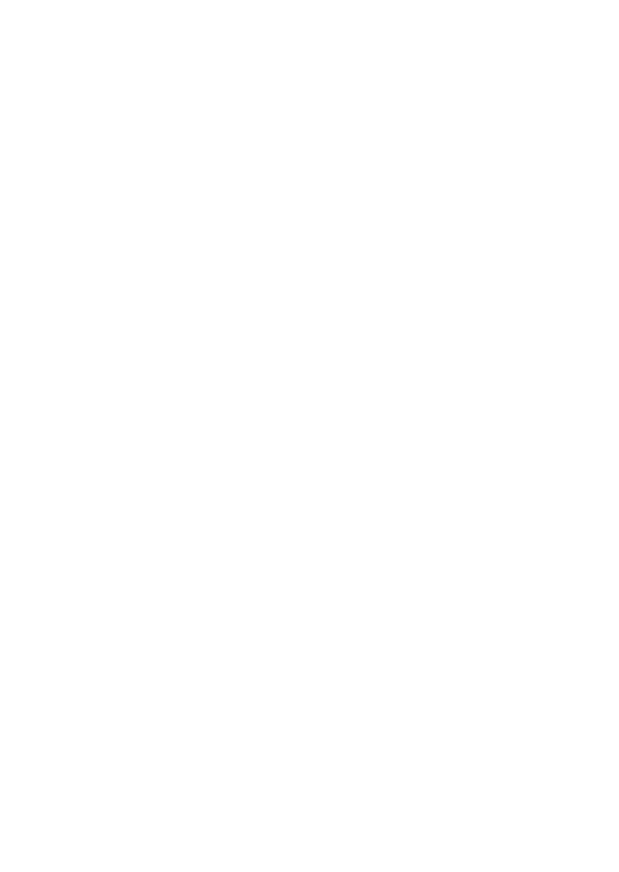}
\caption{Variable properties stack construction when invoking the method $a$}
\end{figure}

We assume that the original \refConcept{variable properties} were $vp_G$ right before calling the method $a$. When the method $a$ is called, a list of variable properties $[\refConcept[square]{\square} \circ_L ([c\mapsto r_c],[],[],[c\mapsto\type{int},@retVal\mapsto\type{int}])]$ is appended to $vp_G$: $$[vp_G\ \circ_G\ [\refConcept[square]{\square} \circ_L ([c\mapsto r_c],[],[],[c\mapsto\type{int},@retVal\mapsto\type{int}])]]$$ (see Figure \ref{fig:vps-0}). In this appended list, method parameters values are passed to the called method; in our case, the method parameter $c$ refers to the memory as set by the reference $r_c$.

On line 1, a new block is started, therefore a new \refConcept{empty variable properties tuple} $\lozenge$ is appended: $$[vp_G \circ_G [[\refConcept[square]{\square} \circ_L ([c\mapsto r_c],[],[],[c\mapsto\type{int},@retVal\mapsto\type{int}])]\ \circ_L\ \lozenge]]$$ (see Figure \ref{fig:vps-1}).
On the next line, a variable $b$ is declared, hence the head element of the inner list is modified: $$[vp_G \circ_G [[\refConcept[square]{\square} \circ_L ([c\mapsto r_c],[],[],[c\mapsto\type{int},@retVal\mapsto\type{int}])] \circ_L ([b\mapsto \none],[],[],[b\mapsto\type{bool}])]]$$ (see Figure \ref{fig:vps-2}).
On line 3, $b$ is assigned a value {\bf true} which is stored in the memory in a place referred by a reference $r_b$. This modifies $f_{var}$ element of the appropriate \refConcept{variable properties} tuple: $$[vp_G \circ_G [[\refConcept[square]{\square} \circ_L ([c\mapsto r_c],[],[],[c\mapsto\type{int},@retVal\mapsto\type{int}])]\ \circ_L ([b\mapsto r_b],[],[],[b\mapsto\type{bool}])]]$$ (see Figure \ref{fig:vps-3}).
On line 4, a new block is started, therefore again a new \refConcept{empty variable properties tuple} $\lozenge$ is appended: $$[vp_G \circ_G [[[\refConcept[square]{\square} \circ_L ([c\mapsto r_c],[],[],[c\mapsto\type{int},@retVal\mapsto\type{int}])]\ \circ_L ([b\mapsto r_b],[],[],[b\mapsto\type{bool}])] \circ_L \lozenge]]$$ (see Figure \ref{fig:vps-4}).
On line 5, a new integer variable $i$ is declared what is reflected in the inner list head \refConcept{variable properties tuple}: $$\begin{array}{l}
[vp_G \circ_G [[[\refConcept[square]{\square} \circ_L ([c\mapsto r_c],[],[],[c\mapsto\type{int},@retVal\mapsto\type{int}])]\ \circ_L ([b\mapsto r_b],[],[],[b\mapsto\type{bool}])]\ \circ_L\\
\hfill ([i \mapsto \none],[],[],[i\mapsto\type{int}])]]\end{array}$$ (see Figure \ref{fig:vps-5}).
On line 6, the block ends, hence the appropriate \refConcept{variable properties tuple} is discarded: $$[vp_G \circ_G [[\refConcept[square]{\square} \circ_L ([c\mapsto r_c],[],[],[c\mapsto\type{int},@retVal\mapsto\type{int}])]\ \circ_L ([b\mapsto r_b],[],[],[b\mapsto\type{bool}])]]$$
 (see Figure \ref{fig:vps-6}).
Finally on line 7, the method execution ends, hence all local \refConcept[variable properties tuple]{variable properties tuples} are discarded and the original \refConcept{variable properties} structure is restored: $$vp_G$$ (see Figure \ref{fig:vps-7}).

\subsection{Configuration}
\label{opSem:configuration}

In this subsection, we formally define abstract machine configuration which is later used for the definition of LanQ operational semantics.

A \concept{configuration} of the abstract machine used for the definition of LanQ operational semantics is composed of two basic parts -- {\em global} and {\em local}. The global part of the configuration stores information about resources -- a quantum state of the whole system and a relation between channels and their ends. The local part of the configuration stores information about individual processes -- the state of their local memory, variables, and terms to be evaluated.

A configuration $C$ describing $n$ processes running in parallel is written as: $$C = [gs\,\vert\,ls_1 \parallel\dots\parallel ls_n]$$ where $gs$ is the global part of the configuration and $ls_j$ represents the local process configuration of $j$-th process.

The components of the abstract machine configuration are formally defined as follows:

\begin{itemize}
 \item Global part of the configuration: a pair $(Q, C)$ where:
 \begin{itemize}
  \item $Q$ describes the state of quantum particles and their dimensions.

   In the present paper, we represent the quantum state of the system by a pair $(\rho,L)$ of a finite density matrix $\rho$ and a finite list $L$ of natural numbers. The list $L$ represents dimensions of individual quantum subsystems. The order of the list elements is given by order of quantum system allocations.
  \item $C$ represents the channel part of the configuration.

   Channels and their ends are stored as pairs $(c_0, c_1)$ written as \channel(c_0,c_1) where $c_0$ and $c_1$ represent individual ends of one channel.
 \end{itemize}

  \item Local part of the configuration: it defines state of one process, hence we call it a \concept{local process configuration}. It is a triplet $(lms, vp, ts)$ where:
  \begin{itemize}
   \item \concept[local memory state]{Local memory state} $lms$ is a quadruple of partial functions, $lms = (lms_{Cl}, lms_Q, lms_{Ch}, lms_{ChE})$ which stores the state of classical memory and references to non-duplicable resources available to the process:
	\begin{itemize}
	\item $\concept{lms_{Cl}}: \refConcept{Ref_{Cl}} \rightharpoonup \refConcept{Values}$ is a partial function which returns a (classical) value stored at the given position in memory. The set of all such partial functions is denoted by $\concept{LMS_{Cl}}$,
	\item $\concept{lms_Q}: \refConcept{Ref_Q} \rightharpoonup \refConcept{Ref_{GQ}}$ returns a global reference to quantum systems given by the local quantum reference. The set of all such partial functions is denoted by $\concept{LMS_Q}$,
	\item $\concept{lms_{Ch}}: \refConcept{Ref_{Ch}} \rightharpoonup \refConcept{Ref_{GCh}}$ returns a global reference to the channel given by the local channel reference. The set of all such partial functions is denoted by $\concept{LMS_{Ch}}$,
	\item $\concept{lms_{ChE}}: \refConcept{Ref_{ChE}} \rightharpoonup \refConcept{Ref_{GCh}}$ returns a global reference to the channel corresponding to the given local channel end reference. The set of all such partial functions is denoted by $\concept{LMS_{ChE}}$.
	\end{itemize}
	To simplify the notation, we regard \concept[function $lms$]{$lms$} itself as a partial function. Note that $\refConcept{Ref_{GQ}} \subseteq \refConcept{Values}$ and $\refConcept{Ref_{GCh}} \subseteq \refConcept{Values}$. Now we can define $lms = (lms_{Cl}, lms_Q, lms_{Ch}, lms_{ChE}): \refConcept{Ref_L} \rightharpoonup \refConcept{Values}$ as: $$lms(r) = \begin{cases}
	lms_{Cl}(r) & \text{if } r \in \refConcept{Ref_{Cl}}, \\
	lms_{Q}(r) & \text{if } r \in \refConcept{Ref_Q}, \\
	lms_{Ch}(r) & \text{if } r \in \refConcept{Ref_{Ch}}, \\
	lms_{ChE}(r) & \text{if } r \in \refConcept{Ref_{ChE}}, \\
	\bot	& \text{if } r = \none.
	\end{cases}$$
	The set of all such quadruples $lms$ is denoted by $\concept{LMS}$.

   \item \concept[variable properties]{Variable properties} $vp$ represent various properties of variables while respecting variable scope. They are stored as a \refConcept{list of lists of partial function tuples} $f = (f_{var},f_{ch},f_{qa},f_{type})$ where:

   \begin{itemize}
    \item $f_{var}: \refConcept{Names} \rightharpoonup \refConcept{Ref_L}$ maps a variable name to a \refConcept{local reference},
    \item $f_{ch}: \refConcept{Names} \rightharpoonup \refConcept{Names} \times \refConcept{Names}$ maps a channel variable name to variable names representing ends of the channel,
    \item $f_{qa}: \refConcept{Names} \rightharpoonup \refConcept{Names}_{[]}$ maps a variable name representing a quantum system to variable names that represent its subsystems,
    \item $f_{type}: \refConcept{Names} \rightharpoonup \refConcept{Types}$ maps a variable name to the variable type.
   \end{itemize}

	We call the quadruple $f$ a \concept{variable properties tuple}.

    We define $\concept[VarProp_L]{VarProp_L}$ to be a set of all finite \refConcept[list of partial function tuples]{lists of such partial function} tuples $f$. These lists are built to reflect the block structure of a method as described in Subsection \ref{subsec:varstack}.

    We define $\concept{VarProp}$ to be a set of all finite \refConcept[list of lists of partial function tuples]{lists of lists of such partial function tuples $f$}. These lists are built to reflect the method calls as described in Subsection \ref{subsec:varstack}.

	We define an \concept{empty variable properties tuple} as a partial function tuple $\lozenge$: $$\lozenge = (f_{var},f_{ch},f_{qa},f_{type})$$ where $\dom(f_{var}) = \dom(f_{ch}) = \dom(f_{qa}) = \dom(f_{type}) = \emptyset$.

   \item \concept[term stack]{Term stack} $ts$: stack of terms to be evaluated. For the sake of readability, we use a notation where individual stack items are underlined. An empty term stack is denoted by $\varepsilon$.
  \end{itemize}

\end{itemize}

A \refConcept{configuration} can evolve to a probabilistic mixture of configurations, so called \concept{mixed configuration}. A mixed configuration is written as: $$\boxplus_{i=1}^{q}p_i\bullet [gs_i\,\vert\,ls_{i,1} \parallel\dots\parallel ls_{i,n}] \text{ where } \sum_i^q p_i = 1.$$ It represents configurations of $q$ different computational branches, each of them running with probability $p_i$. A \refConcept{configuration} is a special case of a \refConcept{mixed configuration} where $q = 1$ and $p_1 = 1$.

\subsection{Variable properties handling functions}

In this subsection, we define functions for \refConcept{variable properties} handling.

First we define functions for retrieving information about variables using \refConcept{variable properties}. The defined functions are designed so that they only work with variables accessible from the actually running method. We achieve this by inspecting the structure of the \refConcept{variable properties}. If the \refConcept{variable properties} given as one of the function arguments are represented by a \refConcept{list of lists of partial function tuples} $L = [L_1 \circ_G K]$ then we consider only its second element -- the \refConcept{list of partial function tuples} $K$. We do not consider the variable properties from $L_1$ as they are inaccessible to the current method (as explained in Subsection \ref{subsec:varstack}).

We then walk through the obtained \refConcept{list of partial function tuples} $[K \circ_L \refConcept[variable properties]{f}]$. We attempt to get the requested information about requested variable using appropriate partial function from the \refConcept{variable properties tuple}. If the requested information about the variable cannot be obtained from the actual tuple, we repeat this procedure with the \refConcept{list of partial function tuples} $K$. This procedure is designed so that it respects block scope of variables (as in more detail explained in Subsection \ref{subsec:varstack}).

\begin{definition}We define a partial function $\concept{varRef}: \refConcept{Names} \times \refConcept{VarProp} \rightarrow \refConcept{Ref_L}$ for getting a \refConcept{local reference} from a variable name and \refConcept{variable properties} as:
\begin{eqnarray*}
varRef(x, [L_1 \circ_G K]) &\isdef &varRef_{L}(x, K)
\end{eqnarray*}
where $\concept{varRef_L}: \refConcept{Names} \times \refConcept{VarProp_L} \rightharpoonup \refConcept{Ref_L}$ is a partial function for getting a \refConcept{local reference} from a variable name and a \refConcept{list of partial function tuples}:
\begin{eqnarray*}
varRef_{L}(x, [K \circ_L \refConcept[variable properties]{(f_{var},f_{ch},f_{qa},f_{type})}]) &\isdef &\begin{cases}
 f_{var}(x) & \text{if } f_{var}(x) \text{ is defined} \\
 varRef_{L}(x, K) & \text{otherwise}
\end{cases}
\end{eqnarray*}
\end{definition}

\begin{definition}
We define a partial function $\concept{chanEnds}: \refConcept{Names} \times \refConcept{VarProp} \rightharpoonup \refConcept{Names} \times \refConcept{Names}$ for getting variable names that represent individual ends of given channel from a name of the channel and \refConcept{variable properties}:
\begin{eqnarray*}
chanEnds(x, [L_1 \circ_G K]) &\isdef &chanEnds_{L}(x, K)
\end{eqnarray*}
where $\concept{chanEnds_L}: \refConcept{Names} \times \refConcept{VarProp_L} \rightharpoonup \refConcept{Names} \times \refConcept{Names}$ is a partial function for getting variable names that represent individual ends of given channel from a name of the channel and a \refConcept{list of partial function tuples}:
\begin{eqnarray*}
chanEnds_{L}(x, [K \circ_L \refConcept[variable properties]{(f_{var},f_{ch},f_{qa},f_{type})}]) &\isdef &\begin{cases}
 f_{ch}(x) & \text{if } f_{ch}(x) \text{ is defined} \\
 chanEnds_{L}(x, K) & \text{otherwise}
\end{cases}
\end{eqnarray*}
\end{definition}

\begin{definition}
We define a partial function $\concept{aliasSubsyst}: \refConcept{Names} \times \refConcept{VarProp} \rightharpoonup \refConcept[Names]{Names_{[]}}$ for getting a list of variable names that represent individual parts of a compound system from a name of the compound system and \refConcept{variable properties}:
\begin{eqnarray*}
aliasSubsyst(x, [L_1 \circ_G K]) &\isdef &aliasSubsyst_{L}(x, K)
\end{eqnarray*}
where $\concept{aliasSubsyst_L}: \refConcept{Names} \times \refConcept{VarProp_L} \rightharpoonup \refConcept{Names}_{[]}$ is a partial function for getting a list of variable names that represent individual parts of a compound system from a name of the compound system and a \refConcept{list of partial function tuples}:
\begin{eqnarray*}
aliasSubsyst_{L}(x, [K \circ_L \refConcept[variable properties]{(f_{var},f_{ch},f_{qa},f_{type})}]) &\isdef &\begin{cases}
 f_{qa}(x) \qquad \hfill \text{if } f_{qa}(x) \text{ is defined} \\
 aliasSubsyst_{L}(x, K) \qquad \text{otherwise}
\end{cases}
\end{eqnarray*}
\end{definition}

\begin{definition}
We define a partial function $\concept{typeOf}: \refConcept{Names} \times \refConcept{VarProp} \rightharpoonup \refConcept{Types}$ for getting a variable type from a name of a variable and \refConcept{variable properties}:
\begin{eqnarray*}
typeOf(x, [L_1 \circ_G K]) &\isdef &typeOf_{L}(x, K)
\end{eqnarray*}
where $typeOf_L: \refConcept{Names} \times \refConcept{VarProp_L} \rightharpoonup \refConcept{Types}$ is a partial function for getting a type from a name of the variable and a \refConcept{list of partial function tuples}:
\begin{eqnarray*}
typeOf_{L}(x, [K \circ_L \refConcept[variable properties]{(f_{var},f_{ch},f_{qa},f_{type})}]) &\isdef &\begin{cases}
 f_{type}(x) & \text{if } f_{type}(x) \text{ is defined} \\
 typeOf_{L}(x, K) & \text{otherwise}
\end{cases}
\end{eqnarray*}
\end{definition}


\subsection{Local memory handling functions}

Next we define functions for \refConcept{local memory state} handling.

LanQ allows the programmer to create multiple processes. These processes can communicate with each other, namely a process can send some resource, {\it ie.} a quantum system or a channel, to another process. In that case, the language must assure that the sent resource becomes unavailable to the sending process.

For this reason we define a function $unmap_{nd}$\footnote{The name $unmap_{nd}$ should be read as ``unmap (a reference to a) {\it n}on{\it d}uplicable (value)''.} which invalidates a reference to resource in given \refConcept{local memory state}. By invalidation we mean setting the appropriate \refConcept{local reference} to the sent resource to point to $\bot$ in the \refConcept{local memory state}.

This function can be split into three functions: unmapping a \refConcept{memory reference} to a quantum system (this is done by the function $\refConcept{unmap_Q}$\footnote{The name $unmap_{Q}$ should be read as ``unmap (a reference to a) {\it q}uantum (value)''.}), unmapping a \refConcept{memory reference} to a channel ($\refConcept{unmap_{Ch}}$\footnote{The name $unmap_{Ch}$ should be read as ``unmap (a reference to a) {\it ch}annel (value)''.}), and unmapping a \refConcept{memory reference} to a channel end ($\refConcept{unmap_{ChE}}$\footnote{The name $unmap_{ChE}$ should be read as ``unmap (a reference to a) {\it ch}annel {\it e}nd (value)''.}).

The function $\refConcept{unmap_Q}$ is designed to obey the following rule: When unmapping a \refConcept[memory reference]{reference} to a quantum system $\rho$ then any \refConcept{memory reference} which refers to any part of $\rho$ is unmapped too.

The function $\refConcept{unmap_{Ch}}$ is designed to obey the following rule: When unmapping a \refConcept[memory reference]{reference} to a channel $c$ then any \refConcept{memory reference} to its ends is unmapped too. The reason is that when a process sends away a channel, it also loses control over both its ends.

The function $\refConcept{unmap_{ChE}}$ is designed to obey the following rule: When unmapping a \refConcept[memory reference]{reference} to a channel end $c$ then we unmap any \refConcept{memory reference} to the corresponding channel. The justification is that when a process sends away a part of a channel, it also loses control over the whole channel.

\begin{definition}
We define a function $\concept{unmap_{nd}}: \refConcept{Ref_L} \times \refConcept{LMS} \rightarrow \refConcept{LMS}$ for unmapping a reference to a non-duplicable value from the local memory:
$$unmap_{nd}((refType, n), lms) \isdef \begin{cases}
\refConcept{unmap_Q}(n, lms) & \text{if } refType = Quantum, \\
\refConcept{unmap_{Ch}}(n, lms) & \text{if } refType = Channel, \\
\refConcept{unmap_{ChE}}(i, n, lms) & \text{if } refType = ChannelEnd_i, \\
lms & \text{otherwise.}
\end{cases}$$
where
\begin{itemize}
 \item 
	Function $\concept{unmap_Q}: \reclist{(\N_\bot)} \times \refConcept{LMS} \rightarrow \refConcept{LMS}$ is defined as:
	$$unmap_{Q}(n, (lms_{Cl}, lms_Q, lms_{Ch}, lms_{ChE})) = (lms_{Cl}, lms'_Q, lms_{Ch}, lms_{ChE}).$$
	where $lms'_Q$ is defined as:
	$$\begin{array}[t]{l}
	lms'_{Q}((Quantum, l)) \isdef \begin{cases}
\refConcept{lms_Q}((Quantum, l)) &\text{if } \recfunc{\set}(n) \cap \recfunc{\set}(l) \subseteq \{\bot\}, \\
\bot &\text{otherwise,}
\end{cases}
	\end{array}$$

 \item 
	Function $\concept{unmap_{Ch}}: \N \times \refConcept{LMS} \rightarrow \refConcept{LMS}$ is defined as:
	$$unmap_{Ch}(n, (lms_{Cl}, lms_Q, lms_{Ch}, lms_{ChE})) = (lms_{Cl}, lms_Q, lms'_{Ch}, lms'_{ChE}).$$
	where \begin{array}[t]{l}
	lms'_{Ch} \isdef lms_{Ch}[(Channel, n) \mapsto \bot], \\
	lms'_{ChE} \isdef lms_{ChE}[(ChannelEnd_0, n) \mapsto \bot,(ChannelEnd_1, n) \mapsto \bot], \\
	\end{array}

 \item 
	Function $\concept{unmap_{ChE}}: \{0,1\} \times \N \times \refConcept{LMS} \rightarrow \refConcept{LMS}$ is defined as:
	$$unmap_{ChE}(i, n, (lms_{Cl}, lms_Q, lms_{Ch}, lms_{ChE})) = (lms_{Cl}, lms_Q, lms'_{Ch}, lms'_{ChE}).$$
	where \begin{array}[t]{l}
	lms'_{Ch} \isdef lms_{Ch}[(Channel, n) \mapsto \bot], \\
	lms'_{ChE} \isdef lms_{ChE}[(ChannelEnd_i, n) \mapsto \bot]. \\
	\end{array}
\end{itemize}
\end{definition}

We extend the function $unmap_{nd}$ so that we can also use any set of references as the first argument. 

\begin{definition}
For any $R = \{r_1,\dots,r_k\} \subseteq \refConcept{Ref_L}$ we define: $$unmap_{nd}(R,lms) \isdef unmap_{nd}(r_1, unmap_{nd}(r_2, \dots unmap_{nd}(r_k, lms)\dots)).$$
\end{definition}

Next, we define a function for updating updating a \refConcept{local memory state} $lms = (lms_{Cl}, lms_Q, lms_{Ch}, lms_{ChE})$. We use the existing concept of partial function \refConcept{update} and extend this concept to \refConcept[local memory state]{local memory states}. The extended function updates appropriate element of the quadruple according to \refConcept{memory reference} type:

\begin{definition}
Let $\refConcept[local memory state]{lms = (lms_{Cl}, lms_Q, lms_{Ch}, lms_{ChE})}$ be a \refConcept{local memory state}. For $r \in \refConcept{Ref_L}$ and $v \in \refConcept{Values}$, we define \concept{local memory state update} $lms[r\mapsto v]_+$ as:
$$lms[r\mapsto v]_+ \isdef\begin{cases}
(lms_{Cl}[r\mapsto v]_+, lms_Q, lms_{Ch}, lms_{ChE}) & \text{if } r \in \refConcept{Ref_{Cl}}, \\
(lms_{Cl}, lms_Q[r\mapsto v]_+, lms_{Ch}, lms_{ChE}) & \text{if } r \in \refConcept{Ref_Q}, \\
(lms_{Cl}, lms_Q, lms_{Ch}[r\mapsto v]_+, lms_{ChE}) & \text{if } r \in \refConcept{Ref_{Ch}}, \\
(lms_{Cl}, lms_Q, lms_{Ch}, lms_{ChE}[r\mapsto v]_+) & \text{if } r \in \refConcept{Ref_{ChE}}. \\
\end{cases}$$
\end{definition}

\subsection{Functions for handling aliasfor constructs}

Quantum algorithms often use multipartite quantum systems. LanQ allows definition of compound quantum systems using \baseterm{aliasfor} construct: if $q_0,\dots,q_n$ are quantum variables then the declaration: $$q\ \baseterm{aliasfor}\ \baseterm{[}q_0\baseterm{,}\dots\baseterm{,}q_n\baseterm{];}$$ declares a new quantum variable $q$ which specifies a quantum system composed from subsystems referred by the variables $q_0,\dots,q_n$. In this subsection, we define functions needed to handle this construct.

To simplify the description of the functions in the following text, we first define two concepts: 
We call a quantum variable which was declared using the \baseterm{aliasfor} construct an \concept{aliased quantum variable}. The quantum variables not declared using the \baseterm{aliasfor} construct are called \concept[proper quantum variable]{proper quantum variables}.

In the following text, we require that all quantum variables that any \refConcept{aliased quantum variable} is composed of are \refConcept[proper quantum variable]{proper quantum variables}. This requirement is later imposed by the semantic rule \semRule{OP-VarDeclAlF}.

Handling the {\bf aliasfor} construct is a little complicated. Two cases must be handled when assigning a \refConcept[local reference]{reference} to a quantum system to a quantum variable $q$:
\begin{enumerate}
 \item[(1)] The variable $q$ is a \refConcept{proper quantum variable}. Hence it can be used in several \refConcept[aliased quantum variable]{aliased quantum variables}. Then this variable (1a) and all the \refConcept[aliased quantum variable]{aliased quantum variables} which use this variable as their subsystem (1b) must be updated,
 \item[(2)] The variable $q$ is an \refConcept{aliased quantum variable}. Then all its subsystems must be appropriately modified. However, the subsystems can be also used in several other \refConcept[aliased quantum variable]{aliased quantum variables} as subsystems. Then all these \refConcept[aliased quantum variable]{aliased quantum variables} must be updated too.
\end{enumerate}

We define several auxiliary functions which help us with handling the assignment to a quantum variable. These functions modify \refConcept{variable properties} and a \refConcept{local memory state} parts of the \refConcept{local process configuration}. Therefore all these auxiliary functions take the unmodified \refConcept{variable properties} $vp$ and \refConcept{local memory state} $lms$ parts as arguments and yield the modified ones. Moreover, all these functions take the name $name$ of the quantum variable being assigned and the assigned reference $ref$ as its additional arguments.

The simplest case is the case (1a). For this case, we define a function $f_{assignQSystemDirect}$ which performs the following operation:
\begin{itemize}
 \item If the assigned reference $ref = (Quantum, q)$ is invalid (this is checked by the condition $\linearize_\bot(q) = \bot$), the reference $ref$ is unmapped from the \refConcept{local process memory}.

 Otherwise the reference $ref$ is set to map from the \refConcept{local process memory} to a \refConcept{global reference} to the corresponding quantum systems in the \refConcept{system memory},
 
 \item The assigned quantum variable is mapped to the reference $ref$.
\end{itemize}

\begin{definition}
We define a function $\concept{f_{assignQSystemDirect}}: \refConcept{LMS_Q} \times \refConcept{VarProp} \times \refConcept{Names} \times \refConcept{Ref_Q} \rightarrow \refConcept{LMS_Q} \times \refConcept{VarProp}$ as:
$$f_{assignQSystemDirect}(lms_Q,vp,name,ref) \isdef (lms_{Q,ret}, vp_{ret})$$
where $\begin{array}[t]{r@{\ }c@{\ }l}
 lms_{Q,ret} &= &lms_{Q}[(Quantum,q) \mapsto gq], \\
 vp_{ret} &= &vp[name \mapsto ref]_{var},\\
 \end{array}$\\
given that $\begin{array}[t]{r@{\ }c@{\ }l}
 ref &= &(Quantum, q), \\
 gq &= &\begin{cases}
          \bot & \text{if } \linearize_\bot(q) = \bot, \\
          (GQuantum, \linearize_\bot(q)) & \text{otherwise.}
        \end{cases}
\end{array}$
\end{definition}

To update all the \refConcept[aliased quantum variable]{aliased quantum variables} which use the assigned \refConcept{proper quantum variable} as their subsystems (the case (1b)), we first define an auxiliary function $f_{assignQSystemInAlias}$ which updates one subsystem reference in the \refConcept{aliased quantum variable}. This function takes one more argument -- the index $index$ of the updated subsystem. It then proceeds as follows:

\begin{itemize}
 \item It takes the original reference of the \refConcept{aliased quantum variable} $\refConcept{varRef}(name)$ and modifies its $index$-th element to point to the newly assigned system (specified by the recursive $\N$-list $q$). Individual elements of the recursive $\N$-list specifying the new reference are denoted by $v'_j$,
 \item It unmaps the original reference from the \refConcept{local process memory},
 \item If any of the systems in the newly assigned reference is invalid (checked by the condition $\exists j: lms_Q((Quantum, v'_j)) = \bot$), the newly assigned reference is unmapped from the \refConcept{local process memory}.
 
 Otherwise the reference $ref$ is set to map from the \refConcept{local process memory} to a \refConcept{global reference} to the corresponding quantum systems in the \refConcept{system memory},
 
 \item The assigned quantum variable is mapped to the new reference $(Quantum, [v'_1,\dots,v'_k])$.
 \end{itemize}

\begin{definition}
We define a function $\concept{f_{assignQSystemInAlias}}: \refConcept{LMS_Q} \times \refConcept{VarProp} \times \refConcept{Names} \times \refConcept{Ref_Q} \times \N \rightarrow \refConcept{LMS_Q} \times \refConcept{VarProp}$ as:
$$f_{assignQSystemInAlias}(lms_Q,vp,name,ref,index) \isdef (lms_{Q,ret}, vp_{ret})$$
where $\begin{array}[t]{r@{\ }c@{\ }l}
lms_{Q,ret} &= &lms_Q[(Quantum, [v_1,\dots,v_k]) \mapsto \bot][(Quantum, [v'_1,\dots,v'_k]) \mapsto gq]  \\
vp_{ret} &= &vp[name \mapsto (Quantum, [v'_1,\dots,v'_k])]_{var} \\
v'_i &= &\begin{cases}
            q &\text{if } i = index  \\
            v_i &\text{otherwise}
          \end{cases} 
\end{array}$ \\
given that $\begin{array}[t]{r@{\ }c@{\ }l}
ref &= &(Quantum, q) \\
\refConcept{varRef}(name, vp) &= &(Quantum, [v_1,\dots,v_k]) \\
gq &= &\begin{cases}
          \bot \text{\qquad \qquad \qquad \quad if } \exists j: lms_Q((Quantum, v'_j)) = \bot \\
          (GQuantum, \refConcept[linearize_bot]{\linearize_\bot}([v'_1,\dots,v'_k])) \text{\qquad otherwise}
        \end{cases}
\end{array}$
\end{definition}

Before we define a function that handles the whole case (1), we must define one more auxiliary function $localAliasedVars$ that returns all the \refConcept[aliased quantum variable]{aliased quantum variables} available to the currently running method.

\begin{definition}
We define a function $\concept{localAliasedVars}: \refConcept{VarProp} \rightarrow \P(\refConcept[Names]{Names})$ as:
\begin{eqnarray*}
localAliasedVars([L_1 \circ_G K]) &\isdef &localAliasedVars_{L}(K) \\
localAliasedVars(\refConcept[blacksquare]{\blacksquare}) &\isdef &\emptyset
\end{eqnarray*}
where $localAliasedVars_{L}: \refConcept{VarProp_L} \rightarrow \P(\refConcept[Names]{Names})$ is a function for getting a set of all names of variables representing compound systems in the local \refConcept{variable properties} list from a \refConcept{list of partial function tuples}:
$$\begin{array}{rcl}
localAliasedVars_{L}([K \circ_L \refConcept[variable properties]{(f_{var},f_{ch},f_{qa},f_{type})}]) &\isdef & dom(f_{qa}) \cup localAliasedVars_L(K) \\
localAliasedVars_{L}(\refConcept[square]{\square}) &\isdef &\emptyset \\
\end{array}$$
\end{definition}

Now we are ready to define a function $f_{assignQSystem}$ that performs assignment to one \refConcept{proper quantum variable} while correctly adjusting all \refConcept[aliased quantum variable]{aliased quantum variables} which use the variable being assigned (case (1)). The function operates as follows:

\begin{itemize}
 \item It first uses the function $\refConcept{f_{assignQSystemDirect}}$ to perform the assignment to the \refConcept{proper quantum variable},
 \item Then it adjusts all the \refConcept[aliased quantum variable]{aliased quantum variables} which use the variable being assigned (the set of such variables is denoted as $AQV$) using the function $\refConcept{f_{assignQSystemInAlias}}$.
\end{itemize}

\begin{definition}
We define a function $\concept{f_{assignQSystem}}: \refConcept{LMS_Q} \times \refConcept{VarProp} \times \refConcept{Names} \times \refConcept{Ref_Q} \rightarrow \refConcept{LMS_Q} \times \refConcept{VarProp}$ as: $$f_{assignQSystem}(lms_Q,vp,name,ref) \isdef (lms_{Q,ret}, vp_{ret})$$
where $\begin{array}[t]{l}
(lms_{Q,0}, vp_0) = \refConcept{f_{assignQSystemDirect}}(lms_Q,vp,name,ref),\\
(lms_{Q,i}, vp_i) = \refConcept{f_{assignQSystemInAlias}}(lms_{Q,i-1},vp_{i-1},qcs_i,ref,l_i) \forall i: 1 \leq i \leq k, \\
lms_{Q,ret} = lms_{Q,k},\ vp_{ret} = vp_k,
\end{array}$ \\
given that $\begin{array}[t]{l}
AQV = \{ aqv\in \refConcept{localAliasedVars}(vp) |\ name\in \set(\refConcept{aliasSubsyst}(aqv,vp) \}, \\
AQV \text{ is indexed by numbers } i \in \N: 1 \leq i \leq k, \\
aqv_i \in AQV, \\
\refConcept{aliasSubsyst}(aqv_i, vp) = [aqv_{i,1},\dots,aqv_{i,m_i}], \\
aqv_{i,l_i} = name.
\end{array}$
\end{definition}

Last, we define a function $f_{assignQAlias}$ that performs an assignment to an \refConcept{aliased quantum variable}. This function operates very straightforwardly -- it takes each \refConcept[proper quantum variable]{proper quantum variables} which the \refConcept{aliased quantum variable} is composed of and applies the function $\refConcept{f_{assignQSystem}}$ onto it. We assume that the number of subsystems of the \refConcept{aliased quantum variable} corresponds to the structure of the assigned reference (the length of the $\N$-list in the reference).

\begin{definition}
We define a function $\concept{f_{assignQAlias}}: \refConcept{LMS_Q} \times \refConcept{VarProp} \times \refConcept{Names} \times \refConcept{Ref_Q} \rightarrow \refConcept{LMS_Q} \times \refConcept{VarProp}$ as:
$$f_{assignQAlias}(lms_Q,vp,name,ref) \isdef (lms_{Q,ret}, vp_{ret})$$
where $\begin{array}[t]{l}
lms_{Q,0} = lms_Q,\ vp_0 = vp, \\
(lms_{Q,i}, vp_i) = \refConcept{f_{assignQSystem}}(lms_{Q,i-1},vp_{i-1},q_i,(Quantum, v_i)) \text{ for all } 1 \leq i \leq k, \\
lms_{Q,ret} = lms_{Q,k},\ vp_{ret} = vp_k,
\end{array}$ \\
given that $\begin{array}[t]{l}
\refConcept{aliasSubsyst}(name, vp) = [q_1,\dots,q_k], \\
ref = (Quantum,[v_1,\dots,v_k]).
\end{array}$
\end{definition}

\subsection{Internal values}

Expressions evaluate to references and values which in turn are possibly references to global memory. Operational semantics uses both values and references so we define \concept{internal value} to be a triplet $(ref,val,\type{T}) \in \refConcept{Ref_L} \times \refConcept{Values}\times \refConcept{Types}$ written as $\intVal{ref}{val}{T}$.

\subsection{Transitions}

We define operational semantics in terms of the following relations:

\begin{itemize}
 \item $\longrightarrow_v$ -- Transitions of expressions to \refConcept[internal value]{internal values},
 \item $\longrightarrow_e$ -- Transitions of expressions to expressions -- the order of evaluation is encoded here,
 \item $\longrightarrow_s$ -- Transitions of statements,
 \item $\longrightarrow_{ret}$ -- Transitions of \baseterm{return} statements,
 \item $\longrightarrow_{rte}$ -- Transitions of runtime errors,
 \item $\longrightarrow_r$ -- Transitions of statements to statements, used to rewrite an abbreviated statement to the unabbreviated form,
 \item $\longrightarrow_p$ -- Transitions of processes,
 \item $\stackrel{p}{\longrightarrow}$ -- This defines probabilistic transitions of processes, $0\leq p \leq 1$ is the probability of the transition.
\end{itemize}

We define relation $\longrightarrow$ as:
$$\longrightarrow = \longrightarrow_v \cup \longrightarrow_e \cup \longrightarrow_s \cup \longrightarrow_{ret} \cup \longrightarrow_{rte} \cup \longrightarrow_r \cup \longrightarrow_p.$$

The relations $\longrightarrow_v, \longrightarrow_e, \longrightarrow_s$, $\longrightarrow_{ret}$, $\longrightarrow_{rte}$, $\longrightarrow_r$ and $\stackrel{p}{\longrightarrow}$ define deterministic and probabilistic single process evolution. Nondeterminism is introduced by parallel evolution of processes -- a choice which process gets evolved is nondeterministic. However, there is no nondeterminism in the evolution of individual processes even when they are run in parallel with other processes. This is an improvement over existing quantum process algebras \cite{JorLal04,GayNag04}.

In these algebras, there is a nondeterminism arising from resource sharing. Although there is no nondeterminism arising from quantum resource control, there is one arising from channel resources: it is possible for three or more processes to share one channel. When these processes use the channel simultaneously, the resulting behaviour is nondeterministic.%
\footnote{Note however that we can take advantage of the nondeterministic behaviour: It can be used {\it eg.} to simply catch server environment serving requests from multiple clients where it is used to resolve which request came from which client.} We avoid this type of nondeterminism by using channel ends for communication instead of the channels themselves and imposing a constraint that one channel end is owned by exactly one process at one time. This approach was also studied in the context of $\pi$-calculus in \cite{KobPieTur96,GayHol05}.

When probabilistic and nondeterministic choice are to be evaluated simultaneously, we must decide which choice is resolved first. We have taken the same approach as other authors ({\it eg.} \cite{GayNag04}): the nondeterministic choice is resoved first.

When we get to the situation when no rule is applicable to the configuration, the configuration becomes \concept{stuck}. Because LanQ is a typed language, the errors arising from invalid typing can be avoided. The formal proof that a language does not suffer from typing errors lies in proving standard lemmata in the style of Wright and Felleisen \cite{WriFel94}. For LanQ, this is done in Section \ref{sec:type-soundness}.

However, there exist some unavoidable cases caused by the by-reference handling of variables. For example, a process $P$ can send away a qubit referred by variable $\psi$. If $P$ later tries to measure a qubit referred by $\psi$, there is none. Such cases are handled by \refConcept[runtime error]{runtime errors} described in the next subsection.

\hypertarget{opSem:runtimeErrors}{}
\subsection{Runtime errors}

There exist errors that cannot be recognized during compile time and can occur during the run time. For that reason. we define special stack symbols representing such \concept[runtime error]{runtime errors}:

\begin{itemize}
 \item \concept[rte:UV]{}\displayRTEuninitVarUsage: an error representing an {\it u}ninitialized {\it v}ariable usage. An example method invoking this error is in Figure \ref{fig:rte-a} ($U$ is a unitary operation there).
 In this example, the variable $q$ is sent away, hence not initialized. The attempt to perform $U(q)$ therefore invokes a runtime error \displayRTEuninitVarUsage.
 
 \item \concept[rte:OQV]{}\displayRTEqVarOverlap: an error representing {\it o}verlapping {\it q}uantum {\it v}ariable usage. An example method invoking this error is shown in Figure \ref{fig:rte-b} ($U$ is a two-qubit unitary operation there).
 In this example, variables $p$ and $q$ refer to the same quantum system. An attempt to perform $U(p,q)$ therefore invokes a runtime error \displayRTEqVarOverlap.

 \item \concept[rte:ISQV]{}\displayRTEassignIncompatibleQuant: an error representing an assignment to an {\it i}ncompatibly {\it s}tructured {\it q}uantum {\it v}ariable. An example method invoking this error is shown in Figure \ref{fig:rte-c}.
 In this example, variables $p$ and $q$ refer to qubit systems, $r$ refers to a system composed of the two qubits. An attempt to assign one four-dimensional quantum system to $r$ fails as this assignment must also appropriately set the two systems $p$ and $q$. Hence this assignment invokes a runtime error \displayRTEassignIncompatibleQuant.
\end{itemize}

\begin{figure}[h!]
\subfigure[]{\label{fig:rte-a}\begin{minipage}[t]{0.32\textwidth}
\method{void}{ex1}{\type{channelEnd[qbit]} c}{%
 \varQ{q};\\
 q = \new \type{qbit}(); \\
 $\send(c,q);$ \\
 $U(q);$
 }\end{minipage}
}
\hfill
\subfigure[]{\label{fig:rte-b}\begin{minipage}[t]{0.3\textwidth}
\method{void}{ex2}{}{%
 \varQ{p, q};\\
 $q = \new\ \type{qbit}();$\\
 $p = q;$\\
 $U(p,q);$
 }\end{minipage}
}
\hfill
\subfigure[]{\label{fig:rte-c}\begin{minipage}[t]{0.3\textwidth}
\method{void}{ex3}{}{%
 \varQ{p,q};\\
 $r$ \aliasfor [$p,q$]; \\
 $r = \new\ \type{Q_4}();$
 }\end{minipage}
}

\caption{Example methods invoking runtime errors.}
\label{fig:rte}
\end{figure}

\subsection{Processes and configurations}

In this subsection, we define special configurations, processes and relations between them.

We define a special \refConcept{configuration} {\bf start} (a starting configuration for any LanQ program execution) and a set {\bf 0$_c$} of silent \refConcept[local process configuration]{local process configurations} as:
$${\bf start}~\isdef~[(((1), []), [])\,\vert\,(\refConcept[local memory state]{([],[],[],[])}, \refConcept[blacksquare]{\blacksquare}, \si{main()})]$$
$${\bf 0_c}~\isdef~\{(lms, vp, \varepsilon)\ \vert\ lms \in \refConcept{LMS}, vp \in \refConcept{VarProp}\}$$

The \concept{terminal configuration} is defined as $$[gs\,\vert\,(lms_1,vp_1, v_1) \parallel \dots \parallel (lms_n,vp_n, v_n)]$$ where $gs$ is a global part of the configuration, and for all $i$, $lms_i \in \refConcept{LMS}$, $vp_i \in \refConcept{VarProp}$, and $v_i$ is either $\varepsilon$, \refConcept{runtime error} $\refConcept{RTErr}$, or an \refConcept{internal value} \value. If some of $v_i$ is a \refConcept{runtime error}, then we call this configuration \concept[errorneous configuration]{errorneous}.

\subsubsection{Structural congruence}

In this subsection, we define structurally congruent processes. A process is fully characterized by a \refConcept{local process configuration}, therefore the relation is defined on these \refConcept[local process configuration]{local process configurations}.

Any process is structurally congruent to a process running in parallel with a silent process (rule \semRule{SC-Nil}). Order of processes in the configuration does not matter (\semRule{SC-Comm}) as well as grouping of processes within the configuration (\semRule{SC-Assoc}).

\begin{center}
$\begin{array}{ll ll}
\hline\\

\raisebox{\normalbaselineskip}[0pt]{\hypertarget{semRule:SC-Nil}{}}\textsc{SC-Nil} &
\multicolumn{3}{l}{
P \parallel 0 \equiv  P \text{\qquad for any }0 \in {\bf 0_c}}
\\[0.5\normalbaselineskip]
\raisebox{\normalbaselineskip}[0pt]{\hypertarget{semRule:SC-Comm}{}}\textsc{SC-Comm} &
P \parallel Q \equiv Q \parallel P &
\qquad
\raisebox{\normalbaselineskip}[0pt]{\hypertarget{semRule:SC-Assoc}{}}\textsc{SC-Assoc} &
(P \parallel Q) \parallel R \equiv P \parallel (Q \parallel R)\\
\\ \hline
\end{array}$
\end{center}

\subsubsection{Nondeterminism and parallelism}

In this subsection, we define behaviour related to nondeterminism and parallelism.

The rule \semRule{NP-PropagProb} states that evolution of a process $P$ leaves all other processes running in parallel with $P$ in their original state while propagating the probability distribution on configurations to the top level. We can exchange congruent processes without any impact on the resulting behaviour (rule \semRule{NP-Cong}). A probabilistic configuration consisting of two or more probabilistic alternatives must resolve a probabilistic choice (rule \semRule{NP-ProbEvol}).

\begin{center}
$\begin{array}{lc}
\hline\\
\raisebox{\normalbaselineskip}[0pt]{\hypertarget{semRule:NP-PropagProb}{}}\textsc{NP-PropagProb} &
\infer%
{[gs\,\vert\,P \parallel Q] \longrightarrow \boxplus_i\ p_i \bullet[gs_i\,\vert\,{P_i} \parallel Q]}%
{[gs\,\vert\,P] \longrightarrow \boxplus_i\ p_i \bullet[gs_i\,\vert\,{P_i}]} \\
\\
\raisebox{\normalbaselineskip}[0pt]{\hypertarget{semRule:NP-Cong}{}}\textsc{NP-Cong} &
\infer%
{[gs\,\vert\,P'] \longrightarrow \boxplus_i\ p_i \bullet[gs_i\,\vert\,P'_i]}%
{[gs\,\vert\,P] \longrightarrow \boxplus_i\ p_i \bullet[gs_i\,\vert\,P_i] \quad P \equiv P' \quad P_i \equiv P'_i\text{ for all $i$}} \\
\\
\raisebox{\normalbaselineskip}[0pt]{\hypertarget{semRule:NP-ProbEvol}{}}\typesetSemRule{NP-ProbEvol} &
\boxplus_{i=1}^q\ p_i \bullet[gs_i\,\vert\,{P_i}] \stackrel{p_i}{\longrightarrow} [gs_i\,\vert\,{P_i}] \qquad \text{ for } q > 1\\
\\ \hline
\end{array}$
\end{center}

\subsection{Evaluation}
In this subsection, we define the transition rules of individual processes.

\begin{operSem}[Basic rules]{
The first four rules define configuration change on a skip statement (rule \semRule{OP-Skip}), a variable (\semRule{OP-Var}) and bracketed expression (\semRule{OP-Bracket}). Next rule (\semRule{OP-BlockHead}) is used to evaluate sequence of statements from first to last. Last two rules (\semRule{OP-SubstE} and \semRule{OP-SubstS}) defines substitution of evaluated expressions.
}
\opSem[s]{OP-Skip}{;}{[gs\,\vert\,(lms, vp, ts)]} \\
\\
\opSem[v]{OP-Var}{x}{} \\
\multicolumn{4}{r}{[gs\,\vert\,(lms, vp, \si{\intVal{ref}{\refConcept[function $lms$]{lms}(ref)}{\mathit{\refConcept{typeOf}(x,vp)}}}\ ts)]} \\
&\multicolumn{3}{r}{\text{where } ref=varRef(x,vp)} \\
\\
\opSem[r]{OP-Bracket}{(E)}{[gs\,\vert\,(lms, vp, \si{E}\ ts)]} \\
\\
\opSem[r]{OP-BlockHead}{\overline{Be}}{[gs\,\vert\,(lms, vp, \si{head(\overline{Be})}\ \si{tail(\overline{Be})}\ ts)]} \\
\\
\opSemVerb[e]{OP-SubstE}{[gs\,\vert\,(lms, vp, \si{\value}\ \si{Ec}\ ts)]}{[gs\,\vert\,(lms, vp, \si{Ec[\value]}\ ts)]} \\
\\
\opSemVerb[e]{OP-SubstS}{[gs\,\vert\,(lms, vp, \si{\value}\ \si{Sc}\ ts)]}{[gs\,\vert\,(lms, vp, \si{Sc[\value]}\ ts)]} \\
\end{operSem}

\begin{operSem}[Promotable expressions]{
Promotable expressions are expressions that can be turned into statements by appending a semicolon. The expression is evaluated (rule \semRule{OP-PromoExpr}) but the resulting value is then forgotten (rule \semRule{OP-PromoForget}).
}
 
\opSem[e]{OP-PromoExpr}{PE \baseterm{;}}{[gs\,\vert\,(lms, vp, \si{PE}\ \si{\bullet \baseterm{;}}\ ts)]} \\
\\
\opSem[s]{OP-PromoForget}{\si{\value} \baseterm{;}}{[gs\,\vert\,(lms, vp, ts)]} \\
\\
\end{operSem}

\begin{operSem}[Allocation]{
Allocating a resource is performed by an evaluation of expression ``{\bf new} $\type{T}$()'' where \type{T} is a type of the resource, {\it ie.} a type of a channel or a quantum system. Type of quantum systems of dimension $d$ are denoted by $\type{Q_d}$, {\it eg.} \type{qbit} = $\type{Q_2}$. 

Allocation of a channel resource is handled by rule \semRule{OP-AllocC}, quantum resource allocation is handled by rule \semRule{OP-AllocQ}.

}
\opSem[v]{OP-AllocQ}{{\bf new}\ \type{Q_d}\baseterm{()}}{\quad}\\
& \multicolumn{3}{r}{[gs'\,\vert\,(lms', vp, \si{\intVal{(Quantum, [l])}{(GQuantum, [l])}{Q_d}}\ ts)]} \\
\multicolumn{4}{r}{\begin{array}{l}
\text{where} \begin{array}[t]{l}
l = |L|\\
gs' = ((\rho \otimes (\frac{1}{d} I_d), L \cdot [d]), C)\\
 lms' = (lms_{Cl}, lms'_Q, lms_{Ch}, lms_{ChE}) \\
 lms'_{Q} = lms_{Q}[(Quantum,[l]) \mapsto (GQuantum, [l])] \\
\end{array} \\
\text{ given that }\begin{array}[t]{l}
gs=((\rho,L), C) \\
lms=(lms_{Cl}, lms_Q, lms_{Ch}, lms_{ChE})\\
\end{array}
\end{array}
}\\
\\
\opSem[v]{OP-AllocC}{{\bf new}\ \type{channel[T]}\baseterm{()}}{\quad}\\
& \multicolumn{3}{r}{[gs'\,\vert\,(lms', vp, \si{\intVal{(Channel, l)}{(GChannel, l)}{channel[T]}}\ ts)]} \\
&\multicolumn{3}{r}{\begin{array}{l}
\text{where} \begin{array}[t]{l}
l = |C| \\
gs' = (Q, C \cdot [ \channel(c_0,c_1)]) \\
 lms' = (lms_{Cl}, lms_Q, lms'_{Ch}, lms'_{ChE}) \\
 lms'_{Ch} = lms_{Ch}[(Channel,l) \mapsto (GChannel, l)] \\
 lms'_{ChE} = lms_{ChE}[(ChannelEnd_0,l) \mapsto (GChannel, l), \\
 \qquad\qquad\qquad\qquad\qquad (ChannelEnd_1,l) \mapsto (GChannel, l)] \\
\end{array}\\
\text{ given that }\begin{array}[t]{l}
gs=(Q, C) \\
lms=(lms_{Cl}, lms_Q, lms_{Ch}, lms_{ChE})\\
\end{array}
\end{array}
}\\
\end{operSem}

\begin{operSem}[Variable declaration]{
Variable declaration is an addition of a variable to the innermost list of mappings of variable names to references. We consider any variable declaration of multiple variables of the same type: $\type{T}\ a,b,c;$ to be an abbreviation of $\type{T}\ a;\ \type{T}\ b;\ \type{T}\ c;$.

For declaration of a quantum compound system a construction $q\ {\bf aliasfor}\ [q_0,\dots,q_n]$ is used where $q_0,\dots,q_n$ are names of quantum variables. Some of them can again be compound systems. To deal with this feature, all variables from $\{q_0,\dots,q_n\}$ that represent compound systems are expanded. This can be seen from the following example -- we require the declarations on the left and right side to be equivalent:

\begin{center}
\begin{minipage}{0.4\linewidth}
$\begin{array}{l}
  \type{qbit}\ q_0,q_1,p; \\
  q\ {\bf aliasfor}\ [q_0,q_1]; \\
  r\ {\bf aliasfor}\ [p,q];
\end{array}$
\end{minipage}
\quad
\begin{minipage}{0.4\linewidth}
$\begin{array}{l}
  \type{qbit}\ q_0,q_1,p; \\
  q\ {\bf aliasfor}\ [q_0,q_1]; \\
  r\ {\bf aliasfor}\ [p,q_0,q_1];
\end{array}$
\end{minipage}
\end{center}
}
\opSemVerb[r]{OP-VarDeclMulti}{[gs\,\vert\,(lms, vp, \si{T\ x, \listOf{I}\baseterm{;}}\ ts)]}{}\\
&\multicolumn{3}{r}{[gs\,\vert\,(lms, vp, \si{T\ x;}\ \si{T\ \listOf{I}\baseterm{;}}\ ts)]} \\
\\


\opSemVerb[s]{OP-VarDecl}{[gs\,\vert\,(lms, [vp_G \circ_G [vp \circ_L vp_L]], \si{T\ x\baseterm{;}}\ ts)]}{}\\
&\multicolumn{3}{r}{[gs\,\vert\,(lms, [vp_G \circ_G [vp \circ_L vp_L']], ts)]} \\
&\multicolumn{3}{r}{\quad \text{ where $vp_L' = vp_L[x \mapsto \none]_{+,var}[x \mapsto T]_{+,type}$ \quad}} \\
\\


\hypertarget{semRule:OP-VarDeclChE}{}\typesetSemRule{OP-VarDeclChE} \\
\multicolumn{2}{r}{[gs\,\vert\,(lms, [vp_G \circ_G [vp \circ_L vp_L]], \si{\type{channel[T]}\ c\ {\bf withends}\baseterm{[}c_0\baseterm{,}c_1\baseterm{];}}\ ts)]} & \longrightarrow_s \\
&\multicolumn{3}{r}{[gs\,\vert\,(lms, [vp_G \circ_G [vp \circ_L vp_L']], ts)]} \\
\multicolumn{4}{r}{\quad \text{ where $vp_L' = vp_L[c \mapsto \none]_{+,var}[c_0 \mapsto \none]_{+,var}[c_1 \mapsto \none]_{+,var}[c \mapsto (c_0,c_1)]_{+,ch}$ \quad\quad}} \\
\multicolumn{4}{r}{\quad \text{$[c \mapsto \type{channel[T]}]_{+,type}[c_0 \mapsto \type{channelEnd[T]}]_{+,type}[c_1 \mapsto \type{channelEnd[T]}]_{+,type}$ \quad}} \\
\\


\opSemVerb[s]{OP-VarDeclAlF}{[gs\,\vert\,(lms, [vp_G \circ_G [vp \circ_L vp_L]], \si{q\ {\bf aliasfor}\ \baseterm{[}q_0\baseterm{,}\dots\baseterm{,}q_n]\baseterm{;}}\ ts)]}{}\\
&\multicolumn{3}{r}{[gs\,\vert\,(lms, [vp_G \circ_G [vp \circ_L vp_L']], ts)]} \\
\multicolumn{4}{r}{\quad \begin{array}[t]{r}
\text{ where $vp_L' = vp_L[q \mapsto (Quantum,[l_0,\dots,l_n])]_{+,var}[q \mapsto [q'_0,\dots,q'_n]]_{+,qa}$ \quad\quad} \\
\left[q \mapsto \bigotimes_{i=0}^n \refConcept{typeOf}(q_n,vp)\right]_{+,type} \quad \\
\text{ given that } \begin{array}[t]{l}
l_i = \begin{cases}
l & \text{if } \refConcept{varRef}_L(q_i, [vp \circ_L vp_L]) = (Quantum,l) \\
\bot & \text{otherwise}
\end{cases} \\
 q'_i = 
 \begin{cases}
  p_0,\dots,p_k & \text{if } \refConcept{aliasSubsyst_L}(q_i,[vp \circ_L vp_L']) = [p_0,\dots,p_k] \\
  q_i & \text{otherwise}
 \end{cases} \\
 \refConcept{varRef}_L(q_i, [vp \circ_L vp_L']) \text{ is defined for } 0 \leq i \leq n
\end{array}
\end{array}}\\
\end{operSem}

\begin{operSem}[Assignment]{
Assignment command $x = E$ has to be divided into two rules: one where expression $e$ is evaluated (\semRule{OP-AssignExpr}) and the other where the result of evaluation of $e$ is bound to variable $x$ and possibly stored into memory (rules \semRule{OP-AssignNewValue} and \semRule{OP-AssignValue}). The value is stored into memory if it was not there yet what is indicated by reference part of the internal value equal to \none{}.

Assigning a quantum system to a variable can be complicated when the variable was declared using the {\bf aliasfor} construct. For example, let $\psi$ be a variable that represents a quantum system composed of systems $\psi_A$ and $\psi_B$ (it was declared as: $\psi\ {\bf aliasfor}\ [\psi_A,\psi_B]$). Assigning a value to $\psi$ must appropriately modify both $\psi_A$ and $\psi_B$ and can be only performed if the assigned value represents a compound system made of two subsystems (rule \semRule{OP-AssignQAValue}). Similarly, assigning a value to $\psi_A$ must also modify $\psi$ (rule \semRule{OP-AssignQValue}). If the structure of assigned system is not compatible with the structure of the assigned variable then a runtime error \RTEassignIncompatibleQuant{} occurs (rule \semRule{OP-AssignQAValueBad}).
}
\opSem[e]{OP-AssignExpr}{x = E}{[gs\,\vert\,(lms, vp, \si{E}\ \si{x = \bullet}\ ts)]} \\
\\
\opSemVerb[v]{OP-AssignNewValue}{[gs\,\vert\,(lms, vp, \si{x = \value}\ ts)]}{[gs\,\vert\,(lms', vp', \si{\intVal{lr'}{lv}{T}}\ ts)]} \\
\multicolumn{4}{r}{\begin{array}[t]{l}\text{where }
\begin{array}[t]{l}
lr' = (Classical, nc) \\
lms'_{Cl} = lms_{Cl}[lr' \mapsto lv]_+ \\
vp' = vp[x \mapsto lr']_{var}
\end{array}
\\
\text{given that \begin{tabular}[t]{l}
$\value = \intVal{lr}{lv}{T}$ \\
$nc \in \N_0$ is such that $\refConcept[function $lms$]{lms}((Classical, nc))$ is not defined \\
$lms' = lms[lr' \mapsto lv]_+$ \\
$lr = \none \wedge lv \neq \bot$ \\
\end{tabular}}
\end{array}}\\
\\
\opSem[v]{OP-AssignQValue}{x = \value}{[gs\,\vert\,(lms', vp', \si{\value}\ ts)]} \\
\multicolumn{4}{r}{\begin{array}[t]{l}
\text{where }(lms'_Q, vp') = \refConcept{f_{assignQSystem}}(lms_Q, vp, x, lr) \\

\text{given that }\begin{array}[t]{l}
\value = \intVal{lr}{ lv}{T} \\
lr = (Quantum, q) \text{ and } \refConcept{aliasSubsyst}(x,vp)\text{ is not defined}\\
lms = (lms_{Cl}, lms_Q, lms_{Ch}, lms_{ChE}) \\
lms' = (lms_{Cl}, lms'_Q, lms_{Ch}, lms_{ChE}) \\
\end{array}
\end{array}}\\
\\
\opSem[v]{OP-AssignQAValue}{x = \value}{[gs\,\vert\,(lms', vp', \si{\value}\ ts)]} \\
\multicolumn{4}{r}{\begin{array}[t]{l}
\text{where }(lms'_Q, vp') = \refConcept{f_{assignQAlias}}(lms_Q, vp, x, lr) \\

\text{given that }\begin{array}[t]{l}
\value = \intVal{lr}{ lv}{T} \\
lr = (Quantum, q) \text{ and } \refConcept{aliasSubsyst}(x,vp) \text{ is defined} \\
lms = (lms_{Cl}, lms_Q, lms_{Ch}, lms_{ChE}) \\
lms' = (lms_{Cl}, lms'_Q, lms_{Ch}, lms_{ChE}) \\
\refConcept{typeOf}(q) = \bigotimes_{i=0}^n Q_i, \type{T} = \bigotimes_{j=0}^m T_j \\
\text{\qquad and } m = n \text{ and } \forall i: Q_i \cong T_i
\end{array}
\end{array}}\\
\\
\opSem[rte]{OP-AssignQAValueBad}{x = \value}{[gs\,\vert\,(lms, vp, \RTEassignIncompatibleQuant)]} \\
\multicolumn{4}{r}{
\text{given that }\begin{array}[t]{l}
\refConcept{typeOf}(q) = \bigotimes_{i=0}^n Q_i, \type{T} = \bigotimes_{j=0}^m T_j \\
\text{\qquad and } m \neq n \text{ or } \exists i: Q_i \ncong T_i
\end{array}}\\
\\
%
\opSem[v]{OP-AssignValue}{x = \value}{[gs\,\vert\,(lms, vp', \si{\value}\ ts)]} \\
\multicolumn{4}{r}{\begin{array}[t]{l}\text{where }
vp' = \begin{cases}
vp[x \mapsto lr]_{var}[x_0 \mapsto (ChannelEnd_0,i)]_{var}[x_1 \mapsto (ChannelEnd_1,i)]_{var} \\
\hfill\text{if $lr = (Channel, i)$ and $chanEnds(x, vp) = (x_0,x_1)$}\\
vp[x \mapsto lr]_{var} \qquad \text{\quad otherwise}\\
\end{cases}\\
\text{given that }\begin{array}[t]{l}
\value = \intVal{lr}{ lv}{T} \\
(lr = \none \wedge lv = \bot) \vee (lr \neq (Quantum, q)) \\
\end{array}
\end{array}}\\
\end{operSem}

\begin{operSem}[Block]{
Block command is used to limit scope of variables and to execute multiple statements:
}

\opSemVerb[s]{OP-Block}{[gs\,\vert\,(lms, [vp_G \circ_G vp], \si{\{ B \}}\ ts)]}{}\\
&\multicolumn{3}{r}{[gs|(lms, [vp_G \circ_G [vp \circ_L \refConcept[square]{\square}]], \si{B}\ \si{\circL}\ ts)]} \\
\\
\opSemVerb[s]{OP-BlockEnd}{[gs\,\vert\,(lms, [vp_G \circ_G [vp \circ_L vp_L]], \si{\circL}\ ts)]}{}\\
&\multicolumn{3}{r}{[gs|(lms, [vp_G \circ_G vp], ts)]} \\

\end{operSem}

\begin{operSem}[Conditional statement -- if]{
Conditional expression ${\bf if}\ (E)\ S_1\ {\bf else}\ S_2$ has to be split into three rules: one where the condition is evaluated (\semRule{OP-IfExpr}) and rules for reduction when the condition evaluates to {\bf true} (\semRule{OP-IfTrue}) and {\bf false} (\semRule{OP-IfFalse}).
}

\opSem[e]{OP-IfExpr}{{\bf if}\ (E)\ S_1\ {\bf else}\ S_2}{(lms, vp, \si{E}\ \si{{\bf if}\ (\bullet)\ S_1\ {\bf else}\ S_2}\ ts)]} \\
\\
\opSem[s]{OP-IfTrue}{{\bf if}\ (\value)\ S_1\ {\bf else}\ S_2}{[gs\,\vert\,(lms, vp, \si{S_1}\ ts)]} \\
\multicolumn{4}{r}{\text{if $\value = \intVal{r}{\mathbf{true}}{bool}$}} \\
\\
\opSem[s]{OP-IfFalse}{{\bf if}\ (\value)\ S_1\ {\bf else}\ S_2}{[gs\,\vert\,(lms, vp, \si{S_2}\ ts)]} \\
\multicolumn{4}{r}{\text{ if $\value = \intVal{r}{\mathbf{false}}{bool}$}} \\
\end{operSem}

\begin{operSem}[Conditional cycle -- while]{
While is syntactically converted to a corresponding {\bf if} statement.
}
\opSem[r]{OP-While}{{\bf while}\ \baseterm{(}E\baseterm{)}\ S}{} \\
&\multicolumn{2}{r}{[gs\,\vert\,(lms, vp, \si{{\bf if}\ \baseterm{(}E\baseterm{)}\ \{S\ {\bf while}\ (E)\ S \}\ {\bf else}\ \baseterm{;}}\ ts)]} \\
\end{operSem}

\begin{operSem}[Method call]{
Call of a method $m$ whose parameters are expressions is rewritten to a call of method $m$ whose parameters are values. Parameters passed to the method are evaluated in the original variable context $vp$ (rule \semRule{OP-MethodCallExpr}).

The call of the method $m$ with value parameters is evaluated in two different ways depending on whether $m$ represents a classical method or a quantum operator.

In the case when $m$ represents a classical method, the call of a method $m$ is rewritten to the unwound body of method $m$ (rule \semRule{OP-DoMethodCallCl}) translated to the internal syntax by \refConcept[method typing context]{$M_B$} taken from \refConcept{method typing context}.

If $m$ represents a quantum operator $\E_m$, the operator $\E_m$ is applied to a quantum subsystem specified by the parameters $\value_1 = \intVal{r_1}{v_1}{T_1},\dots,@\value_n = \intVal{r_n}{v_n}{T_n}$. \refConcept{Values} $v_1,\dots,v_n$ are either global references to quantum storage $(GQuantum, l_{r_1}),$ $\dots, (GQuantum, l_{r_n})$ or $\bot$.  In the case when $\bot$ is referred, a run-time error \RTEuninitVarUsage{} occurs (\semRule{OP-MethodCallQUninit}).

The condition that all manipulated quantum system are physically different can be reformulated as: all the indices in lists $l_{r_1},\dots,l_{r_n}$ are mutually different, {\it ie.} $\recfunc{\set}(l_{r_j})\cap \recfunc{\set}(l_{r_k}) = \emptyset$ and $\recfunc{|l_{r_j}|} = |\recfunc{\set}(l_{r_j})|$ for all $1 \leq j,k \leq n,\ j\neq k$. If this condition is not satisfied, runtime error \RTEqVarOverlap{} is invoked (\semRule{OP-MethodCallQOverlap}).

The list $qsi$ of indices of quantum systems to be measured is given by a concatenation of individual linearized lists: $qsi =l_{r_1}{\cdot}\dots{\cdot}l_{r_n}$, which determines quantum system $q_{qsi}$. Dimension $d_{q_{qsi}}$ of the quantum system $q_l$ is calculated from the global part $((\rho, L),C)$ of the configuration as $$d_{q_{qsi}} = \sum_{i=1}^{|qsi|}L_{qsi_i}.$$ We denote $d$ the order of matrix $\rho$ and $\bar{d}$ the dimension of untouched part of the system, $\bar{d}=d/d_{q_{qsi}}$ (rule \semRule{OP-DoMethodCallQ}).
}
\opSemVerb[e]{OP-MethodCallExpr}{[gs\,\vert\,(lms, vp, \si{m\baseterm{(}\listOf{\value}\baseterm{,}E\baseterm{,} \listOf{E}\baseterm{)}}\ ts)}{} \\
\multicolumn{4}{r}{[gs\,\vert\,(lms, vp, \si{E}\ \si{m\baseterm{(}\listOf{\value}\baseterm{,} \bullet\baseterm{,} \listOf{E}\baseterm{)}}\ ts)]} \\
\\
\opSemVerb[e]{OP-DoMethodCallCl}{[gs\,\vert\,\baseterm{(}lms, vp_G, \si{m\baseterm{(}\value_1\baseterm{,}\dots\baseterm{,}\value_n\baseterm{)}}\ ts)}{} \\
\multicolumn{4}{r}{[gs\,\vert\,(lms', [vp_G \circ_G [\refConcept[square]{\square} \circ_L vp']], \refConcept[method typing context]{M_B}(m)\ \si{\circM}\, ts)]} \\
\multicolumn{4}{r}{\begin{array}[t]{l}
\text{where } \begin{array}[t]{l}
vp' = (\refConcept[variable properties]{[a_1 \mapsto r'_1,\dots,a_n \mapsto r'_n],[],[],[a_1 \mapsto \type{T}_1,\dots,a_n \mapsto \type{T}_n,@retVal \mapsto \type{T}]}) \\
lms' = lms[r'_1 \mapsto v_1]_+\dots[r'_n \mapsto v_n]_+
\end{array}
\\

\text{given that } \begin{array}[t]{l}
 \value_1 = \intVal{r_1}{v_1}{T_1},\dots, \value_n = \intVal{r_n}{v_n}{T_n}\\
 r'_i = \begin{cases}
	(Classical,nc_i)	& \text{if $r_i = \none$} \\
	r_i	& \text{otherwise}
\end{cases} \\
 \text{where all $nc_i$ are mutually different natural numbers} \\
 \text{\qquad such that $\refConcept[function $lms$]{lms}((Classical,nc_i))$ is not defined} \\
 \text{$m$ represents a classical method and} \\
 \type{T}\ m(\type{T}_1\ a_1, \dots \type{T}_n\ a_n) \text{ is a header of method $m$}
\end{array}
\end{array} \text{\quad}}\\
\\
\opSemVerb[rte]{OP-MethodCallQUninit}{[gs\,\vert\,(lms, vp, \si{m\baseterm{(}\value_1\baseterm{,}\dots\baseterm{,}\value_n\baseterm{)}}\ ts)]}{} \\
\multicolumn{4}{r}{[gs\,\vert\,(lms, vp, \RTEuninitVarUsage)]} \\
\multicolumn{4}{r}{\text{given that} \begin{array}[t]{l}
  \value_1 = \intVal{r_1}{v_1}{T_1},\dots, \value_n = \intVal{r_n}{v_n}{T_n}\\
\exists i: v_i = \bot\\
\end{array}} \\
\\
\opSemVerb[rte]{OP-MethodCallQOverlap}{[gs\,\vert\,(lms, vp, \si{m\baseterm{(}\value_1\baseterm{,}\dots\baseterm{,}\value_n\baseterm{)}}\ ts)]}{} \\
\multicolumn{4}{r}{[gs\,\vert\,(lms, vp, \RTEqVarOverlap)]} \\
\multicolumn{4}{r}{\text{given that} \begin{array}[t]{l}
  \value_1 = \intVal{r_1}{v_1}{T_1},\dots, \value_n = \intVal{r_n}{v_n}{T_n}\\
  v_1 = (GQuantum, l_{r_1}), \dots, v_n = (GQuantum, l_{r_n})\\
\recfunc{\set}(l_{r_j})\cap \recfunc{\set}(l_{r_k}) \neq \emptyset \text{ for some } 1 \leq j,k \leq n,\ j\neq k \\
\text{or} \\
\recfunc{|l_{r_j}|} \neq |\recfunc{\set}(l_{r_j})| \text{ for some } 1 \leq j \leq n \\
\end{array}} \\
\\
\end{operSem}

\begin{operSem*}
\opSemVerb[v]{OP-DoMethodCallQ}{[gs\,\vert\,(lms, vp, \si{m\baseterm{(}\value_1\baseterm{,}\dots\baseterm{,}\value_n\baseterm{)}}\ ts)]}{} \\
\multicolumn{4}{r}{[gs'\,\vert\,(lms, vp, \si{\intVal{\none}{\bot}{void}}\ ts)]} \\
\multicolumn{4}{r}{\begin{array}[t]{l}
 \text{where } \begin{array}[t]{l}
  gs' = ((\rho',L), C) \\
  \rho' = \Pi^T(\E_m\otimes I_{\bar{d}}(\Pi\rho\Pi^T))\Pi \\
 \end{array} \\
 \text{given that} \begin{array}[t]{l}
  \value_1 = \intVal{r_1}{v_1}{T_1},\dots, \value_n = \intVal{r_n}{v_n}{T_n}\\
  gs = ((\rho,L), C) \\
  v_1 = (GQuantum, l_{r_1}), \dots, v_n = (GQuantum, l_{r_n})\\
  \recfunc{\set}(l_{r_j})\cap \recfunc{\set}(l_{r_k}) = \emptyset \text{ for all } 1 \leq j,k \leq n,\ j\neq k \\
  \recfunc{|l_{r_j}|} = |\recfunc{\set}(l_{r_j})| \text{ for all } 1 \leq j \leq n \\
  \Pi \text{ is a permutation matrix which places affected quantum} \\
  \text{\hspace{1cm}systems to the head of $\rho$ in the order given by }\value_1,\dots,\value_n \\
  \text{$m$ represents a quantum operator $\E_m$}
 \end{array}
\end{array} \text{\quad}}\\
\end{operSem*}

\begin{operSem}[Returning from a method]{
When a method evaluation finishes, the control is passed back to its caller. The place where the called method was invoked by the caller is marked by the $\circM$ symbol (see \semRule{OP-DoMethodCallCl}). If a method returns no value, it can either end without \return statement just by evaluating the last statement in the method (handled by \semRule{OP-ReturnVoidImpl}) or by explicit \baseterm{return;} statement. In that case the \baseterm{return;} statement pops everything from the term stack until it finds the symbol $\circM$ (\semRule{OP-ReturnVoid}). When the method returns a value, the return value is evaluated first (\semRule{OP-ReturnExpr}) and then the return value is then left on top of the stack after popping all symbols up to $\circM$ from the stack (\semRule{OP-ReturnValue}).
}
\opSemVerb[ret]{OP-ReturnVoid}{[gs\,\vert\,(lms, [vp_G \circ_G vp], \si{\bf return\baseterm{;}}\ ts_M \,\si{\circM}\, ts)]}{\qquad} \\
& \multicolumn{3}{r}{[gs\,\vert\,(lms, vp_G, \si{\intVal{\none}{\bot}{void}}\ ts)] \qquad\ } \\
& \multicolumn{3}{r}{\text{where $ts_M$ does not contain $\circM$}} \\
\\
\opSemVerb[s]{OP-ReturnVoidImpl}{[gs\,\vert\,(lms, [vp_G \circ_G vp], \,\si{\circM}\, ts)]}{\qquad} \\
& \multicolumn{3}{r}{[gs\,\vert\,(lms, vp_G, \si{\intVal{\none}{\bot}{void}}\ ts)] \qquad\ } \\
\\
\opSemVerb[e]{OP-ReturnExpr}{[gs\,\vert\,(lms, vp, \si{{\bf return}\ E\baseterm{;}}\ ts)]}{} \\
& \multicolumn{3}{r}{[gs\,\vert\,(lms, vp, \si{E}\ \si{{\bf return}\ \bullet\baseterm{;}}\ ts) \qquad\ } \\
\\
\opSemVerb[ret]{OP-ReturnValue}{[gs\,\vert\,(lms, [vp_G \circ_G vp], \si{{\bf return}\ \value\baseterm{;}}\ ts_M \,\si{\circM}\, ts)] \qquad}{}\\
& \multicolumn{3}{r}{[gs\,\vert\,(lms, vp_G, \si{\value}\ ts)]\qquad\ } \\
& \multicolumn{3}{r}{\text{where $ts_M$ does not contain $\circM$}} \\
\end{operSem}

\begin{operSem}[Forking]{
Forking creates a new process which is started from given method. 
As {\bf fork} contains a method call construct, the rule \semRule{OP-ForkExpr} for evaluation of arguments is similar to the rule \semRule{OP-MethodCallExpr}. In the rule \semRule{OP-DoFork}, a new process is started, values passed as parameters to the forked method are copied to the new process memory. The non-duplicable values passed as parameters to the forked method are unmapped from the parent process memory.
}
\opSemVerb[e]{OP-ForkExpr}{[gs\,\vert\,(lms, vp, \si{{\bf fork}\ m\baseterm{(}\listOf{\value}\baseterm{,} E\baseterm{,} \listOf{E}\baseterm{);}}\ ts)]}{} \\
\multicolumn{4}{r}{[gs\,\vert\,(lms, vp, \si{E}\ \si{{\bf fork}\ m\baseterm{(}\listOf{\value}\baseterm{,} \bullet\baseterm{,} \listOf{E}\baseterm{);}}\ ts)]} \\
\\
\opSemVerb[p]{OP-DoFork}{[gs\,\vert\,(lms, vp, \si{{\bf fork}\ m\baseterm{(}\value_1\baseterm{,}\dots\baseterm{,}\value_n\baseterm{);}}\ ts)]}{} \\
\multicolumn{4}{r}{[gs\,\vert\,(lms'_1, vp, ts)\parallel (lms'_2, \refConcept[blacksquare]{\blacksquare}, \si{m(\value'_1,\dots,\value'_n)})]} \\
\multicolumn{4}{r}{\begin{array}[t]{l}\text{where }\begin{array}[t]{l}
 lms'_1 = \refConcept{unmap_{nd}}(\{r_1,\dots,r_n\}, lms_1) \\
 lms'_2 = ([],[],[],[])[r'_1 \mapsto v_1]_+\dots[r'_n \mapsto v_n]_+ \\
 \value'_1 = \intVal{r'_1}{v_1}{T_1},\dots, \value'_n = \intVal{r'_n}{v_n}{T_n}\\
\end{array} \text{\quad}
\\
\text{given that }\begin{array}[t]{l} 
  \value_1 = \intVal{r_1}{v_1}{T_1},\dots, \value_n = \intVal{r_n}{v_n}{T_n}\\
 r'_i = \begin{cases}
	(Classical,nc_i)	& \text{if $r_i = \none$} \\
	r_i	& \text{otherwise}
\end{cases} \\
 \text{where all $nc_i$ are mutually different natural numbers} \\
 \text{\qquad such that $lms_{1}((Classical,nc_i))$ is not defined} \\
\end{array}
\end{array}}\\
\end{operSem}

\begin{operSem}[Measurement]{
Measurement is performed when ${\bf measure}\baseterm{(}b,e_1\baseterm{,}\dots\baseterm{,}e_n\baseterm{)}$ primitive method is evaluated. Its first argument $b$ determines measurement basis, the other arguments determine quantum systems that are to be simultaneously measured. Arguments $e_1,\dots,e_n$ evaluate to internal values $\value_1 = \intVal{r_1}{v_1}{T_1}, \dots, \value_n = \intVal{r_n}{v_n}{T_n}$. \refConcept{Values} $v_1,\dots,v_n$ are either global references to quantum storage $(GQuantum, l_{r_1}),$ $\dots, (GQuantum, l_{r_n})$ or $\bot$. In the case when $\bot$ is referred, a run-time error \RTEuninitVarUsage occurs (\semRule{OP-MeasureUninit}).

The condition that all the measured system are physically different can be reformulated as: all the indices in lists $l_{r_1},\dots,l_{r_n}$ are mutually different, {\it ie.} $\recfunc{\set}(l_{r_j})\cap \recfunc{\set}(l_{r_k}) = \emptyset$ and $\recfunc{|l_{r_j}|} = |\recfunc{\set}(l_{r_j})|$ for all $1 \leq j,k \leq n,\ j\neq k$. If this condition is not satisfied, runtime error \RTEqVarOverlap{} is invoked (\semRule{OP-MeasureOverlap}).

The list $qsi$ of indices of quantum systems to be measured is given by a concatenation of individual linearized lists: $qsi =l_{r_1}{\cdot}\dots{\cdot}l_{r_n}$, which determines quantum system $q_{qsi}$. Dimension $d_{q_{qsi}}$ of the quantum system $q_l$ is calculated from the global part $((\rho, L),C)$ of the configuration as $$d_{q_{qsi}} = \sum_{i=1}^{|qsi|}L_{qsi_i}.$$ We denote $d$ the order of matrix $\rho$ and $\bar{d}$ the dimension of unmeasured part of the system, $\bar{d}=d/d_{q_{qsi}}.$

In quantum mechanics, the possible results of the measurement are eigenvalues of the observable. We assign to each eigenvalue an index in a list of all eigenvalues. This index is returned as a result of evaluated {\bf measure} expression. Indeed, it is possible that two or more eigenvalues are the same (they are called {\em degenerate} eigenvalues). In this case, the obtained result is the first index of corresponding eigenvalue in the list. The list is indexed from zero.

Now we can formulate rules for measurement:
}
\opSemVerb[e]{OP-MeasureExpr}{[gs\,\vert\,(lms, vp, \si{{\bf measure}\baseterm{(}\listOf{\value}\baseterm{,} E\baseterm{,} \listOf{E}\baseterm{)}}\ ts]}{} \\
\multicolumn{4}{r}{[gs\,\vert\,(lms, vp, \si{E}\ \si{{\bf measure}\baseterm{(}\listOf{\value}\baseterm{,} \bullet\baseterm{,} \listOf{E}\baseterm{)}}\ ts)]} \\
\\
\opSemVerb[rte]{OP-MeasureUninit}{[gs\,\vert\,(lms, vp, \si{{\bf measure}(\value_b\baseterm{,} \value_1\baseterm{,}\dots\baseterm{,}\value_n\baseterm{)}}\ ts)]}{} \\
\multicolumn{4}{r}{[gs\,\vert\,(lms, vp, \RTEuninitVarUsage)]} \\
\multicolumn{4}{r}{\text{given that} \begin{array}[t]{l}
  \value_1 = \intVal{r_1}{v_1}{T_1},\dots, \value_n = \intVal{r_n}{v_n}{T_n}\\
\exists i: v_i = \bot\\
\end{array}} \\
\\
\opSemVerb[rte]{OP-MeasureOverlap}{[gs\,\vert\,(lms, vp, \si{{\bf measure}\baseterm{(}\value_b, \value_1\baseterm{,}\dots\baseterm{,}\value_n\baseterm{)}}\ ts)]}{} \\
\multicolumn{4}{r}{[gs\,\vert\,(lms, vp, \RTEqVarOverlap)]} \\
\multicolumn{4}{r}{\text{given that} \begin{array}[t]{l}
  \value_1 = \intVal{r_1}{v_1}{T_1},\dots, \value_n = \intVal{r_n}{v_n}{T_n}\\
  v_1 = (GQuantum, l_{r_1}), \dots, v_n = (GQuantum, l_{r_n})\\
\recfunc{\set}(l_{r_j})\cap \recfunc{\set}(l_{r_k}) \neq \emptyset \text{ for some } 1 \leq j,k \leq n,\ j\neq k \\
\text{or} \\
\recfunc{|l_{r_j}|} \neq |\recfunc{\set}(l_{r_j})| \text{ for some } 1 \leq j \leq n \\
\end{array}} \\
\\
%
\opSemVerb[v]{OP-DoMeasure}{[gs\,\vert\,(lms, vp, \si{{\bf measure}\baseterm{(}\value_b\baseterm{,} \value_1\baseterm{,}\dots\baseterm{,}\value_n\baseterm{)}}\ ts)]}{} \\
\multicolumn{4}{r}{\boxplus_i\ p_i \bullet [gs_i\,\vert\,(lms, vp, \si{\intVal{\none}{fdi(i)}{int}}\ ts)]} \\
\multicolumn{4}{r}{\begin{array}[t]{l}\text{where} \begin{array}[t]{l}
gs_i = ((\rho_i,L), C) \\
p_i = \Tr [\Pi^T(P_i \otimes I_{\bar{d}})\Pi \rho \Pi^T(P_i \otimes I_{\bar{d}})^\dagger\Pi] \\
{\displaystyle \rho_i = \frac{\Pi^T(P_i \otimes I_{\bar{d}})\Pi \rho \Pi^T(P_i \otimes I_{\bar{d}})^\dagger\Pi}{p_i}} \\
fdi(i)\text{ is the first index of $i$-th eigenvalue in the list of }\\
\text{\qquad observable eigenvalues (for the case of degenerate eigenvalues)} \\
\end{array}
\\
\text{given that} \begin{array}[t]{l}
  \value_1 = \intVal{r_1}{v_1}{T_1},\dots, \value_n = \intVal{r_n}{v_n}{T_n}\\
v_1 = (GQuantum, l_{r_1}), \dots, v_n = (GQuantum, l_{r_n})\\
gs = ((\rho,L), C) \\
\recfunc{\set}(l_{r_j})\cap \recfunc{\set}(l_{r_k}) = \emptyset \text{ for all } 1 \leq j,k \leq n,\ j\neq k \\
\recfunc{|l_{r_j}|} = |\recfunc{\set}(l_{r_j})| \text{ for all } 1 \leq j \leq n \\
\sum_i \lambda_i P_i \text{ is a spectral decomposition of a measurement in } \\
\text{\hspace{1cm}the basis given by } \value_b\\
\Pi \text{ is a permutation matrix which places measured quantum systems} \\
\text{\hspace{1cm}to the head of $\rho$ in the order given by }\value_1,\dots,\value_n \\
\end{array}
\end{array}}
 \\
\end{operSem}

\begin{operSem}[Communication]{
Communication is performed when there is one process sending a value over a channel end and another process waiting to receive a value over a channel end provided that both channel ends belong to the same channel. This condition is equivalent to the condition that both channel ends refer to the same channel. First three rules (\semRule{OP-SendExpr1}, \semRule{OP-SendExpr2} and \semRule{OP-RecvExpr}) are to evaluate statement arguments and the rule \semRule{OP-SendRecv} performs the communication. When either the sending channel end or the sent value is undefined, or a receiving process attempts to receive from uninitialized channel end, a runtime error \RTEuninitVarUsage{} occurs (rules \semRule{OP-SendUninit} and \semRule{OP-RecvUninit}).

Unique ownership of resources (both quantum and channel) is ensured by unmapping them from the local memory of the sender process using the function $\refConcept{unmap_{nd}}$.
}
\opSemVerb[e]{OP-SendExpr1}{[gs\,\vert\,(lms, vp, \si{{\bf send}(E_1, E_2);}\ ts)]}{} \\
\multicolumn{4}{r}{[gs\,\vert\,(lms, vp, \si{E_1}\ \si{{\bf send}(\bullet, E_2);}\ ts]} \\
\\
\opSemVerb[e]{OP-SendExpr2}{[gs\,\vert\,(lms, vp, \si{{\bf send}(\value_c, E);}\ ts)]}{} \\
\multicolumn{4}{r}{[gs\,\vert\,(lms, vp, \si{E}\ \si{{\bf send}(\value_c, \bullet);}\ ts]} \\
\\
\opSem[rte]{OP-SendUninit}{\send(\value_1, \value_2);}{[gs\,\vert\,(lms, vp, \RTEuninitVarUsage]} \\
\multicolumn{4}{r}{\text{given that } \value_1 = \intVal{r_1}{v_1}{T_1} \text{ and } v_1=\bot \text{ or } \value_2 = \intVal{r_2}{v_2}{T_2} \text{ and } v_2=\bot)} \\
\\
\opSem[e]{OP-RecvExpr}{{\bf recv}(E)}{[gs\,\vert\,(lms, vp, \si{E}\ \si{{\bf recv}(\bullet)}\ ts]} \\
\\
\opSem[rte]{OP-RecvUninit}{{\bf recv}(\value)}{[gs\,\vert\,(lms, vp, \RTEuninitVarUsage]} \\
& \multicolumn{3}{l}{\text{given that } \value = \intVal{r}{v}{T} \text{ and } v=\bot} \\
\\
\hypertarget{semRule:OP-SendRecv}{}\typesetSemRule{OP-SendRecv} & \multicolumn{3}{l}{[gs\,\vert\,(lms_1, vp_1, \si{{\bf send}(\value_{c_1},\value_{e});}\ ts_1)\,\parallel (lms_2, vp_2, \si{{\bf recv}(\value_{c_2})}\ ts_2)] \longrightarrow_p } \\
\multicolumn{4}{r}{[gs\,\vert\,(lms'_1, vp_1, ts_1)\,\parallel\,(lms'_2, vp_2, \si{\intVal{lr'_2}{ lv'_2}{T}}\ ts_2)]} \\
\multicolumn{4}{r}{\begin{array}[t]{l}
\text{where} 
\begin{array}[t]{l}
 lms'_1 = \refConcept{unmap_{nd}}(sentRef, lms_1) \\
 lr'_2 = \begin{cases}
 lr_1 &\text{if } sentRef \in \refConcept{Ref_{nd}}\\
 \none &\text{otherwise}
 \end{cases} \\
 lv'_2 = sentVal \\
 lms_2'= \begin{cases}
 lms_2[sentRef \mapsto sentVal] &\text{if } sentRef \in \refConcept{Ref_{nd}} \\
 lms_2 &\text{otherwise}
 \end{cases}
\end{array}\\
\text{given that}
\begin{array}[t]{l}
 \value_e = \intVal{sentRef}{ sentVal}{T} \\
 \value_{c_1} = \intVal{c_1Ref}{ c_1Val}{T} \\
 \value_{c_2} = \intVal{c_2Ref}{ c_2Val}{T} \\
 c_1Ref \neq \none \text{ and } c_2Ref \neq \none \\
 c_1Val = c_2Val \text{\quad (both ends refer to the same channel)}
\end{array}
\end{array}
}\\
\end{operSem}

\subsection{Rule index}

\begin{description}
\item[$\longrightarrow_e$:]
\semRule{OP-AssignExpr}, \semRule{OP-DoMethodCallCl}, \semRule{OP-ForkExpr}, \semRule{OP-IfExpr}, \semRule{OP-MeasureExpr}, \semRule{OP-MethodCallExpr}, \semRule{OP-PromoExpr}, \semRule{OP-RecvExpr}, \semRule{OP-ReturnExpr}, \semRule{OP-SendExpr1}, \semRule{OP-SendExpr2}, \semRule{OP-SubstE}, \semRule{OP-SubstS}

\item[$\longrightarrow_p$:]
\semRule{OP-DoFork}, \semRule{OP-SendRecv}

\item[$\longrightarrow_r$:]
\semRule{OP-BlockHead}, \semRule{OP-Bracket}, \semRule{OP-VarDeclMulti}, \semRule{OP-While}

\item[$\longrightarrow_{ret}$:]
\semRule{OP-ReturnValue}, \semRule{OP-ReturnVoid}

\item[$\longrightarrow_{rte}$:]
\semRule{OP-AssignQAValueBad}, \semRule{OP-MeasureUninit}, \semRule{OP-MeasureOverlap}, \semRule{OP-MethodCallQUninit}, \semRule{OP-MethodCallQOverlap}, \semRule{OP-RecvUninit}, \semRule{OP-SendUninit}

\item[$\longrightarrow_s$:]
\semRule{OP-Block}, \semRule{OP-BlockEnd}, \semRule{OP-IfFalse}, \semRule{OP-IfTrue}, \semRule{OP-PromoForget}, \semRule{OP-ReturnVoidImpl}, \semRule{OP-Skip}, \semRule{OP-VarDecl}, \semRule{OP-VarDeclAlF}, \semRule{OP-VarDeclChE}

\item[$\longrightarrow_v$:]
\semRule{OP-AllocC}, \semRule{OP-AllocQ}, \semRule{OP-AssignNewValue}, \semRule{OP-AssignQAValue}, \semRule{OP-AssignQValue}, \semRule{OP-AssignValue}, \semRule{OP-DoMeasure}, \semRule{OP-DoMethodCallQ}, \semRule{OP-Var}
\end{description}

\section{Type soundness}
\label{sec:type-soundness}

In this section, we prove type soundness (see {\it eg.} \cite{WriFel94}) for LanQ.

To prove type soundness (also known as Subject reduction) theorem, we first define typing on \refConcept[configuration]{configurations}. Then in series of progress lemmata proofs, we prove that any typable configuration can be reduced by some semantic rule. After this proof, we prove series of type preservation lemmata stating that if a \refConcept{configuration} gets reduced to another \refConcept{configuration}, either a \refConcept{runtime error} occurs or the type of the \refConcept{configuration} is preserved. These lemmata straightforwardly imply type soundness of the language.

However, we cannot prove the type soundness property for the unrestricted language because it is possible that a program gets to a \refConcept{stuck} \refConcept{configuration} during the evaluation. This can happen because the \send{}and \recv{}constructs are blocking and synchronizing actions. Consider a process which attempts to send a value over a channel where no other process is receiving the values from the other end of the channel. Then no semantic rule can be applied to the sending process. Symmetrically, it is indeed possible to define a process which attempts to receive a value from a channel where no other process sends a value over this channel. Such a process also cannot evolve, therefore the evaluation can get to a \refConcept{stuck} \refConcept{configuration}.

In other words, we can prove type soundness only for the noncommunicating part of the language. Nevertheless, if to each \baseterm{send} statement there is a corresponding \baseterm{recv} expression, then it can be proved that the evaluation of a well-typed \refConcept{configuration} never gets \refConcept{stuck}, hence type soundness can be proved for the unrestricted language.

\subsection{Typing of configurations}

To prove LanQ type soundness, we follow the approach of \cite{BieParPit2004}. Before proving type preservation, we define typing of \refConcept[configuration]{configurations} in this subsection in Figures \ref{fig:typRules:termStack} and \ref{fig:typRules:conf}.

Any \refConcept{configuration} $C = [gs\,\vert\,ls_1\parallel\dots\parallel ls_n]$ is assigned a type $T$, written as $C:T$ which is a cartesian product of types of \refConcept[local process configuration]{local process configurations} $ls_1,\dots,ls_n$. If the type of $ls_i$ is $T_i$ then $T = (T_1,\dots,T_n)$ (see the typing rule \typeRule{T-Config}).

We call a configuration which is assigned a type a \concept{well-typed configuration}.

The typing rules provide rules for well-formedness of \refConcept[configuration]{configurations}, hence also for \refConcept[local process configuration]{local process configurations}. This indeed means that structure of \refConcept{variable properties} and a \refConcept{term stack} are tightly connected: To any block end mark $\circL$ on the \refConcept{term stack}, the \refConcept{variable properties} must contain a nonempty list of variable properties $[vp \circ_L vp_L] \in \refConcept{VarProp_L}$ (see rule \typeRule{TC-BlockEnd}); to any method call mark $\circM$ on the \refConcept{term stack}, the \refConcept{variable properties} must contain a nonempty list of lists of variable properties $[vp_G \circ_G vp] \in \refConcept{VarProp}$ (see rules \typeRule{TC-RetHole}, \typeRule{TC-RetExpr}, \typeRule{TC-RetVoid} and \typeRule{TC-RetImpl}).

Typing of \refConcept[local process configuration]{local process configurations} as depicted in Figure \ref{fig:typRules:termStack} needs a deeper explanation. Contrary to usual typing of configurations known {\it eg.} from $\lambda$-calculi where one configuration contains only one expression, we deal with the situation where there are many ``expressions'' in one configuration. These expressions are in our case \refConcept{term stack} elements and they altogether form a \refConcept{term stack} of a \refConcept{local process configuration}.

\begin{figure}[hpt]
\small
\hspace{-1em}
\begin{tabular*}{\linewidth}{lc}
\typeRuleDef{TC-Empty}{$\overline{M;\Gamma \vdash_C (lms, \refConcept[blacksquare]{\blacksquare}, \varepsilon) : \type{\tau \rightarrow \tau}}$} \\
\\[-0.1\normalbaselineskip]

\typeRuleDef{TC-Runtime}{$\overline{M;\Gamma \vdash_C (lms, vp, \si{RTErr}) : \type{\sigma \rightarrow \tau}}$} \\
\\[-0.1\normalbaselineskip]

\typeRuleDef{TC-ExprHole}{\infer{M;\Gamma \vdash_C (lms, vp, \si{Ec}\ ts) : \type{\sigma \rightarrow \tau}}{M;\Gamma, \refConcept{vpContext}(vp), \bullet:\type{\sigma} \vdash_T Ec : \type{\tau'} \quad M;\Gamma \vdash_C (lms,vp, ts) : \type{\tau' \rightarrow \tau}}} \\
\\[-0.1\normalbaselineskip]
\typeRuleDef{TC-StatHole}{\infer{M;\Gamma \vdash_C (lms, vp, \si{Sc}\ ts) : \type{\sigma \rightarrow \tau}}{M;\Gamma, \refConcept{vpContext}(vp), \bullet:\type{\sigma} \vdash_T Sc : \type{void} \quad M;\Gamma \vdash_C (lms,vp, ts) : \type{void \rightarrow \tau}}} \\
&\multicolumn{1}{r}{if $Sc \neq \return\ \bullet\baseterm;$} \\
\\[-0.1\normalbaselineskip]

\typeRuleDef{TC-RetHole}{\infer{M;\Gamma \vdash_C (lms, [vp_G\circ_G vp], \si{\return\ \bullet\baseterm;}\ {\dots}\ \si{\circM}\ ts) : \type{\sigma \rightarrow \tau}}{M;\Gamma \vdash_C (lms,vp_G, ts) : \refConcept{typeOf}(@retVal, [vp_G \circ_G vp]) \rightarrow \type{\tau}}} \\
\\[-0.1\normalbaselineskip]

\typeRuleDef{TC-ExprClo}{\infer{M;\Gamma \vdash_C (lms, vp, \si{E}\ ts) : \type{\sigma \rightarrow \tau}}{M;\Gamma, \refConcept{vpContext}(vp) \vdash_T E : \type{\tau'} \quad M;\Gamma \vdash_C (lms,vp, ts) : \type{\tau' \rightarrow \tau}}} \\
\\[-0.1\normalbaselineskip]
\typeRuleDef{TC-StatClo}{\infer{M;\Gamma \vdash_C (lms, vp, \si{S}\ ts) : \type{\sigma \rightarrow \tau}}{M;\Gamma, \refConcept{vpContext}(vp) \vdash_T S : \type{void} \quad M;\Gamma \vdash_C (lms,vp, ts) : \type{void \rightarrow \tau}}} \\
&\multicolumn{1}{r}{if $S \neq \return\ E\baseterm;$, $S \neq \return\baseterm;$} \\
\\[-0.1\normalbaselineskip]

\typeRuleDef{TC-RetExpr}{\infer{M;\Gamma \vdash_C (lms, [vp_G \circ_G vp], \si{\return\ E\baseterm;}\ {\dots}\ \si{\circM}\ ts) : \type{\sigma \rightarrow \tau}}{
\begin{array}{l}
 M;\Gamma, \refConcept{vpContext}([vp_G \circ_G vp]) \vdash_T E : \refConcept{typeOf}(@retVal,[vp_G \circ_G vp]), \\
 M;\Gamma \vdash_C (lms,vp_G, ts) : \refConcept{typeOf}(@retVal,[vp_G \circ_G vp])\type{ \rightarrow \tau}
\end{array}}} \\
\\[-0.1\normalbaselineskip]
\typeRuleDef{TC-RetVoid}{\infer{M;\Gamma \vdash_C (lms, [vp_G \circ_G vp], \si{\return\baseterm;}\ {\dots}\ \si{\circM}\ ts) : \type{\sigma \rightarrow \tau}}{M;\Gamma \vdash_C (lms,vp_G, ts) : \type{\void \rightarrow \tau}}} \\
\\[-0.1\normalbaselineskip]
\typeRuleDef{TC-RetImpl}{\infer{M;\Gamma \vdash_C (lms, [vp_G \circ_G vp], \si{\circM}\ ts) : \type{\sigma \rightarrow \tau}}{M;\Gamma \vdash_C (lms,vp_G, ts) : \refConcept{typeOf}(@retVal,[vp_G \circ_G vp]) \rightarrow \tau}} \\
\\[-0.1\normalbaselineskip]
\typeRuleDef{TC-BlockHead}{\infer
 {M;\Gamma \vdash_C (lms, vp, \si{Be_0\ Be_1\ \dots\ Be_n}\ ts) : \type{\sigma \rightarrow \tau}}
 {M;\Gamma \vdash_C (lms, vp, \si{Be_0}\ \si{Be_1\ \dots\ Be_n}\ ts) : \type{void \rightarrow \tau}}} \\
\\[-0.1\normalbaselineskip]

\typeRuleDef{TC-BlockEnd}{\infer
 {M;\Gamma \vdash_C (lms, [vp_G \circ_G [vp \circ_L vp_L]], \si{\circL}\ ts) : \type{void \rightarrow \tau}}
 {M;\Gamma \vdash_C (lms,[vp_G \circ_G vp], ts) : \type{void \rightarrow \tau}}} \\
\\[-0.1\normalbaselineskip]

\typeRuleDef{TC-VarDeclMulti}{\infer
 {M;\Gamma \vdash_C (lms, vp, \si{T\ I_0,I_1,\dots,I_n\baseterm;}\ ts) : \type{\sigma \rightarrow \tau}}
 {M;\Gamma \vdash_C (lms, vp, \si{T\ I_0\baseterm;}\ \si{T\ I_1,\dots,I_n\baseterm;}\ ts) : \type{void \rightarrow \tau}}} \\
\\[-0.1\normalbaselineskip]

\typeRuleDef{TC-VarDeclOne}{\infer
 {M;\Gamma \vdash_C (lms, vp, \si{T\ I\baseterm;}\ ts) : \type{\sigma \rightarrow \tau}}
 {\begin{array}{c}
\refConcept{varRef}(I, vp) \text{ is undefined} \\
 M;\Gamma \vdash_C (lms, vp', ts) : \type{void \rightarrow \tau}
  \end{array}}} \\
\multicolumn{2}{r}{\quad \text{ where $vp' = vp[I \mapsto \type{T}]_{+,type}$}} \\
\\[-0.1\normalbaselineskip]

\typeRuleDef{TC-VarDeclChE}{\infer
 {M;\Gamma \vdash_C (lms, vp, \si{\type{channel[}T\type{]}\ I_0\ \baseterm{withends}\baseterm[I_1\baseterm,I_2\baseterm]\baseterm;}\ ts) : \type{\sigma \rightarrow \tau}}
 {
 \begin{array}{l}
   \refConcept{varRef}(I_i, vp) \text{ is undefined for $i = 0,..2$}, \\
    I_0, I_1, I_2 \text{ are mutually different}, \\
 M;\Gamma\vdash_C (lms, vp', ts) : \type{void \rightarrow \tau}
 \end{array}}} \\
\multicolumn{2}{r}{\quad \text{ where $vp' = vp[c \mapsto \type{channel[T]}]_{+,type}[c_0 \mapsto \type{channelEnd[T]}]_{+,type}[c_1 \mapsto \type{channelEnd[T]}]_{+,type}$}} \\
\\[-0.1\normalbaselineskip]

\typeRuleDef{TC-VarDeclAlF}{\infer
 {M;\Gamma \vdash_C (lms, vp, \si{I_0\ \baseterm{aliasfor}\ \baseterm[I_1\baseterm,\dots\baseterm,I_n\baseterm]\baseterm;}\ ts) : \type{\sigma \rightarrow \tau}}
 {\begin{array}{c}
   \refConcept{varRef}(I_0, vp) \text{ is undefined}, \\
   M;\Gamma \vdash_T I_i : \type{T_i} \quad \text{$\type{T_i}$ is a quantum type for $i = 1,..n$}, \\
   M;\Gamma \vdash_C (lms,vp', ts) : \type{void \rightarrow \tau}
  \end{array}}
} \\
\multicolumn{2}{r}{\quad \text{ where $vp' = vp[I_0 \mapsto \bigotimes_{\type{i}=1}^n\type{T_i}]_{+,type}$}} \\

\end{tabular*}
\caption{Typing rules for \refConcept[local process configuration]{local process configurations}}
\label{fig:typRules:termStack}
\end{figure}

\begin{figure}[htp]
$\begin{array}{m{0.3\linewidth}c}
\typeRuleDef{T-MixedConf}{\infer{M;\Gamma \vdash \boxplus_{i=1}^{q}p_i\bullet [GS_i\,\vert\,LS_{i,1} \parallel\dots\parallel LS_{i,n}]}{\forall i \in 1,\dots,q:\ M;\Gamma\vdash [GS_i\,\vert\,LS_{i,1} \parallel\dots\parallel LS_{i,n}]: \tau_i}} \\

\quad \\

\typeRuleDef{T-Config}{\infer{M;\Gamma \vdash [GS\,\vert\,LS_{1} \parallel\dots\parallel LS_{n}]: \prod_{i=1}^n \tau_i}{\forall i \in 1,\dots,n:\ M;\Gamma \vdash_C LS_i : \void \rightarrow \tau_i}} \\
\end{array}$
\caption{Typing rules for \refConcept[configuration]{configurations}}
\label{fig:typRules:conf}
\end{figure}


The type of a \refConcept{local process configuration} is defined as $\sigma \rightarrow \tau$ for types $\sigma,\tau$. The type $\sigma$ specifies the type of the hole in the top \refConcept{term stack} element, $\tau$ specifies the type of the result value. When there is no hole in the top \refConcept{term stack} element, $\sigma$ is \void.

The top \refConcept{term stack} element $TE$ can contain at most one hole. As the hole type $\sigma$ is always known and the \refConcept{variable typing context} can be deduced from \refConcept{variable properties}, the type $\tau'$ of $TE$ is derivable from the typing rules defined in Section \ref{sec:typing}.

If $TE$ is the only \refConcept{term stack} element in the \refConcept{local process configuration}, its type determines the type of the result value $\tau$. Otherwise, the type $\tau'$ is the type of a hole in the \refConcept{term stack} element right beneath $TE$ and we can continue with typing the \refConcept{local process configuration} where $TE$ is popped from the \refConcept{term stack}.

To derive \refConcept{variable typing context} from \refConcept{local process configuration} \refConcept{variable properties}, we define a function $vpContext$. This function is defined as follows:

\begin{definition}
We define a function $\concept{vpContext}: \refConcept{VarProp} \rightarrow (\refConcept{Names} \rightharpoonup \refConcept{Types})$ which creates a \refConcept{variable typing context} from \refConcept{variable properties} as:
\noindent\begin{eqnarray*}
\hline\\
vpContext([L_1 \circ_G K]) &\isdef &vpContext_{L}(K) \\
vpContext(\refConcept[blacksquare]{\blacksquare}) &\isdef &[] \\
\hline
\end{eqnarray*}

\noindent where $vpContext_{L}: \refConcept{VarProp_L} \rightarrow (\refConcept{Names} \rightharpoonup \refConcept{Types})$ is a function for getting a \refConcept{variable typing context} from \refConcept[variable properties]{local variable properties} defined as:
\begin{eqnarray*}
\hline
vpContext_{L}([L_1 \circ_L (f_{var},f_{ch},f_{qa},f_{type})]) &\isdef & 
vpContext_L(L_1) \refConcept[g*f]{*} f_{type} \\
vpContext_{L}(\refConcept[square]{\square}) &\isdef &[] \\
\hline
\end{eqnarray*}
\end{definition}

\subsection{Progress}

In this subsection, we prove series of progress lemmata, {\it ie.} assertions claiming that any \refConcept{well-typed configuration} which is not \refConcept[terminal configuration]{terminal} can be reduced by some semantic rule. It also follows from the proof that the choice of the semantic rule is unique, hence the semantics is deterministic.

\begin{lemma}[Progress Lemma for probabilism]
If $C_0 = \boxplus_{i=1}^{q} p_i \bullet [gs_i\,\vert\, ls_{i,0}]$, $q>1$ and $\vdash C_0:\tau$ then there exists a \refConcept{configuration} $C_1$ such that $C_0 \xrightarrow{p} C_1$.
\end{lemma}

\begin{proof}
Such a \refConcept{mixed configuration} $C_0$ is reduced by the rule \semRule{NP-ProbEvol}.
\end{proof}

\begin{lemma}[Progress Lemma for local processes]
If $C_0 = [gs_0\,\vert\,(lms_0,vp_0,\si{TE}\ ts_0)]$ is not \refConcept[terminal configuration]{terminal}, $TE \neq \recv\baseterm(\value\baseterm)$, $TE \neq \send\baseterm(\value_1\baseterm,\value_2\baseterm)\baseterm;$, and $\vdash C_0:\tau$ then there exists a \refConcept{mixed configuration} $C_1$ such that $C_0 \longrightarrow C_1$.
\end{lemma}

\begin{proof}
By case analysis of all possibilities of the top \refConcept{term stack} element $TE$:

\begin{description}
\item[Case $TE = {\bf new}\ \type{T}()$:] As $C_0$ is well-typed, we know that \type{T} is either $\type{Q}_d$ or $\type{channel[T]}$ (from the rule \typeRule{T-Alloc}). If $\type{T}=\type{Q}_d$, then the \refConcept{configuration} $C_0$ is reduced by the rule \semRule{OP-AllocQ}. Otherwise $\type{T}=\type{channel[T]}$ and the \refConcept{configuration} $C_0$ is reduced by the rule \semRule{OP-AllocC}.

\item[Case $TE = I = E$:] As $C_0$ is well typed, we know that types of $I$ and $E$ match (from the rule \typeRule{T-Assign}). If $E = \value = {\intVal{lr}{lv}{T}}$ then one of the following rules is applied:
	\begin{itemize}
	 \item \semRule{OP-AssignNewValue} if $lv \neq \bot$ and $lr = \none$,
	 \item \semRule{OP-AssignQValue} if $lr = (Quantum, q)$ and $\refConcept{aliasSubsyst}(I,vp_0)$ is not defined,
	 \item \semRule{OP-AssignQAValue} if $lr = (Quantum, q)$ and $\refConcept{aliasSubsyst}(I,vp_0)$ is defined,
	 \item \semRule{OP-AssignValue} otherwise.
	\end{itemize}
Otherwise, the \refConcept{configuration} $C_0$ is reduced by the rule \semRule{OP-AssignExpr}.

\item[Case $TE = I(\listOf{E})$:] As $C_0$ is well typed, we know that $I$ denotes either a quantum operator or a classical method (from the rule \typeRule{T-MethodCall}). If $\listOf{E} = \listOf{\value}$ then one of the following rules is applied:
	\begin{itemize}
	 \item \semRule{OP-DoMethodCallCl} if $I$ represents a classical method,
	 \item \semRule{OP-DoMethodCallQ} if $I$ represents a quantum operator.
	\end{itemize}
Otherwise the \refConcept{configuration} $C_0$ is reduced by the rule \semRule{OP-MethodCallExpr}.

\item[Case $TE = {\bf measure}\baseterm(\listOf{E}\baseterm)$:] If $\listOf{E} = \listOf{\value}$ then the \refConcept{configuration} $C_0$ is reduced by the rule \semRule{OP-DoMeasure}, otherwise it is reduced by the rule \semRule{OP-MeasureExpr}.

\item[Case $TE = {\bf recv}\baseterm(E\baseterm)$:] As $E$ cannot be $\value$ (from assumptions) the \refConcept{configuration} $C_0$ is reduced by the rule \semRule{OP-RecvExpr}.

\item[Case $TE = I$:] As $C_0$ is well typed, we know that $\refConcept{varRef}(I, vp_0)$ is defined. Configuration $C_0$ is reduced by the rule \semRule{OP-Var}.

\item[Case $TE = \value$:] 
	If $\vert ts_0 \vert = 1$ then $C_0$ is a \refConcept{terminal configuration}.
	Otherwise let $UT$ be the first symbol under the top element of the stack. As the \refConcept{configuration} is well-typed, we know that $UT$ must be a symbol containing a hole. Now:
	\begin{description}
	 \item[Case $UT = Ec$:] Configuration $C_0$ is reduced by the rule \semRule{OP-SubstE}.
	 \item[Case $UT = Sc$:] Configuration $C_0$ is reduced by the rule \semRule{OP-SubstS}.
	\end{description}

\item[Case $TE = \baseterm(E\baseterm)$:] Configuration $C_0$ is reduced by the rule \semRule{OP-Bracket}.

\item[Case $TE = T\ \listOf{I}\baseterm;$:] If $TE = T\ I;$ then the \refConcept{configuration} $C_0$ is reduced by the rule \semRule{OP-VarDecl}. Otherwise it is reduced by the rule \semRule{OP-VarDeclMulti}.

\item[Case $TE = \type{channel[T]}\ I\ \baseterm{withends}\baseterm[I\baseterm,I\baseterm{]}\baseterm;$:] Configuration $C_0$ is reduced by the rule \semRule{OP-VarDeclChE}.
\item[Case $TE = q\ \baseterm{aliasfor}\ \baseterm[q_0\baseterm,\dots\baseterm,q_n\baseterm{]}\baseterm;$:] Configuration $C_0$ is reduced by the rule \semRule{OP-VarDeclAlF}.

\item[Case $TE = \baseterm;$:] Configuration $C_0$ is reduced by the rule \semRule{OP-Skip}.

\item[Case $TE = PE \baseterm{;}$:] If $PE = \value$ then the \refConcept{configuration} $C_0$ is reduced by the rule \semRule{OP-PromoForget}, otherwise it is reduced by \semRule{OP-PromoExpr}.

\item[Case $TE = \circ_L$:] Configuration $C_0$ is reduced by the rule \semRule{OP-BlockEnd}.

\item[Case $TE = \overline{Be}$:] Configuration $C_0$ is reduced by the rule \semRule{OP-BlockHead}.

\item[Case $TE = \baseterm\{ B \baseterm\}$:] Configuration $C_0$ is reduced by the rule \semRule{OP-Block}.

\item[Case $TE = {\bf if}\ \baseterm(E\baseterm)\ S_1\ {\bf else}\ S_2$:] If $E = \value$, then the \refConcept{configuration} is reduced by the rule \semRule{OP-IfTrue} (\semRule{OP-IfFalse}) when $\value$ is \textbf{true} (\textbf{false}). Otherwise, the \refConcept{configuration} $C_0$ is reduced by the rule \semRule{OP-IfExpr}.

\item[Case $TE = {\bf while}\ \baseterm(E\baseterm)\ S$:] Configuration $C_0$ is reduced by the rule \semRule{OP-While}.

\item[Case $TE = \circM$:] Configuration $C_0$ is reduced by the rule \semRule{OP-ReturnVoidImpl}.

\item[Case $TE = \baseterm{return;}$:] Configuration $C_0$ is reduced by the rule \semRule{OP-ReturnVoid}.
\item[Case $TE = \baseterm{return}\ E\baseterm;$:] If $E = \value$ then the \refConcept{configuration} $C_0$ is reduced by the rule \semRule{OP-ReturnValue}, otherwise it is reduced by \semRule{OP-ReturnExpr}.

\item[Case $TE = \baseterm{fork}\ m\baseterm(\listOf{E}\baseterm)\baseterm;$:] If $\listOf{E} = \listOf{\value}$ then the \refConcept{configuration} $C_0$ is reduced by the rule \semRule{OP-DoFork}. Otherwise it is reduced by the rule \semRule{OP-ForkExpr}.

\item[Case $TE = \send\baseterm(E_1\baseterm, E_2\baseterm)\baseterm;$:] If $E_1 \neq \value$ then the \refConcept{configuration} $C_0$ is reduced by the rule \semRule{OP-SendExpr1}. If $E_2 \neq \value$, the \refConcept{configuration} $C_0$ is reduced by the rule \semRule{OP-SendExpr2}. Case $E_1 = \value_1$ and $E_2 = \value_2$ is prohibited by the assumptions.
\end{description}
\end{proof}

\begin{lemma}[Progress Lemma for communication]
If $C_0 = [gs_0\,\vert\,(lms_0,vp_0,\si{\recv\baseterm(\value\baseterm)}\ ts_0) \parallel (lms_1,vp_1,\si{\send\baseterm(\value_1\baseterm,\value_2\baseterm)\baseterm;}\ ts_1) \parallel ls_2 \parallel \dots \parallel ls_n]:\tau$ then there exists a \refConcept{configuration} $C_1$ such that $C_0 \longrightarrow C_1$.
\end{lemma}

\begin{proof}
Such a \refConcept{configuration} is not \refConcept[terminal configuration]{terminal} because of the elements $\recv\baseterm(\value\baseterm)$ and $\send\baseterm(\value_1\baseterm,\value_2\baseterm)\baseterm;$ on tops of the process \refConcept[term stack]{term stacks}. It is reduced by the rule \semRule{OP-SendRecv}.
\end{proof}

{\corollary It is possible that a process evolves to a \refConcept{stuck} \refConcept{configuration}. This is the case when one process attempts to send/receive a value over a channel end and there is no matching process receiving/sending over the corresponding channel end.}

\subsection{Type preservation}

\subsubsection{Evaluation theorems}

In the definition of LanQ semantics, we assumed that an expression always evaluates to a value (or diverges or invokes a \refConcept{runtime error}). This assumption was used in all rules which manipulate a subexpression $Sub$ of a statement or an expression: The subexpression $Sub$ is pushed onto the top of the term stack and the place of the awaited result in the original statement/expression is marked with a hole $\bullet$ (see {\it eg.} the rule \semRule{OP-AssignExpr}). We expect that if the evaluation correctly finishes, the subexpression evaluates to an \refConcept{internal value} that in the subsequent step replaces the hole.

However, we have not yet shown that the subexpression evaluation accomplishes this assertion. We have to prove that the statement/expression awaiting the result of the subexpression evaluation is not modified before the evaluation of the subexpression yields an \refConcept{internal value} (unless a \refConcept{runtime error} occurs).

This will be shown in this subsection. We will use the following semantic predicates on \refConcept[configuration]{configurations}:
\begin{itemize}
 \item $\refConcept{ExpOk}$ which is defined for configurations where the top \refConcept{term stack} element of the first \refConcept{local process configuration} is some \refConcept[syntax!expression]{expression} $E$, and
 \item $\refConcept{BFOk}$ and $\refConcept{StkRetOk}$ which is defined for configurations where the top \refConcept{term stack} element of the first \refConcept{local process configuration} is some \refConcept[syntax!block-forming statement]{block-forming statement}$B$.
\end{itemize}
These predicates satisfiability is determined by the future evolution of the \refConcept[configuration]{configurations} in question.

We will further define inductive {\em syntactic} predicates $\refConcept{ExpOk_i}$, $\refConcept{BFOk_i}$ and show the connection between these syntactic predicates and the semantic ones.

The defined predicates are then used in the proof of the statement that if the syntactic predicate $\refConcept{RetOk}$ defined in Section \ref{sec:typing-rules} is satisfied for a \refConcept[syntax!block-forming statement]{block-forming statement} $B$ then any control path of evaluation of $B$ reaches a \baseterm{return;} or \baseterm{return} $E$\baseterm{;} statement, or a \refConcept{runtime error}, or diverges (see Lemma \ref{lemma-eval-retok-to-return}). This proof is later used in proving type preservation lemma for $\longrightarrow_s$ (see Lemma \ref{lemma-type-preservation-s}, proof of the case \semRule{OP-ReturnVoidImpl}).

\begin{definition}
 For an \refConcept[syntax!expression]{expression} $E$, we define a predicate $\concept{ExpOk}(E)$: $ExpOk(E)$ is satisfied iff for any configuration $C_0=[gs_0\,|\,(lms_0,vp_0,\si{E}\ ts_0)]:\tau$, the evaluation of $C_0$ either diverges or reaches one of the following configurations:
\begin{itemize}[itemsep=0pt]
 \item $[gs_n\,|\,(lms_{0,n},vp_{0,n},\si{\value}\ ts_{0}) \parallel \dots \parallel (lms_{k,n},vp_{k,n}, ts_{k,n})], k \geq 0$, or
 \item $[gs_n\,|\,(lms_{0,n},vp_{0,n},\si{\refConcept{RTErr}}) \parallel \dots \parallel (lms_{k,n},vp_{k,n}, ts_{k,n})], k \geq 0$.
\end{itemize}
\end{definition}

\begin{definition}
For a \refConcept[syntax!block-forming statement]{block-forming statement} $B$, we define a predicate $\concept{BFOk}(B)$: $BFOk(B)$ is satisfied iff for any configuration $C_0=[gs_0\,|\,(lms_0,vp_0,\si{B}\ ts_0)]:\tau$, the evaluation either diverges or reaches one of the following configurations:
\begin{itemize}[itemsep=0pt]
 \item $[gs_n\,|\,(lms_{0,n},vp_{0,n}, ts_{0}) \parallel \dots \parallel (lms_{k,n},vp_{k,n}, ts_{k,n})], k \geq 0$, or
 \item $[gs_n\,|\,(lms_{0,n},vp_{0,n}, \si{\return\baseterm{;}}\ ts_{0,n}\ ts_{0}) \parallel \dots \parallel (lms_{k,n},vp_{k,n}, ts_{k,n})], k \geq 0$, $ts_{0,n}$ does not contain $\circM$,  or
 \item $[gs_n\,|\,(lms_{0,n},vp_{0,n}, \si{\return\ \value\baseterm{;}}\ ts_{0,n}\ ts_{0}) \parallel \dots \parallel (lms_{k,n},vp_{k,n}, ts_{k,n})], k \geq 0$, $ts_{0,n}$ does not contain $\circM$, or
 \item $[gs_n\,|\,(lms_{0,n},vp_{0,n},\si{\refConcept{RTErr}}) \parallel \dots \parallel (lms_{k,n},vp_{k,n}, ts_{k,n})], k \geq 0$.
\end{itemize}

We further define a predicate $\concept{StkRetOk}(B)$: $StkRetOk(E)$ is satisfied iff for any configuration $C_0=[gs_0\,|\,(lms_0,vp_0,\si{B}\ \si{\circM}\ ts_0)]:\tau$, the evaluation either diverges or reaches one of the following configurations:
\begin{itemize}[itemsep=0pt]
 \item $[gs_n\,|\,(lms_{0,n},vp_{0,n},\si{\value}\ ts_{0}) \parallel \dots \parallel (lms_{k,n},vp_{k,n}, ts_{k,n})], k \geq 0$, or
 \item $[gs_n\,|\,(lms_{0,n},vp_{0,n},\si{\refConcept{RTErr}}) \parallel \dots \parallel (lms_{k,n},vp_{k,n}, ts_{k,n})], k \geq 0$.
\end{itemize}
\end{definition}

\begin{lemma}\label{lemma-BFOk=>StkRetOk}
Let $B$ be a \refConcept[syntax!block-forming statement]{block-forming statement} such that $\refConcept{BFOk}(B)$. Then $\refConcept{StkRetOk}(B)$.
\end{lemma}

\begin{proof}
The evaluation starts from the following configuration:
$$[gs_0\,|\,(lms_0,vp_0, \si{B}\ \si{\circM}\ ts_0)]$$
From $\refConcept{BFOk}(B)$ we know that evolution of this configuration:
\begin{itemize}
 \item Diverges or gets to a configuration: $$[gs_n\,|\,(lms_n,vp_n, \si{\refConcept{RTErr}}) \parallel \dots \parallel (lms_{k,n},vp_{k,n}, ts_{k,n})]$$ therefore the lemma holds.
 
 \item Gets to a configuration: $$[gs_{n'}\,|\,(lms_{n'},vp_{n'},\si{\return\baseterm{;}}\ ts_{n'}\ \si{\circM}\ ts_0) \parallel \dots \parallel (lms_{k,n'},vp_{k,n'}, ts_{k,n'})]$$
 By application of \semRule{OP-ReturnVoid}, the 0-th process gets to a configuration: $$[gs_{n'}\,|\,(lms_{n'},vp_{n'},\si{\intVal{\none}{\bot}{void}}\ ts_0) \parallel \dots \parallel (lms_{k,n'},vp_{k,n'}, ts_{k,n'})]$$ therefore the lemma holds.

 \item Gets to a configuration: $$[gs_{n'}\,|\,(lms_{n'},vp_{n'},\si{\return\ \value\baseterm{;}}\ ts_{n'}\ \si{\circM}\ ts_0) \parallel \dots \parallel (lms_{k,n'},vp_{k,n'}, ts_{k,n'})]$$
 By application of \semRule{OP-ReturnValue}, the 0-th process gets to a configuration: $$[gs_{n'}\,|\,(lms_{n'},vp_{n'},\si{\value}\ ts_0) \parallel \dots \parallel (lms_{k,n'},vp_{k,n'}, ts_{k,n'})]$$ therefore the lemma holds.
\end{itemize}
\end{proof}

We want to prove that for any expression $E$ the predicate $\refConcept{ExpOk}(E)$ is satisfied. We do this by defining inductive predicates $ExpOk_i$ and $BFOk_i$ with respect to the structure of $E$ and $B$, respectively. The dependency is the following ($a \leftarrow b$ means $a$ is dependent on $b$):
$$\xymatrix{
ExpOk_0	& 	\ar@{->}[l] BFOk_0 & 	\ar@{->}[l] ExpOk_1 & \ar@{.>}[l]\dots & \ar@{.>}[l] ExpOk_i	&	\ar@{->}[l] BFOk_i	& \ar@{.>}[l]\cdots
}$$


\begin{definition}
Let $E$ be an \refConcept[syntax!expression]{expression}, $B$ be a \refConcept[syntax!block-forming statement]{block-forming statement}. We define predicates $\concept{ExpOk_0}(E)$ and $\concept{BFOk_0}(B):$
\begin{eqnarray*}
\refConcept{ExpOk_0}(E) &\Longleftrightarrow &\text{$E$ does not contain $I(\listOf{E})$ as its subexpression.} \\
BFOk_0(B) &\Longleftrightarrow &\text{$B$ contains only such subexpressions $E$ which satisfy $\refConcept{ExpOk_0}(E)$.}
\end{eqnarray*}

For any $i \in \N_0$, we further define predicates $\concept[ExpOk_i]{ExpOk_{i+1}}(E)$ and $\concept[BFOk_i]{BFOk_{i+1}}(B):$
\begin{eqnarray*}
ExpOk_{i+1}(E) &\Longleftrightarrow &\text{For any subexpression $E_0$ of $E$, $ExpOk_i(E_0)$ is satisfied} \\
& & \text{or $E_0 = I(\listOf{E})$ and $BFOk_i(\refConcept[method typing context]{M_B}(I))$ is satisfied.} \\
BFOk_{i+1}(B) &\Longleftrightarrow &\text{$B$ contains only such subexpressions $E$ which satisfy $\refConcept{ExpOk_i}_{+1}(E)$.}
\end{eqnarray*}
\end{definition}

\begin{lemma}
\label{lemma-eval-expr-base}
Let $E$ be an \refConcept[syntax!expression]{expression} such that $\refConcept{ExpOk_0}(E)$. Then $\refConcept{ExpOk}(E)$ is satisfied.
\end{lemma}

\begin{proof}
By induction on the structure of $E$. Let $C_0=[gs_0\,|\,(lms_0,vp_0,\si{E}\ ts_0)]:\tau$ be any configuration. Base case:

\begin{description}
 \item[Case $E = \value$:] $\refConcept{ExpOk}(\value)$ trivially.
 
 \item[Case $E = I$:] By application of \semRule{OP-Var} we reach configuration $[gs\,|\,(lms,vp,\si{\value}\ ts_0)]$, hence $\refConcept{ExpOk}(E)$.
 
 \item[Case $E = \new\ \type{T}\baseterm{()}$:] By application of \semRule{OP-AllocC}/\semRule{OP-AllocQ} reaches configuration $[gs\,|\,(lms,vp,\si{\value}\ ts_0)]$, hence $\refConcept{ExpOk}(E)$.
 
 \item[Case $E = I = \value$:] By application of \semRule{OP-AssignNewValue}/\semRule{OP-AssignQAValue}/\semRule{OP-AssignQValue}/\semRule{OP-AssignValue} we reach configuration $[gs\,|\,(lms,vp,\si{\value}\ ts_0)]$, hence $\refConcept{ExpOk}(E)$; by application of \semRule{OP-AssignQAValueBad} we reach configuration $[gs\,|\,(lms,vp, \si{\RTEassignIncompatibleQuant})]$, hence $\refConcept{ExpOk}(E)$.

 \item[Case $E = \recv\baseterm(\value\baseterm)$:] By application of \semRule{OP-SendRecv} we reach configuration $[gs\,|\,(lms,vp,\si{\value}\ ts_0)]$, hence $\refConcept{ExpOk}(E)$; by application of \semRule{OP-RecvUninit} we reach configuration $[gs\,|\,(lms,vp,\si{\RTEuninitVarUsage})]$, hence $\refConcept{ExpOk}(E)$.
 
 \item[Case $E = {\bf measure}\baseterm(\listOf{\value}\baseterm)$:] By subsequent application of rules \semRule{OP-DoMeasure} and \semRule{NP-ProbEvol} we reach configuration $[gs\,|\,(lms,vp,\si{\value}\ ts_0)]$, hence $\refConcept{ExpOk}(E)$; by application of \semRule{OP-MeasureUninit} we reach configuration $[gs\,|\,(lms,vp,\si{\RTEuninitVarUsage})]$, hence $\refConcept{ExpOk}(E)$; by application of \semRule{OP-MeasureOverlap} we reach configuration $[gs\,|\,(lms,vp,\si{\RTEqVarOverlap})]$, hence $\refConcept{ExpOk}(E)$.
\end{description}

Next we assume that the lemma holds for all subexpressions $E_0$ of $E$ (inductive hypotesis, IH). Then:

\begin{description}
 \item[Case $E = \baseterm(E_0\baseterm)$:] By application of \semRule{OP-Bracket} we reach configuration $[gs\,|\,(lms,vp,\si{E_0}\ ts_0)]$ for which the theorem holds by the inductive hypothesis. Hence $\refConcept{ExpOk}(E)$.
 
\item[Case $E = I = E_0$:] We get the following evolution:

$\begin{array}{rcl}
{}[gs_0\,|\,(lms_0,vp_0,\si{I = E_0}\ ts_0)] & \xrightarrow{\semRule{OP-AssignExpr}} & [gs_0\,|\,(lms_0,vp_0,\si{E_0}\ \si{I = \bullet}\ ts_0)] \\
 & \xrightarrow{IH}\!^* & [gs_{n'}\,|\,(lms_{n'},vp_{n'},\si{\value}\ \si{I = \bullet}\ ts_0)] \\
 & \xrightarrow{\semRule{OP-SubstE}} & [gs_{n'}\,|\,(lms_{n'},vp_{n'},\si{I = \value}\ ts_0)] \\
\end{array}$

The last configuration is one of the base cases for which the lemma holds. Hence $\refConcept{ExpOk}(E)$.

\item[Case $E = \recv(E_0)$:] We get the following evolution:

$\begin{array}{rcl}
{}[gs_0\,|\,(lms_0,vp_0,\si{\recv\baseterm(E_0\baseterm)}\ ts_0)] & \xrightarrow{\semRule{OP-RecvExpr}} & [gs_0\,|\,(lms_0,vp_0,\si{E_0}\ \si{\recv\baseterm(\bullet\baseterm)}\ ts_0)] \\
 & \xrightarrow{IH}\!^* & [gs_{n'}\,|\,(lms_{n'},vp_{n'},\si{\value}\ \si{\recv\baseterm(\bullet\baseterm)}\ ts_0)] \\
 & \xrightarrow{\semRule{OP-SubstE}} & [gs_{n'}\,|\,(lms_{n'},vp_{n'},\si{\recv\baseterm(\value\baseterm)}\ ts_0)] \\
\end{array}$

The last configuration is one of the base cases for which the lemma holds. Hence $\refConcept{ExpOk}(E)$.

\item[Case $E = \Measure\baseterm(E_0\baseterm,\dots\baseterm,E_e\baseterm)$:] We get the following evolution:

$\begin{array}{rcl}
\multicolumn{3}{l}{[gs_0\,|\,(lms_0,vp_0,\si{\Measure\baseterm(E_0\baseterm,\dots\baseterm,E_e\baseterm)}\ ts_0)]} \\ 
\hspace{1cm} & \xrightarrow{\semRule{OP-MeasureExpr}} & [gs_0\,|\,(lms_0,vp_0,\si{E_0}\ \si{\Measure\baseterm(\bullet\baseterm,\dots\baseterm,E_e\baseterm)}\ ts_0)] \\
 & \xrightarrow{IH}\!^* & [gs_{n'}\,|\,(lms_{n'},vp_{n'},\si{\value_0}\ \si{\Measure\baseterm(\bullet\baseterm,\dots\baseterm,E_e\baseterm)}\ ts_0)] \\
 & \xrightarrow{\semRule{OP-SubstE}} & [gs_{n'}\,|\,(lms_{n'},vp_{n'},\si{\Measure\baseterm(\value_0\baseterm,\dots\baseterm,E_e\baseterm)}\ ts_0)] \\

 & \vdots \\

 & \xrightarrow{\semRule{OP-MeasureExpr}} & [gs_{n''}\,|\,(lms_{n''},vp_{n''},\si{E_e}\ \si{\Measure\baseterm(\value_0\baseterm,\dots\baseterm,\bullet\baseterm)}\ ts_0)] \\
 & \xrightarrow{IH}\!^* & [gs_{n'''}\,|\,(lms_{n'''},vp_{n'''},\si{\value_e}\ \si{\Measure\baseterm(\value_0\baseterm,\dots\baseterm,\bullet\baseterm)}\ ts_0)] \\
 & \xrightarrow{\semRule{OP-SubstE}} & [gs_{n'''}\,|\,(lms_{n'''},vp_{n'''},\si{\Measure\baseterm(\value_0\baseterm,\dots\baseterm,\value_e\baseterm)}\ ts_0)] \\
\end{array}$

The last configuration is one of the base cases for which the lemma holds. Hence $\refConcept{ExpOk}(E)$.

\end{description}
\end{proof}


\begin{lemma}
\label{lemma-eval-b-base}
Let $B$ be a \refConcept[syntax!block-forming statement]{block-forming statement} such that $\refConcept{BFOk_0}(B)$. Then $\refConcept{BFOk}(B)$ is satisfied.
\end{lemma}

\begin{proof}
By induction on the structure of $B$. Let $C_0=[gs_0\,|\,(lms_0,vp_0,\si{B}\ ts_0)]:\tau$ be any configuration. Base case:

\begin{description}
 \item[Case $B = \return\ \value\baseterm{;}$:] $\refConcept{BFOk}(\return\ \value\baseterm{;})$ trivially.
 
 \item[Case $B = \return\ E\baseterm{;}$:] We get the following evolution:

$\begin{array}{rcl}
\multicolumn{3}{l}{[gs_a\,|\,(lms_a,vp_a, \si{\return\ E\baseterm;}\ ts_0)]} \\ 
\hspace{1cm} & \xrightarrow{\semRule{OP-ReturnExpr}} & [gs_a,|,(lms_a,vp_a,\si{E}\ \si{\return\ \bullet\baseterm;}\ ts_0)] \\
 & \xrightarrow{\text{Lemma \ref{lemma-eval-expr-base}}}\!^* & [gs_{n'}\,|\,(lms_{n'},vp_{n'},\si{\refConcept{RTErr}})] \text{ {\it ie.} the lemma holds} \\
 & \hfill\text{\it or} & [gs_{n'}\,|\,(lms_{n'},vp_{n'},\si{\value}\ \si{\return\ \bullet\baseterm;}\ ts_0)] \\
 & \xrightarrow{\semRule{OP-SubstS}} & [gs_{n'}\,|\,(lms_{n'},vp_{n'},\si{\return\ \value\baseterm;}\ ts_0)] \\
\end{array}$

In the last step we see that $\refConcept{BFOk}(B)$.
 
 \item[Case $B = \return\baseterm{;}$:]  $\refConcept{BFOk}(\return\baseterm{;})$ trivially.
 
 \item[Case $B = \baseterm{;}$:] By application of \semRule{OP-Skip} we reach configuration $[gs_a\,|\,(lms_a,vp_a, ts_0)]$, hence $\refConcept{BFOk}(B)$.
 
 \item[Case $B = \value\baseterm{;}$:] By application of \semRule{OP-PromoForget} we reach configuration $[gs_a\,|\,(lms_a,vp_a, ts_0)]$, hence $\refConcept{BFOk}(B)$.
 
 \item[Case $B = PE\baseterm{;}$:] We get the following evolution:

$\begin{array}{rcl}
\multicolumn{3}{l}{[gs_a\,|\,(lms_a,vp_a, \si{PE\baseterm;}\ ts_0)]} \\ 
\hspace{1cm} & \xrightarrow{\semRule{OP-PromoExpr}} & [gs_a,|,(lms_a,vp_a,\si{PE}\ \si{\bullet\baseterm;}\ ts_0)] \\
 & \xrightarrow{\text{Lemma \ref{lemma-eval-expr-base}}}\!^* & [gs_{n'}\,|\,(lms_{n'},vp_{n'},\si{\refConcept{RTErr}})] \text{ {\it ie.} the lemma holds} \\
 & \hfill\text{\it or} & [gs_{n'}\,|\,(lms_{n'},vp_{n'},\si{\value}\ \si{\bullet\baseterm;}\ ts_0)] \\
 & \xrightarrow{\semRule{OP-SubstS}} & [gs_{n'}\,|\,(lms_{n'},vp_{n'},\si{\value\baseterm;}\ ts_0)] \\
\end{array}$

The last configuration is exactly the previous case for which the lemma holds.
 
 \item[Case $B = \circL$:] By application of \semRule{OP-BlockEnd} we reach configuration $[gs_a\,|\,(lms_a,vp_a, ts_0)]$, hence $\refConcept{BFOk}(B)$.
 
 \item[Case $B = \fork\ I\baseterm(\value_0\baseterm,\dots\baseterm,\value_e\baseterm)\baseterm{;}$:] By application of \semRule{OP-DoFork} we reach configuration $[gs_a\,|\,(lms_{0,a},vp_{0,a}, ts_0) \parallel (lms_{1,a},\refConcept[blacksquare]{\blacksquare}, I\baseterm(\value_0\baseterm,\dots\baseterm,\value_e\baseterm)]$, hence $\refConcept{BFOk}(B)$.
 
 \item[Case $B = \fork\ I\baseterm(E_0\baseterm,\dots\baseterm,E_e\baseterm)\baseterm{;}$:] We get the following evolution:

{\small$\begin{array}{rcl}
\multicolumn{3}{l}{[gs_a\,|\,(lms_a,vp_a, \si{\fork\ I\baseterm(E_0\baseterm,\dots\baseterm,E_e\baseterm)\baseterm{;}}\ ts_0)]} \\ 
\hspace{1cm} & \xrightarrow{\semRule{OP-ForkExpr}} & [gs_a,|,(lms_a,vp_a,\si{E_0}\ \si{\fork\ I\baseterm(\bullet\baseterm,\dots\baseterm,E_e\baseterm)\baseterm{;}}\ ts_0)] \\
 & \xrightarrow{\text{Lemma \ref{lemma-eval-expr-base}}}\!^* & [gs_{n'}\,|\,(lms_{n'},vp_{n'},\si{\refConcept{RTErr}})] \text{ {\it ie.} the lemma holds} \\
 & \hfill\text{\it or} & [gs_{n'}\,|\,(lms_{n'},vp_{n'},\si{\value_0}\ \si{\fork\ I\baseterm(\bullet\baseterm,\dots\baseterm,E_e\baseterm)\baseterm{;}}\ ts_0)] \\
 & \xrightarrow{\semRule{OP-SubstS}} & [gs_{n'}\,|\,(lms_{n'},vp_{n'},\si{\fork\ I\baseterm(\value_0\baseterm,\dots\baseterm,E_e\baseterm)\baseterm{;}}\ ts_0)] \\

 & \vdots \\

 & \xrightarrow{\semRule{OP-ForkExpr}} & [gs_{a'},|,(lms_{a'},vp_{a'},\si{E_e}\ \si{\fork\ I\baseterm(\value_0\baseterm,\dots\baseterm,\bullet\baseterm)\baseterm{;}}\ ts_0)] \\
 & \xrightarrow{\text{Lemma \ref{lemma-eval-expr-base}}}\!^* & [gs_{n''}\,|\,(lms_{n''},vp_{n''},\si{\refConcept{RTErr}})] \text{ {\it ie.} the lemma holds} \\
 & \hfill\text{\it or} & [gs_{n'''}\,|\,(lms_{n'''},vp_{n'''},\si{\value_e}\ \si{\fork\ I\baseterm(\value_0\baseterm,\dots\baseterm,\bullet\baseterm)\baseterm{;}}\ ts_0)] \\
 & \xrightarrow{\semRule{OP-SubstS}} & [gs_{n'''}\,|\,(lms_{n'''},vp_{n'''},\si{\fork\ I\baseterm(\value_0\baseterm,\dots\baseterm,\value_e\baseterm)\baseterm{;}}\ ts_0)] \\
 
\end{array}$}

The last configuration is exactly the previous case for which the lemma holds.
 
 \item[Case $B = \send\baseterm(\value_c\baseterm,\value_v\baseterm)\baseterm{;}$:] By application of \semRule{OP-SendRecv} we reach configuration $[gs_a\,|\,(lms_{0,a},vp_{0,a}, ts_0)]$, hence $\refConcept{BFOk}(B)$; by application of \semRule{OP-SendUninit} we reach configuration $[gs_a\,|\,(lms_{0,a},vp_{0,a}, \RTEuninitVarUsage)]$, hence $\refConcept{BFOk}(B)$.
 
 \item[Case $B = \send\baseterm(E_c\baseterm,E_v\baseterm)\baseterm{;}$:] We get the following evolution:

$\begin{array}{rcl}
\multicolumn{3}{l}{[gs_a\,|\,(lms_a,vp_a, \si{\send\baseterm(E_c\baseterm,E_v\baseterm)\baseterm{;}}\ ts_0)]} \\ 
\hspace{1cm} & \xrightarrow{\semRule{OP-SendExpr1}} & [gs_a,|,(lms_a,vp_a,\si{E_c}\ \si{\send(\bullet\baseterm,E_v\baseterm)\baseterm{;}}\ ts_0)] \\
 & \xrightarrow{\text{Lemma \ref{lemma-eval-expr-base}}}\!^* & [gs_{n'}\,|\,(lms_{n'},vp_{n'},\si{\refConcept{RTErr}})] \text{ {\it ie.} the lemma holds} \\
 & \hfill\text{\it or} & [gs_{n'}\,|\,(lms_{n'},vp_{n'},\si{\value_c}\ \si{\send\baseterm(\bullet\baseterm,E_v\baseterm)\baseterm{;}}\ ts_0)] \\
 & \xrightarrow{\semRule{OP-SubstS}} & [gs_{n'}\,|\,(lms_{n'},vp_{n'},\si{\send\baseterm(\value_c\baseterm,E_v\baseterm)\baseterm{;}}\ ts_0)] \\

 & \xrightarrow{\semRule{OP-SendExpr2}} & [gs_{a'},|,(lms_{a'},vp_{a'},\si{E_v}\ \si{\send\baseterm(\value_c\baseterm,\bullet\baseterm)\baseterm{;}}\ ts_0)] \\
 & \xrightarrow{\text{Lemma \ref{lemma-eval-expr-base}}}\!^* & [gs_{n''}\,|\,(lms_{n''},vp_{n''},\si{\refConcept{RTErr}})] \text{ {\it ie.} the lemma holds} \\
 & \hfill\text{\it or} & [gs_{n''}\,|\,(lms_{n''},vp_{n''},\si{\value_v}\ \si{\send\baseterm(\value_c\baseterm,\bullet\baseterm)\baseterm{;}}\ ts_0)] \\
 & \xrightarrow{\semRule{OP-SubstS}} & [gs_{n''}\,|\,(lms_{n''},vp_{n''},\si{\send\baseterm(\value_c\baseterm,\value_v\baseterm)\baseterm{;}}\ ts_0)] \\
\end{array}$

The last configuration is exactly the previous case for which the lemma holds.

 \item[Case $B = T\ \listOf{I}\baseterm{;}$:] By possibly multiple application of \semRule{OP-VarDeclMulti} and \semRule{OP-VarDecl} we reach configuration $[gs_a\,|\,(lms_a,vp_a, ts_0)]$, hence $\refConcept{BFOk}(B)$.

 \item[Case $B = \type{channel[}T\type{]}\ I\ {\bf withends}{[}I\baseterm,I{]}\baseterm{;}$:] By application of \semRule{OP-VarDeclChE} we reach configuration $[gs_a\,|\,(lms_a,vp_a, ts_0)]$, hence $\refConcept{BFOk}(B)$.
 
 \item[Case $B = I\ \baseterm{aliasfor}\ \baseterm{[}\listOf{I}\baseterm{]}\baseterm{;}$:] By application of \semRule{OP-VarDeclAlF} we reach configuration $[gs_a\,|\,(lms_a,vp_a, ts_0)]$, hence $\refConcept{BFOk}(B)$.
\end{description}

Next we assume that the lemma holds for all substatements $Be_0$ of $B$ (inductive hypotesis, IH). Then:

\begin{description}
\item[Case $B = Be_0\ Be_1\ \dots\ Be_m$:] We get the following evolution:

\hspace*{-5mm}$\begin{array}{rcl}
\multicolumn{3}{l}{[gs_a\,|\,(lms_a,vp_a,\si{Be_0\ Be_1\ \dots\ Be_m}\ ts_0)]} \\
\hspace{1cm} & \xrightarrow{\semRule{OP-BlockHead}} & [gs_a\,|\,(lms_a,vp_a,\si{Be_0}\ \si{Be_1\ \dots\ Be_m}\ ts_0)] \\
 & \xrightarrow{\text{IH}}\!^* & [gs_{n'}\,|\,(lms_{n'},vp_{n'},\si{\refConcept{RTErr}})] \text{ {\it ie.} the lemma holds} \\
 & \hfill\text{\it or} & [gs_{n'}\,|\,(lms_{n'},vp_{n'},\si{\return\baseterm{;}}\ \si{Be_1\ \dots\ Be_m}\ ts_0)] \text{ {\it ie.} the lemma holds} \\
 & \hfill\text{\it or} & [gs_{n'}\,|\,(lms_{n'},vp_{n'},\si{\return\ \value\baseterm{;}}\ \si{Be_1\ \dots\ Be_m}\ ts_0)] \text{ {\it ie.} the lemma holds} \\
 & \hfill\text{\it or} & \text{diverges {\it ie.} the lemma holds} \\
 & \hfill\text{\it or} & [gs_{n'}\,|\,(lms_{n'},vp_{n'}, \si{Be_1\ \dots\ Be_m}\ ts_0)] \\
 
 &  & \vdots \\
 
 & \xrightarrow{\semRule{OP-BlockHead}} & [gs_{a'}\,|\,(lms_{a'},vp_{a'}, \si{Be_{m-1}}\ \si{Be_m}\ ts_0)] \\
 & \xrightarrow{\text{IH}}\!^* & [gs_{n''}\,|\,(lms_{n''},vp_{n''},\si{\refConcept{RTErr}})] \text{ {\it ie.} the lemma holds} \\
 & \hfill\text{\it or} & [gs_{n''}\,|\,(lms_{n''},vp_{n''},\si{\return\baseterm{;}}\ \si{Be_{m}}\ ts_0)] \text{ {\it ie.} the lemma holds} \\
 & \hfill\text{\it or} & [gs_{n''}\,|\,(lms_{n''},vp_{n''},\si{\return\ \value\baseterm{;}}\ \si{Be_{m}}\ ts_0)] \text{ {\it ie.} the lemma holds} \\
 & \hfill\text{\it or} & \text{diverges {\it ie.} the lemma holds} \\
 & \hfill\text{\it or} & [gs_{n''}\,|\,(lms_{n''},vp_{n''}, \si{Be_m}\ ts_0)] \\

 & \xrightarrow{\text{IH}}\!^* & [gs_{n'''}\,|\,(lms_{n'''},vp_{n'''},\si{\refConcept{RTErr}})] \text{ {\it ie.} the lemma holds} \\
 & \hfill\text{\it or} & [gs_{n'''}\,|\,(lms_{n'''},vp_{n'''},\si{\return\baseterm{;}}\ ts_0)] \text{ {\it ie.} the lemma holds} \\
 & \hfill\text{\it or} & [gs_{n'''}\,|\,(lms_{n'''},vp_{n'''},\si{\return\ \value\baseterm{;}}\ ts_0)] \text{ {\it ie.} the lemma holds} \\
 & \hfill\text{\it or} & \text{diverges {\it ie.} the lemma holds} \\
 & \hfill\text{\it or} & [gs_{n'''}\,|\,(lms_{n'''},vp_{n'''}, ts_0)] \\
\end{array}$

Therefore the lemma holds for this case.

\item[Case $B = \IF\ \baseterm(\value\baseterm)\ S_1\ \ELSE\ S_2$:] Depending on $\value$, the evolution continues by rule \semRule{OP-IfTrue} to configuration $[gs_{n'}\,|\,(lms_{n'},vp_{n'},\si{S_1}\ ts_0)]$, or by rule \semRule{OP-IfFalse} to configuration $[gs_{n'}\,|\,(lms_{n'},vp_{n'},\si{S_2}\ ts_0)]$. By IH, we assume that the lemma holds for both $S_1$ and $S_2$, {\it ie.} the lemma holds for this case.

\item[Case $B = \IF\ \baseterm(E\baseterm)\ S_1\ \ELSE\ S_2$:] We get the following evolution:

$\begin{array}{rcl}
\multicolumn{3}{l}{[gs_a\,|\,(lms_a,vp_a,\si{\IF\ \baseterm(E\baseterm)\ S_1\ \ELSE\ S_2}\ ts_0)]} \\
\hspace{1cm} & \xrightarrow{\semRule{OP-IfExpr}} & [gs_a\,|\,(lms_a,vp_a,\si{E}\ \si{\IF\ \baseterm(\bullet\baseterm)\ S_1\ \ELSE\ S_2}\ ts_0)] \\
 & \xrightarrow{\text{Lemma \ref{lemma-eval-expr-base}}}\!^* & [gs_{n'}\,|\,(lms_{n'},vp_{n'},\si{\refConcept{RTErr}})] \text{ {\it ie.} the lemma holds} \\
 & \hfill\text{\it or} & [gs_{n'}\,|\,(lms_{n'},vp_{n'}, \si{\value}\ \si{\IF\ \baseterm(\bullet\baseterm)\ S_1\ \ELSE\ S_2}\ ts_0)] \\
 & \xrightarrow{\semRule{OP-SubstS}} & [gs_{n''}\,|\,(lms_{n''},vp_{n''},\si{\IF\ \baseterm(\value\baseterm)\ S_1\ \ELSE\ S_2}\ ts_0)] \\
\end{array}$

The last configuration is exactly the previous case for which the lemma holds.

 \item[Case $B = \baseterm\{\ B\ \baseterm\}$:]  We get the following evolution:

$\begin{array}{rcl}
\multicolumn{3}{l}{[gs_a\,|\,(lms_a,vp_a,\si{\{\ B\ \}}\ ts_0)]} \\
\hspace{1cm} & \xrightarrow{\semRule{OP-Block}} & [gs_a\,|\,(lms_a,vp_{a'},\si{B}\ \si{\circL}\ ts_0)] \\
 & \xrightarrow{\text{IH}}\!^* & [gs_{n'}\,|\,(lms_{n'},vp_{n'},\si{\refConcept{RTErr}})] \text{ {\it ie.} the lemma holds} \\
 & \hfill\text{\it or} & [gs_{n'}\,|\,(lms_{n'},vp_{n'},\si{\return\baseterm{;}}\ ts_{n'}\ \si{\circL}\ ts_0)] \text{ {\it ie.} the lemma holds} \\
 & \hfill\text{\it or} & [gs_{n'}\,|\,(lms_{n'},vp_{n'},\si{\return\ \value\baseterm{;}}\ ts_{n'}\ \si{\circL}\ ts_0)] \text{ {\it ie.} the lemma holds} \\
 & \hfill\text{\it or} & \text{diverges {\it ie.} the lemma holds} \\
 & \hfill\text{\it or} & [gs_{n'}\,|\,(lms_{n'},vp_{n'}, \si{\circL}\ ts_0)] \\
 & \xrightarrow{\semRule{OP-BlockEnd}} & [gs_{n'}\,|\,(lms_{n'},vp_a, ts_0)] \\
\end{array}$

Therefore the lemma holds for this case.

 \item[Case $B = \while\ \baseterm(E\baseterm)\ S$:]  We get the following evolution:

$\begin{array}{rcl}
\multicolumn{3}{l}{[gs_a\,|\,(lms_a,vp_a,\si{\while\ \baseterm(E\baseterm)\ S}\ ts_0)]} \\
\hspace{1cm} & \xrightarrow{\semRule{OP-While}} & [gs_a\,|\,(lms_a,vp_a,\si{\IF\ \baseterm(E\baseterm)\ \baseterm\{ S\ \while\ \baseterm(E\baseterm)\ S \baseterm\}\ \ELSE\ \baseterm{;}}\ ts_0)] \\
 & \xrightarrow{\text{see \IF case}}\!^* & [gs_{n'}\,|\,(lms_{n'},vp_{n'},\si{\refConcept{RTErr}})] \text{ {\it ie.} the lemma holds} \\
 & \hfill\text{\it or} & \text{diverges {\it ie.} the lemma holds} \\
 & \hfill\text{\it or} & [gs_{n'}\,|\,(lms_{n'},vp_{n'},\si{\IF\ \baseterm(\value\baseterm)\ \baseterm\{ S\ \while\ \baseterm(E\baseterm)\ S \baseterm\}\ \ELSE\ \baseterm{;}}\ ts_0)] \\
\end{array}$

Depending on $\value$, the evaluation continues either this way:

$\begin{array}{rcl}
\hspace{1cm} & \xrightarrow{\semRule{OP-IfFalse}} & [gs_{n'}\,|\,(lms_{n'},vp_{n'},\si{\baseterm{;}}\ ts_0)] \\
 & \xrightarrow{\semRule{OP-Skip}} & [gs_{n'}\,|\,(lms_{n'},vp_{n'}, ts_0)] \\
\end{array}$

so the lemma holds for this case; or the evaluation continues as follows:

$\begin{array}{rcl}
\hspace{1cm} & \xrightarrow{\semRule{OP-IfTrue}} & [gs_{n'}\,|\,(lms_{n'},vp_{n'},\si{\baseterm\{ S\ \while\ \baseterm(E\baseterm)\ S \baseterm\}}\ ts_0)] \\
 & \xrightarrow{\semRule{OP-Block}} & [gs_{n'}\,|\,(lms_{n'},vp_{n'},\si{S\ \while\ \baseterm(E\baseterm)\ S}\ \si{\circL}\ ts_0)] \\
 & \xrightarrow{\semRule{OP-BlockHead}} & [gs_{n'}\,|\,(lms_{n'},vp_{n'},\si{S}\ \si{\while\ \baseterm(E\baseterm)\ S}\ \si{\circL}\ ts_0)] \\
 & \xrightarrow{\text{IH}}\!^* & [gs_{n''}\,|\,(lms_{n''},vp_{n''},\si{\refConcept{RTErr}})] \text{ {\it ie.} the lemma holds} \\
 & \hfill\text{\it or} & [gs_{n''}\,|\,(lms_{n''},vp_{n''},\si{\return\baseterm{;}}\ ts_{n''}\ \si{\circL}\ ts_0)] \text{ {\it ie.} the lemma holds} \\
 & \hfill\text{\it or} & [gs_{n''}\,|\,(lms_{n''},vp_{n''},\si{\return\ \value\baseterm{;}}\ ts_{n''}\ \si{\circL}\ ts_0)] \text{ {\it ie.} the lemma holds} \\
 & \hfill\text{\it or} & \text{diverges {\it ie.} the lemma holds} \\
 & \hfill\text{\it or} & [gs_{n''}\,|\,(lms_{n''},vp_{n''}, \si{\while\ \baseterm(E\baseterm)\ S}\ \si{\circL}\ ts_0)]
\end{array}$

Therefore it is possible that evaluation of \while statement continues forever. This is the statement of the lemma, hence the lemma holds even for this case.
\end{description}
\end{proof}


\begin{lemma}
\label{lemma-eval-expr-ind-step}
Let $E$ be an \refConcept[syntax!expression]{expression} such that $\refConcept{ExpOk_i}_{+1}(E)$. Then $\refConcept{ExpOk}(E)$ is satisfied.
\end{lemma}

\begin{proof}
The proof proceeds similarly to the proof of Lemma \ref{lemma-eval-expr-base}. We must only add two cases:

\begin{description}
 \item[Case $E = I\baseterm(\listOf{\value}\baseterm)$:] We know that the predicate $\refConcept{BFOk_i}(\refConcept[method typing context]{M_B}(I))$ is satisfied, hence by Lemmata \ref{lemma-eval-b-ind-step} for $\refConcept{BFOk_i}$ and \ref{lemma-BFOk=>StkRetOk}, the predicate $\refConcept{StkRetOk}(\refConcept[method typing context]{M_B}(I))$ is satisfied ($\dagger$). We get the following evolution:

$\begin{array}{rcl}
\multicolumn{3}{l}{[gs_a\,|\,(lms_a,vp_a,\si{I\baseterm(\listOf{\value}\baseterm)}\ ts_a]} \\ 
\hspace{1cm} & \xrightarrow{\semRule{OP-DoMethodCallCl}} & [gs_{n'}\,|\,(lms_{n'},vp_{n'}, \si{\refConcept[method typing context]{M_B}(I)}\ \si{\circM}\ ts_0)] \\

 & \xrightarrow{(\dagger)}\!^* & [gs_{n''}\,|\,(lms_{n''},vp_{n''},\si{\refConcept{RTErr}}) \parallel \dots \parallel (lms_{k,n''},vp_{k,n''}, ts_{k,n''})] \\
 & \hfill\text{\it or} & [gs_{n''}\,|\,(lms_{0,n''},vp_{0,n''},\si{\value}\ ts_{0}) \parallel \dots \parallel (lms_{k,n''},vp_{k,n''}, ts_{k,n''})]\\
 & \hfill\text{\it or} & \text{diverges} \\
\end{array}$

Therefore the lemma holds, $\refConcept{ExpOk}(E)$.

\item[Case $E = I\baseterm(E_0\baseterm,\dots\baseterm,E_e\baseterm)$:] By induction on the structure of $E$: Assume that the lemma holds for all $E_0,\dots,E_e$ (inductive hypotesis, IH). We get the following evolution:

$\begin{array}{rcl}
\multicolumn{3}{l}{[gs_0\,|\,(lms_0,vp_0,\si{I\baseterm(E_0\baseterm,\dots\baseterm,E_e\baseterm)}\ ts_0)]} \\ 
\hspace{1cm} & \xrightarrow{\semRule{OP-MethodCallExpr}} & [gs_0\,|\,(lms_0,vp_0,\si{E_0}\ \si{I\baseterm(\bullet\baseterm,\dots\baseterm,E_e\baseterm)}\ ts_0)] \\
 & \xrightarrow{IH}\!^* & [gs_{n'}\,|\,(lms_{n'},vp_{n'},\si{\value_0}\ \si{I(\bullet,\dots,E_e)}\ ts_0)] \\
 & \xrightarrow{\semRule{OP-SubstE}} & [gs_{n'}\,|\,(lms_{n'},vp_{n'},\si{I(\value_0,\dots,E_e)}\ ts_0)] \\

 & \vdots \\

 & \xrightarrow{\semRule{OP-MethodCallExpr}} & [gs_{n''}\,|\,(lms_{n''},vp_{n''},\si{E_e}\ \si{I\baseterm(\value_0\baseterm,\dots\baseterm,\bullet\baseterm)}\ ts_0)] \\
 & \xrightarrow{IH}\!^* & [gs_{n'''}\,|\,(lms_{n'''},vp_{n'''},\si{\value_e}\ \si{I\baseterm(\value_0\baseterm,\dots\baseterm,\bullet\baseterm)}\ ts_0)] \\
 & \xrightarrow{\semRule{OP-SubstE}} & [gs_{n'''}\,|\,(lms_{n'''},vp_{n'''},\si{I\baseterm(\value_0\baseterm,\dots\baseterm,\value_e\baseterm)}\ ts_0)] \\
\end{array}$

The last configuration is one of the base cases for which the lemma holds. Hence $\refConcept{ExpOk}(E)$.

\end{description}
\end{proof}

\begin{lemma}
\label{lemma-eval-b-ind-step}
Let $B$ be a \refConcept[syntax!block-forming statement]{block-forming statement} such that $\refConcept{BFOk_i}_{+1}(B)$. Then $\refConcept{BFOk}(B)$ is satisfied.
\end{lemma}

\begin{proof}
The proof is nearly identical to the proof of Lemma \ref{lemma-eval-b-base}. We only exchange all usages of Lemma \ref{lemma-eval-expr-base} for evaluation of expression $E$ by Lemma \ref{lemma-eval-expr-ind-step} for $\refConcept{ExpOk_i}(E)$.
\end{proof}

\begin{corollary}
 For any at most countably derivable \refConcept[syntax!expression]{expression} $E$, $\refConcept{ExpOk}(E)$ is satisfied.

 For any at most countably derivable \refConcept[syntax!block-forming statement]{block-forming statement} $B$, $\refConcept{BFOk}(B)$ is satisfied.
\end{corollary}

Informally, as the programs are always countably derivable, we have proved the following:
\begin{itemize}
 \item Unless a \refConcept{runtime error} occurs, the evaluation of an \refConcept[syntax!expression]{expression} $E$ never modifies \refConcept{term stack} elements under the evaluated expression. If the evaluation does not diverge, the expression $E$ always yields an \refConcept{internal value},
 \item Unless a \refConcept{runtime error} occurs, the evaluation of a \refConcept[syntax!block-forming statement]{block-forming statement} $B$ never modifies \refConcept{term stack} elements under the evaluated statement if it does not get to a \baseterm{return} statement. In that case, it pops all the elements from the \refConcept{term stack} up to the first occurence of the symbol $\circM$.
\end{itemize}

\begin{lemma}\label{lemma-eval-retok-to-return}
Let $B$ be a \refConcept[syntax!block-forming statement]{block-forming statement} such that $\refConcept{RetOk}(B)$. For any configuration $C_0=[gs_0\,|\,(lms_0,vp_0,\si{B}\ ts_0)]:\tau$, the evaluation either diverges or reaches one of the following configurations:
\begin{itemize}[itemsep=0pt]
 \item $[gs_n\,|\,(lms_{0,n},vp_{0,n}, \si{\return\baseterm{;}}\ ts_{0,n}\ ts_{0}) \parallel \dots \parallel (lms_{k,n},vp_{k,n}, ts_{k,n})], k \geq 0$, $ts_{0,n}$ does not contain $\circM$, or
 \item $[gs_n\,|\,(lms_{0,n},vp_{0,n}, \si{\return\ \value\baseterm{;}}\ ts_{0,n}\ ts_{0}) \parallel \dots \parallel (lms_{k,n},vp_{k,n}, ts_{k,n})], k \geq 0$, $ts_{0,n}$ does not contain $\circM$, or
 \item $[gs_n\,|\,(lms_{0,n},vp_{0,n},\si{\refConcept{RTErr}}) \parallel \dots \parallel (lms_{k,n},vp_{k,n}, ts_{k,n})], k \geq 0$.
\end{itemize}
\end{lemma}

\begin{proof}
By induction on the structure of $B$. Let $C_0=[gs_0\,|\,(lms_0,vp_0,\si{B}\ ts_0)]:\tau$ be any configuration. Base case:

\begin{description}
 \item[Case $B = \return\ \value\baseterm{;}$:] The lemma holds trivially.
 \item[Case $B = \return\ E\baseterm{;}$:] We get the following evolution:

$\begin{array}{rcl}
\multicolumn{3}{l}{[gs_a\,|\,(lms_a,vp_a, \si{\return\ E\baseterm;}\ ts_0)]} \\ 
\hspace{1cm} & \xrightarrow{\semRule{OP-ReturnExpr}} & [gs_a,|,(lms_a,vp_a,\si{E}\ \si{\return\ \bullet\baseterm;}\ ts_0)] \\
 & \xrightarrow{\refConcept{ExpOk}(E)}\!^* & [gs_{n'}\,|\,(lms_{n'},vp_{n'},\si{\refConcept{RTErr}})] \text{ {\it ie.} the lemma holds} \\
 & \hfill\text{\it or} & \text{diverges {\it ie.} the lemma holds} \\
 & \hfill\text{\it or} & [gs_{n'}\,|\,(lms_{n'},vp_{n'},\si{\value}\ \si{\return\ \bullet\baseterm;}\ ts_0)] \\
 & \xrightarrow{\semRule{OP-SubstS}} & [gs_{n'}\,|\,(lms_{n'},vp_{n'},\si{\return\ \value\baseterm;}\ ts_0)] \\
\end{array}$

In the last step we see we get requested configuration, hence the lemma holds.
 
 \item[Case $B = \return\baseterm{;}$:] The lemma holds trivially.
\end{description}

Next we assume that the lemma holds for all substatements $Be$ of $B$ such that $\refConcept{RetOk}(Be)$ (inductive hypotesis, IH). Then:

\begin{description}
 \item[Case $B = Be_0\ Be_1\dots Be_m$:] We know that there is at least one $j$ such that $\refConcept{RetOk}(Be_j)$. Let $b$ be smallest such $j$. Moreover, $\refConcept{BFOk}(Be_i)$ for $0 \leq i \leq m$.
 We get the following evolution:

\hspace*{-5mm}$\begin{array}{rcl}
\multicolumn{3}{l}{[gs_a\,|\,(lms_a,vp_a,\si{Be_0\ Be_1\ \dots\ Be_m}\ ts_a)]} \\
\hspace{1cm} & \xrightarrow{\semRule{OP-BlockHead}} & [gs_a\,|\,(lms_a,vp_a,\si{Be_0}\ \si{Be_1\ \dots\ Be_m}\ ts_0)] \\
 & \xrightarrow{\refConcept{BFOk}(Be_0)}\!^* & [gs_{n'}\,|\,(lms_{n'},vp_{n'},\si{\refConcept{RTErr}})] \text{ {\it ie.} the lemma holds} \\
 & \hfill\text{\it or} & [gs_{n'}\,|\,(lms_{n'},vp_{n'},\si{\return\baseterm{;}}\ \si{Be_1\ \dots\ Be_m}\ ts_0)] \text{ {\it ie.} the lemma holds} \\
 & \hfill\text{\it or} & [gs_{n'}\,|\,(lms_{n'},vp_{n'},\si{\return\ \value\baseterm{;}}\ \si{Be_1\ \dots\ Be_m}\ ts_0)] \text{ {\it ie.} the lemma holds} \\
 & \hfill\text{\it or} & \text{diverges {\it ie.} the lemma holds} \\
 & \hfill\text{\it or} & [gs_{n'}\,|\,(lms_{n'},vp_{n'}, \si{Be_1\ \dots\ Be_m}\ ts_0)] \\
 
 &  & \vdots \\
 
 & \xrightarrow{\semRule{OP-BlockHead}} & [gs_{a'}\,|\,(lms_{a'},vp_{a'}, \si{Be_{b}}\ ts_{0,a'}\ ts_0)] \\
\end{array}$

By IH, the lemma holds for the last configuration, therefore it holds for this case.

\item[Case $B = \IF\ \baseterm(\value\baseterm)\ S_1\ \ELSE\ S_2$:] Depending on $\value$, the evolution continues by rule \semRule{OP-IfTrue} to configuration $[gs_{n'}\,|\,(lms_{n'},vp_{n'},\si{S_1}\ ts_0)]$, or by rule \semRule{OP-IfFalse} to configuration $[gs_{n'}\,|\,(lms_{n'},vp_{n'},\si{S_2}\ ts_0)]$. From definition of $\refConcept{RetOk}$, both $\refConcept{RetOk}(S_1)$ and $\refConcept{RetOk}(S_2)$ are satisfied. By IH, the lemma holds for this case.

\item[Case $B = \IF\ \baseterm(E\baseterm)\ S_1\ \ELSE\ S_2$:] We get the following evolution:

$\begin{array}{rcl}
\multicolumn{3}{l}{[gs_a\,|\,(lms_a,vp_a,\si{\IF\ \baseterm(E\baseterm)\ S_1\ \ELSE\ S_2}\ ts_0)]} \\
\hspace{1cm} & \xrightarrow{\semRule{OP-IfExpr}} & [gs_a\,|\,(lms_a,vp_a,\si{E}\ \si{\IF\ \baseterm(\bullet\baseterm)\ S_1\ \ELSE\ S_2}\ ts_0)] \\
 & \xrightarrow{\refConcept{ExpOk}(E)}\!^* & [gs_{n'}\,|\,(lms_{n'},vp_{n'},\si{\refConcept{RTErr}})] \text{ {\it ie.} the lemma holds} \\
 & \hfill\text{\it or} & \text{diverges {\it ie.} the lemma holds} \\
 & \hfill\text{\it or} & [gs_{n'}\,|\,(lms_{n'},vp_{n'}, \si{\value}\ \si{\IF\ \baseterm(\bullet\baseterm)\ S_1\ \ELSE\ S_2}\ ts_0)] \\
 & \xrightarrow{\semRule{OP-SubstS}} & [gs_{n'}\,|\,(lms_{n'},vp_{n'},\si{\IF\ \baseterm(\value\baseterm)\ S_1\ \ELSE\ S_2}\ ts_0)] \\
\end{array}$

The last configuration is exactly the previous case for which the lemma holds.

 \item[Case $B = \baseterm\{\ B\ \baseterm\}$:]  We get the following evolution:

$\begin{array}{rcl}
[gs_a\,|\,(lms_a,vp_a,\si{\baseterm\{\ B\ \baseterm\}}\ ts_0)] & \xrightarrow{\semRule{OP-Block}} & [gs_a\,|\,(lms_a,vp_{a'},\si{B}\ \si{\circL}\ ts_0)]
\end{array}$

 By IH, the lemma holds for the last \refConcept{configuration}, therefore it holds for this case.

\end{description}
\end{proof}

Informally, for any \refConcept[syntax!block-forming statement]{block-forming statement} $B$ such that $\refConcept{RetOk}(B)$, we have proved the following:
\begin{itemize}
 \item Unless a \refConcept{runtime error} occurs, the evaluation of $B$ always leads to a \baseterm{return} statement or diverges.
\end{itemize}

\subsubsection{Type preservation lemmata}

In the previous subsections, we have proved that for any \refConcept{well-typed configuration} $C_0:\tau$, there exists a \refConcept{configuration} $C_1$ such that $C_0 \longrightarrow C_1$.
The next step in proving type soundness is to prove that any such one-step evaluation preserves the type of the \refConcept{configuration}. In other words, that the \refConcept{configuration} $C_1$ is \refConcept[well-typed configuration]{well-typed}, $C_1:\tau'$, and that $\tau' = \tau$ in the case of single process-evolution or $\tau' = \tau \times \theta$ in the case of the fork statement. In all cases, the type of the original processes is preserved.

{\lemma[Type preservation for $\longrightarrow_v$] Let $C_0 = [gs_0\,\vert\,(lms_0, vp_0, \si{TE}\ ts_0)]$, $C_1 = \boxplus_{i=1}^{q} p_i \bullet C_{i,1}$ where $C_{i,1} = [gs_{i,1}\,\vert\,(lms_{i,1}, vp_{i,1}, \si{TE_i}\ ts_{i,1})]$ be two configurations such that $C_0 \longrightarrow_v C_1$. If $C_0:\tau$ then $C_{i,1}:\tau$ for all $i$.}

\begin{proof}
From $C_0:\tau$ we know:
\begin{prooftree}
\small
 \AxiomC{\small\newPremise{}}
   \RightLabel{\small T-rule}
 \UnaryInfC{$M;\refConcept{vpContext}(vp_0) \vdash_T TE : \type{\tau'}$}
             \AxiomC{\newPremise{}}
             \UnaryInfC{$M;\emptyset \vdash_C (lms_0,vp_0, ts_0) : \type{\tau' \rightarrow \tau}$}
       \RightLabel{\small \typeRule{TC-ExprClo}}
    \BinaryInfC{$M;\emptyset \vdash_C (lms_0, vp_0, \si{TE}\ ts_0): \type{void \rightarrow \tau}$}
      \RightLabel{\small\typeRule{T-Config}}
    \UnaryInfC{$M;\emptyset\vdash[gs_0\,\vert\,(lms_0, vp_0, \si{TE}\ ts_0)] : \tau$}
\end{prooftree}

For each $i = 1,\dots,n$, we want to prove $C_{i,1} : \tau$:
\begin{prooftree}
\small
 \AxiomC{\small(premise(T-rule))}
   \RightLabel{\small T-rule}
 \UnaryInfC{$M;\refConcept{vpContext}(vp_{i,1}) \vdash_T TE_i : \type{\tau'}$}
             \AxiomC{\newPremise{}}
             \UnaryInfC{$M;\emptyset \vdash_C (lms_{i,1},vp_{i,1}, ts_{i,1}) : \type{\tau' \rightarrow \tau}$}
       \RightLabel{\small \typeRule{TC-ExprClo}}
    \BinaryInfC{$M;\emptyset \vdash_C (lms_{i,1}, vp_{i,1}, \si{TE_i}\ ts_{i,1}): \type{void \rightarrow \tau}$}
      \RightLabel{\small\typeRule{T-Config}}
    \UnaryInfC{$M;\emptyset\vdash[gs_{i,1}\,\vert\,(lms_{i,1}, vp_{i,1}, \si{TE_i}\ ts_{i,1})] : \tau$}
\end{prooftree}

As $\longrightarrow_v$ rules change the top stack element only, $ts_0$ = $ts_{i,1}$ for all $i$, therefore we must only prove $TE:\tau' \Rightarrow TE_i:\tau'$.
We do this by case examination of relation $\longrightarrow_v$. 
\begin{description}
 \item[\semRule{OP-AllocC}:] $TE : \type{channel[T]}$ from \typeRule{T-Alloc}. $TE_i = \intVal{R}{\value}{channel[T]}$, thus $TE_i : \type{channel[T]}$ from \typeRule{T-Value}.
 
 \item[\semRule{OP-AllocQ}:] $TE : \type{Q_d}$ from \typeRule{T-Alloc}. $TE_i = \intVal{R}{\value}{Q_d}$, thus $TE_i : \type{Q_d}$ from \typeRule{T-Value}.
 
 \item[\semRule{OP-AssignNewValue}, \semRule{OP-AssignQAValue}, \semRule{OP-AssignQValue}, \semRule{OP-AssignValue}:] $TE : \type{T}$ from \typeRule{T-Assign}. $TE_i = \intVal{R}{\value}{T}$, thus $TE_i : \type{T}$ from \typeRule{T-Value}.
 
 \item[\semRule{OP-DoMeasure}:] $TE : \type{int}$ from \typeRule{T-Measurement}. $TE_i = \intVal{\none}{\value_i}{int}$, thus $TE_i : \type{int}$ from \typeRule{T-Value}.
 
 \item[\semRule{OP-DoMethodCallQ}:] As all methods representing quantum operators are regarded as method with \refConcept{return type} \void, $TE : \type{\void}$ from \typeRule{T-MethodCall}. $TE_i = \intVal{\none}{\bot}{\void}$, thus $TE_i : \void$ from \typeRule{T-Value}.
 
 \item[\semRule{OP-Var}:] $TE = x : \type{T}$ from \typeRule{T-Var}, hence $\refConcept{vpContext}(vp_0)(x) = \type{T}$. $TE_i = \intVal{R}{\value_x}{\mathit{\refConcept{typeOf}(x,vp_{i,1})}}$, thus $TE_i : \refConcept{typeOf}(x,vp_{i,1})$ from \typeRule{T-Value}. We know that $vp_0 = vp_{i,1}$ as rule \semRule{OP-Var} does not modify \refConcept{variable properties}. From definition of $\refConcept{vpContext}$, $\refConcept{typeOf}(x,vp_{i,1}) = \type{T} \Leftrightarrow \refConcept{vpContext}(vp_{i,1})(x) = \type{T}$.

\end{description}
\end{proof}

{\lemma[Type preservation for $\longrightarrow_e$] Let $C_0 = [gs_0\,\vert\,(lms_0, vp_0, \si{TE_0}\ ts_0)]$, $C_1 = [gs_1\,\vert\,(lms_1, vp_1, ts_1)]$ be two configurations such that $C_0 \longrightarrow_e C_1$. If $C_0:\tau$ then $C_1:\tau$.}

\begin{proof}
We prove this lemma for each rule of relation $\longrightarrow_e$:

\begin{description}
\item[\semRule{OP-MethodCallExpr}] $TE_0 = m\baseterm(\listOf{\value}\baseterm,E\baseterm,\listOf{E}\baseterm)$. From $C_0:\tau$ we know:
\begin{prooftree}
\small
 \AxiomC{\small \newPremise{e-mc-prem-TE0}}
             \AxiomC{\newPremise{e-mc-ts0}}
             \UnaryInfC{$M;\emptyset \vdash_C (lms_0,vp_0, ts_0) : \type{\tau' \rightarrow \tau}$}
       \RightLabel{\small \typeRule{TC-ExprClo}}
    \BinaryInfC{$M;\emptyset \vdash_C (lms_0, vp_0, \si{m\baseterm(\listOf{\value}\baseterm,E\baseterm,\listOf{E}\baseterm)}\ ts_0): \type{void \rightarrow \tau}$}
      \RightLabel{\small\typeRule{T-Config}}
    \UnaryInfC{$M;\emptyset\vdash[gs_0\,\vert\,(lms_0, vp_0, \si{m\baseterm(\listOf{\value}\baseterm,E\baseterm,\listOf{E}\baseterm)}\ ts_0)] : \tau$}
\end{prooftree}
where \premise{e-mc-prem-TE0} is: 
\begin{prooftree}\small
\small
\AxiomC{\newPremise{e-mc-E-is-sigma-left}} \AxiomC{\newPremise{e-mc-E-is-sigma}}\UnaryInfC{$M;\refConcept{vpContext}(vp_0) \vdash_T E:\sigma$} \AxiomC{\newPremise{e-mc-E-is-sigma-right}}
 \RightLabel{\small\typeRule{T-MethodCall}}
\TrinaryInfC{$M;\refConcept{vpContext}(vp_0) \vdash_T m\baseterm(\listOf{\value}\baseterm,E\baseterm,\listOf{E}\baseterm) : \tau'$}
\end{prooftree}

We want to prove $C_1 : \tau$:
\begin{prooftree}\small
 \AxiomC{\newPremise{e-mc-E-shouldBe-sigma}}
 \UnaryInfC{$M;\refConcept{vpContext}(vp_0) \vdash_T E : \type{\sigma}$}
    \AxiomC{\newPremise{e-mc-ExprClo-Right}}
       \RightLabel{\small \typeRule{TC-ExprClo}}
    \BinaryInfC{$M;\emptyset \vdash_C(lms_0, vp_0, \si{E}\ \si{m\baseterm(\listOf{\value}\baseterm,\bullet\baseterm,\listOf{E}\baseterm)}\ ts_0): \type{void \rightarrow \tau}$}
      \RightLabel{\small \typeRule{T-Config}}
    \UnaryInfC{$M;\emptyset\vdash[gs_0\,\vert\,(lms_0, vp_0, \si{E}\ \si{m\baseterm(\listOf{\value}\baseterm,\bullet\baseterm,\listOf{E}\baseterm)}\ ts_0)] : \tau$}
\end{prooftree}

where \premise{e-mc-ExprClo-Right} is (here we denote $vpc = \refConcept{vpContext}(vp_0)$):

\noindent{\small
 \AxiomC{\newPremise{e-mc-E-shouldBe-sigma-left}}
 \AxiomC{}
 \UnaryInfC{$M;vpc,\bullet:\sigma \vdash_T \bullet:\sigma$}
 \AxiomC{\newPremise{e-mc-E-shouldBe-sigma-right}}
 \TrinaryInfC{$M;vpc,\bullet:\sigma \vdash_T m\baseterm(\listOf{\value}\baseterm,\bullet\baseterm,\listOf{E}\baseterm) : \tau'$}

      \AxiomC{\newPremise{e-mc-c1-ts0}}
       \UnaryInfC{$M;\emptyset \vdash_C (lms_0,vp_0, ts_0) : \type{\tau' \rightarrow \tau}$}
       \RightLabel{\small \typeRule{TC-ExprHole}}
   \BinaryInfC{$M;\emptyset \vdash_C (lms_0,vp_0, \si{m\baseterm(\listOf{\value}\baseterm,\bullet\baseterm,\listOf{E}\baseterm)}\ ts_0) : \type{\sigma \rightarrow \tau}$}
\hspace*{-1cm}\centerline{\DisplayProof}}

Realizing that \premise{e-mc-E-shouldBe-sigma} = \premise{e-mc-E-is-sigma}, \premise{e-mc-E-shouldBe-sigma-left} = \premise{e-mc-E-is-sigma-left}, \premise{e-mc-E-shouldBe-sigma-right} = \premise{e-mc-E-is-sigma-right}, and \premise{e-mc-c1-ts0} = \premise{e-mc-ts0} finishes the proof.


{
\item[\semRule{OP-AssignExpr}, \semRule{OP-IfExpr},  \semRule{OP-MeasureExpr}, \semRule{OP-PromoExpr}, \semRule{OP-RecvExpr},]
\itemsep=-0.25\baselineskip
\item[\semRule{OP-SendExpr1}, \semRule{OP-SendExpr2}] The proof is a straightforward alteration of the proof of \semRule{OP-MethodCallExpr}.
}


\item[\semRule{OP-ForkExpr}] $TE_0 = \fork\ m\baseterm(\listOf{\value}\baseterm,E\baseterm,\listOf{E}\baseterm{);}$. From $C_0:\tau$ we know:
\begin{prooftree}
\small
 \AxiomC{\small \newPremise{e-fork-prem-TE0}}
             \AxiomC{\newPremise{e-fork-ts0}}
             \UnaryInfC{$M;\emptyset \vdash_C (lms_0,vp_0, ts_0) : \type{\void \rightarrow \tau}$}
       \RightLabel{\small \typeRule{TC-StatClo}}
    \BinaryInfC{$M;\emptyset \vdash_C (lms_0, vp_0, \si{\fork\ m\baseterm(\listOf{\value}\baseterm,E\baseterm,\listOf{E}\baseterm{);}}\ ts_0): \type{void \rightarrow \tau}$}
      \RightLabel{\small\typeRule{T-Config}}
    \UnaryInfC{$M;\emptyset\vdash[gs_0\,\vert\,(lms_0, vp_0, \si{\fork\ m\baseterm(\listOf{\value}\baseterm,E\baseterm,\listOf{E})\baseterm{;}}\ ts_0)] : \tau$}
\end{prooftree}
where \premise{e-fork-prem-TE0} is (here we denote $vpc = \refConcept{vpContext}(vp_0)$): 
\begin{prooftree}\small
\small\AxiomC{\newPremise{e-fork-E-is-sigma-left}} \AxiomC{\newPremise{e-fork-E-is-sigma}}\UnaryInfC{$M;vpc \vdash_T E:\sigma$} \AxiomC{\newPremise{e-fork-E-is-sigma-right}}
 \RightLabel{\small\typeRule{T-MethodCall}}
\TrinaryInfC{$M;vpc \vdash_T m\baseterm(\listOf{\value}\baseterm,E\baseterm,\listOf{E}\baseterm{);} : \tau'$}
             \AxiomC{$m$ is a classical method}
 \RightLabel{\small\typeRule{T-Fork}}
 \BinaryInfC{$M;vpc \vdash_T \fork\ m\baseterm(\listOf{\value}\baseterm,E\baseterm,\listOf{E}\baseterm{);} : \void$}
\end{prooftree}

We want to prove $C_1 : \tau$:
\begin{prooftree}\small
 \AxiomC{\newPremise{e-fork-E-shouldBe-sigma}}
 \UnaryInfC{$M;\refConcept{vpContext}(vp_0) \vdash_T E : \type{\sigma}$}
    \AxiomC{\newPremise{e-fork-ExprClo-Right}}
       \RightLabel{\small \typeRule{TC-ExprClo}}
    \BinaryInfC{$M;\emptyset \vdash_C(lms_0, vp_0, \si{E}\ \si{\fork\ m\baseterm(\listOf{\value}\baseterm,\bullet\baseterm,\listOf{E}\baseterm{);}}\ ts_0): \type{void \rightarrow \tau}$}
      \RightLabel{\small \typeRule{T-Config}}
    \UnaryInfC{$M;\emptyset\vdash[gs_0\,\vert\,(lms_0, vp_0, \si{E}\ \si{\fork\ m\baseterm(\listOf{\value}\baseterm,\bullet\baseterm,\listOf{E}\baseterm{);}}\ ts_0)] : \tau$}
\end{prooftree}

where \premise{e-fork-ExprClo-Right} is (here we denote $vpc = \refConcept{vpContext}(vp_0)$):

\noindent{\small
 \AxiomC{\newPremise{e-fork-E-shouldBe-sigma-left}}
 \AxiomC{}
 \UnaryInfC{$M;vpc,\bullet:\sigma \vdash_T \bullet:\sigma$}
 \AxiomC{\newPremise{e-fork-E-shouldBe-sigma-right}}
 \TrinaryInfC{$M;vpc,\bullet:\sigma \vdash_T m\baseterm(\listOf{\value}\baseterm,\bullet\baseterm,\listOf{E}\baseterm{);} : \tau'$}
 \UnaryInfC{$M;vpc,\bullet:\sigma \vdash_T \fork\ m\baseterm(\listOf{\value}\baseterm,\bullet\baseterm,\listOf{E}\baseterm{);} : \void$}
       \AxiomC{\newPremise{e-fork-c1-ts0}}
       \UnaryInfC{$M;\emptyset \vdash_C (lms_0,vp_0, ts_0) : \type{\void \rightarrow \tau}$}
       \RightLabel{\small \typeRule{TC-StatHole}}
   \BinaryInfC{$M;\emptyset \vdash_C (lms_0,vp_0, \si{\fork\ m\baseterm(\listOf{\value}\baseterm,\bullet\baseterm,\listOf{E})\baseterm{;}}\ ts_0) : \type{\sigma \rightarrow \tau}$}
\hspace*{-1.1cm}\centerline{\DisplayProof}}

Realizing that \premise{e-fork-E-shouldBe-sigma} = \premise{e-fork-E-is-sigma}, \premise{e-fork-E-shouldBe-sigma-right} = \premise{e-fork-E-is-sigma-right}, \premise{e-fork-E-shouldBe-sigma-left} = \premise{e-fork-E-is-sigma-left}, and \premise{e-fork-c1-ts0} = \premise{e-fork-ts0} finishes the proof.


\item[\semRule{OP-ReturnExpr}] $TE_0 = \return\ E\baseterm;$. We know that $vp_0 = vp_1 = [vp_G \circ_G vp]$, $ts_0 = \dots \si{\circM}\ ts_r$. Denote by $\rho = \refConcept{typeOf}(@retVal,vp_1)$. From $C_0:\tau$ we know:
\begin{prooftree}
\small
 \AxiomC{\newPremise{e-retExpr-E-is-rho}}
 \UnaryInfC{$M;\refConcept{vpContext}(vp_0) \vdash_T E:\rho$}
             \AxiomC{\newPremise{e-retExpr-prem-void}}
             \UnaryInfC{$M;\emptyset \vdash_C (lms_0,vp_G, ts_r) : \type{\rho \rightarrow \tau}$}
       \RightLabel{\small \typeRule{TC-RetExpr}}
    \BinaryInfC{$M;\emptyset \vdash_C (lms_0, vp_0, \si{\return\ E\baseterm;}\dots \si{\circM}\ ts_r): \type{void \rightarrow \tau}$}
      \RightLabel{\small\typeRule{T-Config}}
    \UnaryInfC{$M;\emptyset\vdash[gs_0\,\vert\,(lms_0, vp_0, \si{\return\ E\baseterm;}\dots \si{\circM}\ ts_r)] : \tau$}
\end{prooftree}

We want to prove $C_1 : \tau$:
\begin{prooftree}\small
\AxiomC{\newPremise{e-retExpr-E-shouldBe-rho}}
 \UnaryInfC{$M;\refConcept{vpContext}(vp_0) \vdash_T E : \type{\rho}$}
      \AxiomC{\newPremise{e-retExpr-ExprClo-Right-2}}
      \RightLabel{\small \typeRule{TC-ExprClo}}
    \BinaryInfC{$M;\emptyset \vdash_C(lms_0, vp_0, \si{E}\ \si{\return\ \bullet\baseterm;}\dots \si{\circM}\ ts_r): \type{void \rightarrow \tau}$}
      \RightLabel{\small \typeRule{T-Config}}
    \UnaryInfC{$M;\emptyset\vdash[gs_0\,\vert\,(lms_0, vp_0, \si{E}\ \si{\return\ \bullet\baseterm;}\dots \si{\circM}\ ts_r)] : \tau$}
\end{prooftree}

where \premise{e-retExpr-ExprClo-Right-2} is:
\begin{prooftree}\small
      \AxiomC{\newPremise{e-retExpr-E-shouldBe-void}}
      \UnaryInfC{$M;\emptyset \vdash_C (lms_0,vp_G, ts_r) : \type{\rho \rightarrow \tau}$}
      \RightLabel{\small \typeRule{TC-RetHole}}
      \UnaryInfC{$M;\emptyset \vdash_C (lms_0,vp_0, \si{\return\ \bullet\baseterm;}\dots \si{\circM}\ ts_r) : \type{void \rightarrow \tau}$}
\end{prooftree}

Realizing that \premise{e-retExpr-E-shouldBe-rho} = \premise{e-retExpr-E-is-rho}, and \premise{e-retExpr-E-shouldBe-void} = \premise{e-retExpr-prem-void} finishes the proof.


\item[\semRule{OP-DoMethodCallCl}] $TE_0 = m\baseterm(\listOf{\value}\baseterm)$. From $C_0:\tau$ we know:
\begin{prooftree}
\small
 \AxiomC{\small \newPremise{e-mcClprem-TE0}}
             \AxiomC{\newPremise{e-mcClts0}}
             \UnaryInfC{$M;\emptyset \vdash_C (lms_0,vp_0, ts_0) : \type{\tau' \rightarrow \tau}$}
       \RightLabel{\small \typeRule{TC-ExprClo}}
    \BinaryInfC{$M;\emptyset \vdash_C (lms_0, vp_0, \si{m\baseterm(\value_0\baseterm,\dots\baseterm,\value_n\baseterm)}\ ts_0): \type{void \rightarrow \tau}$}
      \RightLabel{\small\typeRule{T-Config}}
    \UnaryInfC{$M;\emptyset\vdash[gs_0\,\vert\,(lms_0, vp_0, \si{m\baseterm(\value_0\baseterm,\dots\baseterm,\value_n\baseterm)}\ ts_0)] : \tau$}
\end{prooftree}
where \premise{e-mcClprem-TE0} is inferred by rule \typeRule{T-MethodCall}: 

\noindent{\small
 \AxiomC{$M;\refConcept{vpContext}(vp_0) \vdash_T M_T(m) = \sigma_0,\dots,\sigma_n \rightarrow \tau'$ \newPremise{e-mcClE-is-sigma}}
 \AxiomC{$\infer{M;\refConcept{vpContext}(vp_0) \vdash_T \value_0: \sigma_0}{\newPremise{e-mcClE-v0-is-sigma0}}$ \quad $\cdots$}
 \BinaryInfC{$M;\refConcept{vpContext}(vp_0) \vdash_T m\baseterm(\value_0\baseterm,\dots\baseterm,\value_n\baseterm) : \type{\tau'}$}
\hspace*{-1.1cm}\centerline{\DisplayProof}}

We want to prove $C_1 : \tau$:

\begin{prooftree}
\small
 \AxiomC{\newPremise{e-mcClE-shouldBe-sigma}}
 \RightLabel{\small \typeRule{T-Block}}
 \UnaryInfC{$M;\refConcept{vpContext}([vp_0\circ_G[\refConcept[square]{\square} \circ_L vp'_M]]) \vdash_T M_B(m) : \void$}
    \AxiomC{\newPremise{e-mcClExprClo-Right}}
       \RightLabel{\small \typeRule{TC-StatClo}}
    \BinaryInfC{$M;\emptyset \vdash_C(lms_0, [vp_0\circ_G[\refConcept[square]{\square} \circ_L vp'_M]], \si{M_B(m)}\ \si{\circM}\ ts_0): \type{void \rightarrow \tau}$}
      \RightLabel{\small \typeRule{T-Config}}
    \UnaryInfC{$M;\emptyset\vdash[gs_0\,\vert\,(lms_0, [vp_0\circ_G[\refConcept[square]{\square} \circ_L vp'_M]], \si{M_B(m)}\ \si{\circM}\ ts_0)] : \tau$}
\end{prooftree}

where $M=\refConcept[method typing context]{(M_T,M_H,M_B)}$ and \premise{e-mcClExprClo-Right} is:
\begin{prooftree}\small
   \AxiomC{\newPremise{e-mcClc1-ts0}}
   \UnaryInfC{$M;\emptyset \vdash_C (lms_0,vp_0, ts_0) : \refConcept{typeOf}(@retVal,[vp_0 \circ_G [\refConcept[square]{\square} \circ_L vp'_M]]) \rightarrow \tau$}
   \RightLabel{\small \typeRule{TC-RetImpl}}
   \UnaryInfC{$M;\emptyset \vdash_C (lms_0,[vp_0 \circ_G [\refConcept[square]{\square} \circ_L vp'_M]], \si{\circM}\ ts_0) : \type{\sigma \rightarrow \tau}$}
\end{prooftree}

Method body is always a block. This justifies usage of rule \typeRule{T-Block}. To see that the premises \premise{e-mcClc1-ts0} and \premise{e-mcClts0} specify the same proof, we recall the premise \premise{e-mcClE-is-sigma}: the return type of method $m$ is $\tau'$. From the definition of rule \semRule{OP-DoMethodCallCl} we get that $\refConcept{typeOf}(@retVal,[vp_0 \circ_G [\refConcept[square]{\square} \circ_L vp'_M]]) = \tau'$. This finishes the proof.


\item[\semRule{OP-SubstE}] $TE_0 = \value$, moreover the symbol under the top element is $Ec$. From $C_0:\tau$ we know:
\begin{prooftree}
\small
 \AxiomC{}
 \RightLabel{\typeRule{T-Value}}
 \UnaryInfC{$M;\emptyset \vdash_T \intVal{r}{\value}{T}: \type{T}$\newPremise{e-subste-value-is-T}}
             \AxiomC{\newPremise{e-subste-Ec-ts0}}
       \RightLabel{\small \typeRule{TC-ExprClo}}
    \BinaryInfC{$M;\emptyset \vdash_C (lms_0, vp_0, \si{\value}\ \si{Ec}\ ts_0): \type{void \rightarrow \tau}$}
      \RightLabel{\small\typeRule{T-Config}}
    \UnaryInfC{$M;\emptyset\vdash[gs_0\,\vert\,(lms_0, vp_0, \si{\value}\ \si{Ec}\ ts_0)] : \tau$}
\end{prooftree}
where \premise{e-subste-Ec-ts0} is: 
\begin{prooftree}\small
 \AxiomC{\newPremise{e-subste-ec-is-tau'}}
 \UnaryInfC{$M;\refConcept{vpContext}(vp_0),\bullet:\type{T} \vdash_T Ec : \tau'$}
     \AxiomC{\newPremise{e-subste-ts0-is-tau-to-tau'}}
     \UnaryInfC{$M;\emptyset \vdash_C (lms_0,vp_0, ts_0) : \type{\tau' \rightarrow \tau}$}
 \RightLabel{\small\typeRule{TC-ExprHole}}
 \BinaryInfC{$M;\emptyset \vdash_C (lms_0,vp_0, \si{Ec}\ ts_0) : \type{T \rightarrow \tau}$}
\end{prooftree}

We want to prove $C_1 : \tau$:
\begin{prooftree}
\small
 \AxiomC{\newPremise{e-subste-Ecv-shouldBe-tau'}}
 \UnaryInfC{$M;\refConcept{vpContext}(vp_0) \vdash_T Ec[\value]: \tau'$}
             \AxiomC{\newPremise{e-subste-ts_0-shouldBe-tau'-to-tau}}
             \UnaryInfC{$M;\emptyset \vdash_C (lms_0, vp_0, ts_0): \type{\tau' \rightarrow \tau}$}
       \RightLabel{\small \typeRule{TC-ExprClo}}
    \BinaryInfC{$M;\emptyset \vdash_C (lms_0, vp_0, \si{Ec[\value]}\ ts_0): \type{void \rightarrow \tau}$}
      \RightLabel{\small\typeRule{T-Config}}
    \UnaryInfC{$M;\emptyset\vdash[gs_0\,\vert\,(lms_0, vp_0, \si{Ec[\value]}\ ts_0)] : \tau$}
\end{prooftree}

From the premise \premise{e-subste-value-is-T}, we know $\value:\type{T}$. The substitution of $\value$ in place of $\bullet$ is just an $\alpha$-conversion that does not change type ($\bullet: \type{T}$). Indeed, \premise{e-subste-ts_0-shouldBe-tau'-to-tau} = \premise{e-subste-ts0-is-tau-to-tau'}, and \premise{e-subste-Ecv-shouldBe-tau'} = \premise{e-subste-ec-is-tau'} what finishes the proof.


\item[\semRule{OP-SubstS}] The proof is a straightforward alteration of the proof of \semRule{OP-SubstE}.

\end{description}
\end{proof}

{\lemma[Type preservation for \semRule{OP-DoFork}] Let $C_0 = [gs_0\,\vert\,(lms_0, vp_0, \si{\fork\ m\baseterm(\listOf{\value}\baseterm{);}}\ ts_0)]$, $C_1 = [gs_1\,\vert\,(lms_{1,1}, vp_{1,1}, ts_{1,1}) \parallel (lms_{1,2}, vp_{1,2}, ts_{1,2})]$ be two configurations such that $C_0 \xrightarrow{\semRule{OP-DoFork}}_p C_1$. If $C_0:\tau$ then $C_1:\tau \times \theta$.}

\begin{proof}
From $C_0:\tau$ we know:
\begin{prooftree}
\small
 \AxiomC{\newPremise{p-doFork-prem-meth}}
	\RightLabel{\typeRule{T-MethodCall}}
\UnaryInfC{$M;\emptyset\vdash_T m(\listOf{\value}) : \type{T}$}
	\RightLabel{\typeRule{T-Fork}}
\UnaryInfC{$M;\emptyset\vdash_T \fork\ m\baseterm(\listOf{\value}\baseterm)\baseterm; : \void$}
             \AxiomC{\newPremise{p-doFork-ts0}}
             \UnaryInfC{$M;\emptyset \vdash_C (lms_0,vp_0, ts_0) : \type{\void \rightarrow \tau}$}
       \RightLabel{\small \typeRule{TC-StatClo}}
    \BinaryInfC{$M;\emptyset \vdash_C (lms_0, vp_0, \si{\fork\ m\baseterm(\listOf{\value}\baseterm{);}}\ ts_0): \type{void \rightarrow \tau}$}
      \RightLabel{\small\typeRule{T-Config}}
    \UnaryInfC{$M;\emptyset\vdash[gs_0\,\vert\,(lms_0, vp_0, \si{\fork\ m\baseterm(\listOf{\value}\baseterm{);}}\ ts_0)] : \tau$}
\end{prooftree}

We want to prove $C_1 : \tau \times \theta$:
\begin{prooftree}\small
\AxiomC{\newPremise{p-doFork-ExprClo-Left}}
\UnaryInfC{$M;\emptyset\vdash_C(lms_{1,1}, vp_0, ts_0): \type{\void \rightarrow \tau}$}
		\AxiomC{\newPremise{p-doFork-ExprClo-Right}}
      \RightLabel{\typeRule{T-Config}}
    \BinaryInfC{$M;\emptyset\vdash[gs_0\,\vert\,(lms_{1,1}, vp_0, ts_0) \parallel (lms_{1,2}, vp_{1,2}, ts_{1,2})] : \tau \times \theta$}
\end{prooftree}

where \premise{p-doFork-ExprClo-Right} is:

\begin{prooftree}\small
\AxiomC{\newPremise{p-doFork-mod-meth}}
	\RightLabel{\typeRule{T-MethodCall}}
\UnaryInfC{$M;\emptyset\vdash_T m(\listOf{\value'}) : \type{T}$}
		\AxiomC{}
			\RightLabel{\typeRule{TC-Empty}}
		\UnaryInfC{$M;\emptyset\vdash_C(lms_{1,2}, \refConcept[blacksquare]{\blacksquare}, \varepsilon): \type{T \rightarrow \theta}$}
			\RightLabel{\typeRule{TC-ExprClo}}
\BinaryInfC{$M;\emptyset\vdash_C(lms_{1,2}, \refConcept[blacksquare]{\blacksquare}, \si{m(\listOf{\value'})}): \type{void \rightarrow \theta}$}
\end{prooftree}

By inspecting the definition of rule \semRule{OP-DoFork} we see that transformation from $\listOf{\value}$ to $\listOf{\value'}$ preserves typing of individual values. Hence \premise{p-doFork-mod-meth} = \premise{p-doFork-prem-meth} and $\theta = \type{T}$. Realizing that \premise{p-doFork-ExprClo-Left} = \premise{p-doFork-ts0} finishes the proof.

\end{proof}


{\lemma[Type preservation for \semRule{OP-SendRecv}] Let $C_0 = [gs_0\,\vert\,(lms_{0,1}, vp_{0,1}, \si{\send\baseterm(\value_s\baseterm,\value_v\baseterm)\baseterm;}\ ts_{0,1}) \parallel (lms_{0,2}, vp_{0,2}, \si{\recv\baseterm(\value_r\baseterm)}\ ts_{0,2})]$, $C_1 = [gs_1\,\vert\,(lms_{1,1}, vp_{1,1}, ts_{1,1}) \parallel (lms_{1,2}, vp_{1,2}, ts_{1,2})]$ be two configurations such that $C_0 \xrightarrow{\semRule{OP-SendRecv}}_p C_1$. If $C_0:\tau \times \theta$ then $C_1:\tau \times \theta$.}

\begin{proof}
From $C_0:\tau \times \theta$ by rule \typeRule{T-Config} we know:
\begin{prooftree}
\small
\AxiomC{\newPremise{p-SendRecv-Send-Left}}
\AxiomC{\newPremise{p-SendRecv-Send-Right}}
\UnaryInfC{$M;\emptyset\vdash_C(lms_{0,1}, vp_{0,1}, ts_{0,1}): \type{\void \rightarrow \tau}$}
\RightLabel{\typeRule{TC-StatClo}}
\BinaryInfC{$M;\emptyset\vdash_C(lms_{0,1}, vp_{0,1}, \si{\send\baseterm(\value_s,\value_v\baseterm{);}}\ ts_{0,1}): \type{\void \rightarrow \tau}$}
\end{prooftree}

where \premise{p-SendRecv-Send-Left} is (here we denote $vpc_{0,1} = \refConcept{vpContext}(vp_{0,1})$):

\begin{prooftree}
\small
\AxiomC{}
\RightLabel{\typeRule{T-Value}}
\UnaryInfC{$M;vpc_{0,1}\vdash_T \value_s : \type{channelEnd[\tau_v]}$}
\AxiomC{}
\RightLabel{\typeRule{T-Value}}
\UnaryInfC{$M;vpc_{0,1}\vdash_T \value_v : \type{\tau_v}$}
\RightLabel{\typeRule{T-Send}}
\BinaryInfC{$M;vpc_{0,1}\vdash_T \send\baseterm(\value_s,\value_v\baseterm{);} : \type{\void}$}
\end{prooftree}

and (here we denote $vpc_{0,2} = \refConcept{vpContext}(vp_{0,2})$):
\begin{prooftree}
\small
\AxiomC{}
\RightLabel{\typeRule{T-Value}}
\UnaryInfC{$M;vpc_{0,2}\vdash_T \value_r: \type{channelEnd[\theta']}$}
\RightLabel{\typeRule{T-Recv}}
\UnaryInfC{$M;vpc_{0,2}\vdash_T \recv\baseterm(\value_r\baseterm): \type{\theta'}$}
\AxiomC{\newPremise{p-SendRecv-Recv-Right}}
\UnaryInfC{$M;\emptyset\vdash_C(lms_{0,1}, vp_{0,2}, ts_{0,2}): \type{\theta' \rightarrow \theta}$}
\RightLabel{\typeRule{TC-ExprClo}}
\BinaryInfC{$M;\emptyset\vdash_C(lms_{0,2}, vp_{0,2}, \si{\recv\baseterm(\value_r\baseterm)}\ ts_{0,2}): \type{\void \rightarrow \theta}$}
\end{prooftree}

We want to prove $C_1 : \tau \times \theta$:
\begin{prooftree}\small
\AxiomC{\newPremise{p-SendRecv-ExprClo-Left-2}}
\UnaryInfC{$M;\emptyset\vdash_C(lms_{1,1}, vp_{0,1}, ts_{0,1}): \type{\void \rightarrow \tau}$}
		\AxiomC{\newPremise{p-SendRecv-ExprClo-Right-2}}
      \RightLabel{\typeRule{T-Config}}
    \BinaryInfC{$M;\emptyset\vdash[gs_0\,\vert\,(lms_{1,1}, vp_{0,1}, ts_{0,1}) \parallel (lms_{1,2}, vp_{1,2}, \si{\value_v}\ ts_{1,2})] : \tau \times \theta$}
\end{prooftree}

where \premise{p-SendRecv-ExprClo-Right-2} is (here we denote $vpc_{1,2} = \refConcept{vpContext}(vp_{1,2})$):

\begin{prooftree}\small
\AxiomC{}
	\RightLabel{\typeRule{T-Value}}
\UnaryInfC{$M;vpc_{1,2}\vdash_T \value_v : \type{\theta''}$}
		\AxiomC{\newPremise{p-SendRecv-ExprClo}}
		\UnaryInfC{$M;\emptyset\vdash_C(lms_{1,2}, vp_{0,2}, ts_{0,2}): \type{\theta'' \rightarrow \theta}$}
			\RightLabel{\typeRule{TC-ExprClo}}
\BinaryInfC{$M;\emptyset\vdash_C(lms_{1,2}, vp_{0,2}, \si{\value_v}\ ts_{0,2}): \type{void \rightarrow \theta}$}
\end{prooftree}

By inspecting the definition of rule \semRule{OP-SendRecv} we see that $\value_s$ and $\value_r$ are ends of the same channel, hence their types match. Therefore, $\theta' = \type{\tau_v}$. Indeed, $\theta'' = \type{\tau_v}$. As changes in \refConcept{local memory state} does not change type, \premise{p-SendRecv-ExprClo} = \premise{p-SendRecv-Recv-Right}. Realizing that \premise{p-SendRecv-ExprClo-Left-2} = \premise{p-SendRecv-Send-Right} finishes the proof.

\end{proof}

{\lemma[Type preservation for $\longrightarrow_r$] Let $C_0 = [gs_0\,\vert\,(lms_0, vp_0, \si{TE_0}\ ts_0)]$, $C_1 = [gs_1\,\vert\,(lms_1, vp_1, ts_1)]$ be two configurations such that $C_0 \longrightarrow_r C_1$. If $C_0:\tau$ then $C_1:\tau$.}

\begin{proof}
We prove this lemma for each rule of relation $\longrightarrow_r$:

\begin{description}

\item[\semRule{OP-BlockHead}] $TE_0 = \overline{Be} = Be_0\ Be_1\dots Be_n$, $n \geq 1$. From $C_0:\tau$ we know:

\begin{prooftree}
\small
	\AxiomC{\newPremise{r-bh-ts0}}
	\UnaryInfC{$M;\emptyset \vdash_C (lms_0,vp_0, \si{Be_0}\ \si{Be_1\dots Be_n}\ ts_0) : \type{\void \rightarrow \tau}$}
		\RightLabel{\small\typeRule{TC-BlockHead}}
	\UnaryInfC{$M;\emptyset \vdash_C (lms_0, vp_0, \si{\overline{Be}}\ ts_0): \type{void \rightarrow \tau}$}
		\RightLabel{\small\typeRule{T-Config}}
	\UnaryInfC{$M;\emptyset\vdash[gs_0\,\vert\,(lms_0, vp_0, \si{\overline{Be}}\ ts_0)] : \tau$}
\end{prooftree}

We want to prove $C_1:\tau$:

\begin{prooftree}
\small
	\AxiomC{\newPremise{r-bh-2-ts0}}
	\UnaryInfC{$M;\emptyset \vdash_C (lms_0, vp_0, \si{Be_0}\ \si{Be_1\dots Be_n}\ ts_0): \type{void \rightarrow \tau}$}
		\RightLabel{\small\typeRule{T-Config}}
	\UnaryInfC{$M;\emptyset\vdash[gs_0\,\vert\,(lms_0, vp_0, \si{Be_0}\ \si{Be_1\dots Be_n}\ ts_0)] : \tau$}
\end{prooftree}
Indeed, \premise{r-bh-ts0} = \premise{r-bh-2-ts0}.


\item[\semRule{OP-Bracket}] $TE_0 = \baseterm(E\baseterm)$. From $C_0:\tau$ we know:

\noindent{\small
\small
\AxiomC{\newPremise{r-brk-E-is-tau'}}
\UnaryInfC{$M;\refConcept{vpContext}(vp_0) \vdash_T E : \tau'$}
\RightLabel{\typeRule{T-Bracket}}
\UnaryInfC{$M;\refConcept{vpContext}(vp_0) \vdash_T \baseterm(E\baseterm) : \tau'$}
	\AxiomC{\newPremise{r-brk-ts0}}
	\UnaryInfC{$M;\emptyset \vdash_C (lms_0,vp_0, ts_0) : \type{\tau' \rightarrow \tau}$}
		\RightLabel{\small\typeRule{TC-ExprClo}}
	\BinaryInfC{$M;\emptyset \vdash_C (lms_0, vp_0, \si{\baseterm(E\baseterm)}\ ts_0): \type{void \rightarrow \tau}$}
		\RightLabel{\small\typeRule{T-Config}}
	\UnaryInfC{$M;\emptyset\vdash[gs_0\,\vert\,(lms_0, vp_0, \si{\baseterm(E\baseterm)}\ ts_0)] : \tau$}
\hspace*{-1.1cm}\centerline{\DisplayProof}}

We want to prove $C_1:\tau$:

\begin{prooftree}
\small
\AxiomC{\newPremise{r-brk-E-shouldbe-tau'}}
\UnaryInfC{$M;\refConcept{vpContext}(vp_0) \vdash_T E : \tau'$}
	\AxiomC{\newPremise{r-brk-ts0-2}}
	\UnaryInfC{$M;\emptyset \vdash_C (lms_0,vp_0, ts_0) : \type{\tau' \rightarrow \tau}$}
		\RightLabel{\small\typeRule{TC-ExprClo}}
	\BinaryInfC{$M;\emptyset \vdash_C (lms_0, vp_0, \si{E}\ ts_0): \type{void \rightarrow \tau}$}
		\RightLabel{\small\typeRule{T-Config}}
	\UnaryInfC{$M;\emptyset\vdash[gs_0\,\vert\,(lms_0, vp_0, \si{E}\ ts_0)] : \tau$}
\end{prooftree}

Indeed, \premise{r-brk-E-shouldbe-tau'} = \premise{r-brk-E-is-tau'}, and \premise{r-brk-ts0-2} = \premise{r-brk-ts0}.


\item[\semRule{OP-VarDeclMulti}] $TE_0 = \type{T}\ I_0\baseterm,I_1\baseterm,\dots\baseterm,I_n\baseterm;$. From $C_0:\tau$ we know:

\begin{prooftree}
\small
\AxiomC{\newPremise{r-vdm-ts0}}
\UnaryInfC{$M;\emptyset \vdash_C (lms_0, vp_0, \si{\type{T}\ I_0\baseterm;}\ \si{\type{T}\ I_1\baseterm,\dots\baseterm,I_n\baseterm;}\ ts_0): \type{\void \rightarrow \tau}$}
	\RightLabel{\small\typeRule{TC-VarDeclMulti}}
\UnaryInfC{$M;\emptyset \vdash_C (lms_0, vp_0, \si{\type{T}\ I_0\baseterm,I_1\baseterm,\dots,I_n\baseterm;}\ ts_0): \type{\void \rightarrow \tau}$}
	\RightLabel{\small\typeRule{T-Config}}
\UnaryInfC{$M;\emptyset\vdash[gs_0\,\vert\,(lms_0, vp_0, \si{\type{T}\ I_0\baseterm,I_1\baseterm,\dots\baseterm,I_n\baseterm;}\ ts_0)] : \tau$}
\end{prooftree}

We want to prove $C_1:\tau$:

\begin{prooftree}
\small
\AxiomC{\newPremise{r-vdm-ts0-2}}
\UnaryInfC{$M;\emptyset \vdash_C (lms_0, vp_0, \si{\type{T}\ I_0\baseterm;}\ \si{\type{T}\ I_1\baseterm,\dots\baseterm,I_n\baseterm;}\ ts_0): \type{\void \rightarrow \tau}$}
	\RightLabel{\small\typeRule{T-Config}}
\UnaryInfC{$M;\emptyset\vdash[gs_0\,\vert\,(lms_0, vp_0, \si{\type{T}\ I_0\baseterm;}\ \si{\type{T}\ I_1\baseterm,\dots\baseterm,I_n\baseterm;}\ ts_0)] : \tau$}
\end{prooftree}

Indeed, \premise{r-vdm-ts0-2} = \premise{r-vdm-ts0}.


\item[\semRule{OP-While}] $TE_0 = \mathbf{while}\ \baseterm(E\baseterm)\ S$. From $C_0:\tau$ we know:

\begin{prooftree}
\small
\AxiomC{\newPremise{r-while-left}}
	\AxiomC{\newPremise{r-while-ts0}}
	\UnaryInfC{$M;\emptyset \vdash_C (lms_0,vp_0, ts_0) : \type{\void \rightarrow \tau}$}
		\RightLabel{\small\typeRule{TC-StatClo}}
	\BinaryInfC{$M;\emptyset \vdash_C (lms_0, vp_0, \si{\mathbf{while}\ \baseterm(E\baseterm)\ S}\ ts_0): \type{void \rightarrow \tau}$}
		\RightLabel{\small\typeRule{T-Config}}
	\UnaryInfC{$M;\emptyset\vdash[gs_0\,\vert\,(lms_0, vp_0, \si{\mathbf{while}\ \baseterm(E\baseterm)\ S}\ ts_0)] : \tau$}
\end{prooftree}

where \premise{r-while-left} is (here we denote $vpc = \refConcept{vpContext}(vp_0)$):
\begin{prooftree}
\small
\AxiomC{\newPremise{r-while-E-is-bool}}
\UnaryInfC{$M;vpc \vdash_T E : \type{bool}$}
	\AxiomC{\newPremise{r-while-S-is-void}}
	\UnaryInfC{$M;vpc \vdash_T S : \void$}
		\RightLabel{\typeRule{T-While}}
\BinaryInfC{$M;vpc \vdash_T \mathbf{while}\ \baseterm(E\baseterm)\ S : \void$}
\end{prooftree}

We want to prove $C_1:\tau$:

\begin{prooftree}
\small
\AxiomC{\newPremise{r-while-2-left}}
	\AxiomC{\newPremise{r-while-ts0-2}}
	\UnaryInfC{$M;\emptyset \vdash_C (lms_0,vp_0, ts_0) : \type{\tau' \rightarrow \tau}$}
		\RightLabel{\typeRule{TC-StatClo}}
	\BinaryInfC{$M;\emptyset \vdash_C (lms_0, vp_0, \si{\mathbf{if}\ \baseterm(E\baseterm)\ \baseterm\{S\ \mathbf{while}\ \baseterm(E\baseterm)\ S \baseterm\}\ \mathbf{else}\ \baseterm;}\ ts_0): \type{void \rightarrow \tau}$}
		\RightLabel{\typeRule{T-Config}}
	\UnaryInfC{$M;\emptyset\vdash[gs_0\,\vert\,(lms_0, vp_0, \si{\mathbf{if}\ \baseterm(E\baseterm)\ \baseterm\{S\ \mathbf{while}\ \baseterm(E\baseterm)\ S \baseterm\}\ \mathbf{else}\ \baseterm;}\ ts_0)] : \tau$}
\end{prooftree}

where \premise{r-while-2-left} is (here we denote $vpc = \refConcept{vpContext}(vp_0)$):
\noindent\begin{prooftree}
\small
\AxiomC{\newPremise{r-while-E-shouldBe-bool}}
\UnaryInfC{$M;vpc \vdash_T E : \type{bool}$}
	\AxiomC{\newPremise{r-while-2-center}}
	\UnaryInfC{$M;vpc \vdash_T \baseterm\{S\ \mathbf{while}\ \baseterm(E\baseterm)\ S \baseterm\} : \void$}
		\AxiomC{}
		\RightLabel{\typeRule{T-Skip}}
		\UnaryInfC{$M;vpc \vdash_T \baseterm; : \void$}
		\RightLabel{\typeRule{T-If}}
\TrinaryInfC{$M;vpc \vdash_T \mathbf{if}\ \baseterm(E\baseterm)\ \baseterm\{S\ \mathbf{while}\ \baseterm(E\baseterm)\ S \baseterm\}\ \mathbf{else}\ \baseterm; : \void$}
\end{prooftree}

where \premise{r-while-2-center} is by \typeRule{T-Block} (here we again denote $vpc = \refConcept{vpContext}(vp_0)$):
\noindent\begin{prooftree}
\small
\AxiomC{\newPremise{r-while-S-shouldBe-void}}
\UnaryInfC{$M;vpc \vdash_T S : \void$}
	\AxiomC{\newPremise{r-while-E-shouldBe-bool-2}}
	\UnaryInfC{$M;vpc \vdash_T E : \type{bool}$}
			\AxiomC{\newPremise{r-while-S-shouldBe-void-2}}
			\UnaryInfC{$M;vpc \vdash_T S : \void$}
		\RightLabel{\typeRule{T-While}}
		\BinaryInfC{$\mathbf{while}\ \baseterm(E\baseterm)\ S : \void$}
\RightLabel{\typeRule{T-BlockHead}}
\BinaryInfC{$M;vpc \vdash_T S\ \mathbf{while}\ \baseterm(E\baseterm)\ S : \void$}
\end{prooftree}

Indeed, \premise{r-while-E-shouldBe-bool} = \premise{r-while-E-shouldBe-bool-2} = \premise{r-while-E-is-bool}, \premise{r-while-S-shouldBe-void} = \premise{r-while-S-shouldBe-void-2} = \premise{r-while-S-is-void}, and \premise{r-while-ts0-2} = \premise{r-while-ts0}.

\end{description}

\end{proof}

{\lemma[Type preservation for $\longrightarrow_s$]\label{lemma-type-preservation-s} Let $C_0 = [gs_0\,\vert\,(lms_0, vp_0, \si{TE_0}\ ts_0)]$, $C_1 = [gs_1\,\vert\,(lms_1, vp_1, ts_1)]$ be two configurations such that $C_0 \longrightarrow_s C_1$. If $C_0:\tau$ then $C_1:\tau$.}

\begin{proof}
We prove this lemma for each rule of relation $\longrightarrow_s$:

\begin{description}
\item[\semRule{OP-Block}] $TE_0 = \baseterm\{ B \baseterm\}$. From $C_0:\tau$ we know (here we denote $vpc = \refConcept{vpContext}([vp_G \circ_G vp])$):
\begin{prooftree}
\small
  \AxiomC{\newPremise{s-block-B-is-void}}
 \UnaryInfC{$M;vpc \vdash_T B: \void$}
  \RightLabel{\small\typeRule{T-Block}}
 \UnaryInfC{$M;vpc \vdash_T \baseterm\{ B \baseterm\} : \void$}
             \AxiomC{\newPremise{s-block-ts0}}
             \UnaryInfC{$M;\emptyset \vdash_C (lms_0,[vp_G \circ_G vp], ts_0) : \type{\void \rightarrow \tau}$}
       \RightLabel{\small \typeRule{TC-StatClo}}
    \BinaryInfC{$M;\emptyset \vdash_C (lms_0, [vp_G \circ_G vp], \si{\baseterm\{ B \baseterm\}}\ ts_0): \type{\void \rightarrow \tau}$}
      \RightLabel{\small\typeRule{T-Config}}
    \UnaryInfC{$M;\emptyset\vdash[gs_0\,\vert\,(lms_0, [vp_G \circ_G vp], \si{\baseterm\{ B \baseterm\}}\ ts_0)] : \tau$}
\end{prooftree}

We want to prove $C_1 : \tau$:
\begin{prooftree}\small
 \AxiomC{\newPremise{s-block-B-shouldBe-void}}
 \UnaryInfC{$M;\refConcept{vpContext}([vp_G \circ_G [vp \circ_L \refConcept[square]{\square}]]) \vdash_T B : \void$}
    \AxiomC{\newPremise{s-block-StatClo-Right}}
       \RightLabel{\small \typeRule{TC-StatClo}}
    \BinaryInfC{$M;\emptyset \vdash_C(lms_0, [vp_G \circ_G [vp \circ_L \refConcept[square]{\square}]], \si{B}\ \si{\circL}\ ts_0): \type{void \rightarrow \tau}$}
      \RightLabel{\small \typeRule{T-Config}}
    \UnaryInfC{$M;\emptyset\vdash[gs_0\,\vert\,(lms_0, [vp_G \circ_G [vp \circ_L \refConcept[square]{\square}]], \si{B}\ \si{\circL}\ ts_0)] : \tau$}
\end{prooftree}

where \premise{s-block-StatClo-Right} is:

\begin{prooftree}\small
   \AxiomC{\newPremise{s-block-c1-ts0}}
   \UnaryInfC{$M;\emptyset \vdash_C (lms_0, [vp_G \circ_G vp], ts_0) : \type{\void \rightarrow \tau}$}
   \RightLabel{\small \typeRule{TC-BlockEnd}}
   \UnaryInfC{$M;\emptyset \vdash_C (lms_0, [vp_G \circ_G [vp \circ_L \refConcept[square]{\square}]], \si{\circL}\ ts_0) : \type{\void \rightarrow \tau}$}
\end{prooftree}

Indeed, $\refConcept{vpContext}([vp_G \circ_G [vp \circ_L \refConcept[square]{\square}]]) = \refConcept{vpContext}([vp_G \circ_G vp])$, hence the premises \premise{s-block-B-shouldBe-void} and \premise{s-block-B-is-void} are the same. Realizing that \premise{s-block-c1-ts0} = \premise{s-block-ts0} finishes the proof.


\item[\semRule{OP-BlockEnd}] $TE_0 = \circL$. From $C_0:\tau$ we know:
\begin{prooftree}
\small
 \AxiomC{\newPremise{s-blockEnd-ts0}}
 \UnaryInfC{$M;\emptyset \vdash_C (lms_0,[vp_G \circ_G vp], ts_0) : \type{\void \rightarrow \tau}$}
 \RightLabel{\small \typeRule{TC-BlockEnd}}
 \UnaryInfC{$M;\emptyset \vdash_C (lms_0, [vp_G \circ_G [vp \circ_L vp_L]], \si{\circL}\ ts_0): \type{\void \rightarrow \tau}$}
   \RightLabel{\small\typeRule{T-Config}}
 \UnaryInfC{$M;\emptyset\vdash[gs_0\,\vert\,(lms_0, [vp_G \circ_G [vp \circ_L vp_L]], \si{\circL}\ ts_0)] : \tau$}
\end{prooftree}

We want to prove $C_1 : \tau$:
\begin{prooftree}\small
    \AxiomC{\newPremise{s-blockEnd-ExprClo-Right}}
    \UnaryInfC{$M;\emptyset \vdash_C(lms_0, [vp_G \circ_G vp], ts_0): \type{void \rightarrow \tau}$}
      \RightLabel{\small \typeRule{T-Config}}
    \UnaryInfC{$M;\emptyset\vdash[gs_0\,\vert\,(lms_0, [vp_G \circ_G vp], ts_0)] : \tau$}
\end{prooftree}

Indeed, \premise{s-blockEnd-ExprClo-Right} = \premise{s-blockEnd-ts0}.


\item[\semRule{OP-IfFalse}] $TE_0 = {\bf if}\ \baseterm(\intVal{r}{\baseterm{false}}{bool}\baseterm)\ S_1\ {\bf else}\ S_2$. From $C_0:\tau$ we know:
\begin{prooftree}
\small
 \AxiomC{\newPremise{s-if-prem}}
     \AxiomC{\newPremise{s-if-ts0}}
     \UnaryInfC{$M;\emptyset \vdash_C (lms_0, vp_0, ts_0) : \type{\void \rightarrow \tau}$}
     \RightLabel{\small \typeRule{TC-StatClo}}
   \BinaryInfC{$M;\emptyset \vdash_C (lms_0, vp_0, \si{{\bf if}\ \baseterm(\intVal{r}{\baseterm{false}}{bool}\baseterm)\ S_1\ {\bf else}\ S_2}\ ts_0): \type{\void \rightarrow \tau}$}
   \RightLabel{\small\typeRule{T-Config}}
 \UnaryInfC{$M;\emptyset\vdash[gs_0\,\vert\,(lms_0, vp_0, \si{{\bf if}\ \baseterm(\intVal{r}{\baseterm{false}}{bool}\baseterm)\ S_1\ {\bf else}\ S_2}\ ts_0)] : \tau$}
\end{prooftree}

where \premise{s-if-prem} is (here we denote $vpc = \refConcept{vpContext}(vp_0)$):

\begin{prooftree}
\small
 \AxiomC{\newPremise{s-if-E-is-bool}}
 \UnaryInfC{$M;vpc \vdash_T \intVal{r}{\baseterm{false}}{bool}: \type{bool}$}
   \AxiomC{\newPremise{s-if-S1-is-void}}
   \UnaryInfC{$M;vpc \vdash_T S_1: \void$}
     \AxiomC{\newPremise{s-if-S2-is-void}}
     \UnaryInfC{$M;vpc \vdash_T S_2: \void$}
 \RightLabel{\typeRule{T-If}}
 \TrinaryInfC{$M;vpc \vdash_T {\bf if}\ \baseterm(\intVal{r}{\baseterm{false}}{bool}\baseterm)\ S_1\ {\bf else}\ S_2 : \void$}
\end{prooftree}

We want to prove $C_1 : \tau$:
\begin{prooftree}\small
   \AxiomC{\newPremise{s-if-S2-shouldBe-void}}
   \UnaryInfC{$M;\refConcept{vpContext}(vp_0) \vdash_T \si{S_2}: \void$}
     \AxiomC{\newPremise{s-if-StatClo-Right}}
     \UnaryInfC{$M;\emptyset \vdash_C (lms_0,vp_0, ts_0) : \type{\void \rightarrow \tau}$}
    \RightLabel{\small \typeRule{TC-StatClo}}
  \BinaryInfC{$M;\emptyset \vdash_C(lms_0, vp_0, \si{S_2}\ ts_0): \type{void \rightarrow \tau}$}
   \RightLabel{\small \typeRule{T-Config}}
  \UnaryInfC{$M;\emptyset\vdash[gs_0\,\vert\,(lms_0, vp_0, \si{S_2}\ ts_0)] : \tau$}
\end{prooftree}

Realizing that \premise{s-if-S2-shouldBe-void} = \premise{s-if-S2-is-void}, and \premise{s-if-StatClo-Right} = \premise{s-if-ts0} finishes the proof.


\item[\semRule{OP-IfTrue}] The proof is a straightforward alteration of the proof of \semRule{OP-IfFalse}.


\item[\semRule{OP-PromoForget}] $TE_0 = \value\baseterm;$. From $C_0:\tau$ we know:
\begin{prooftree}
\small
 \AxiomC{\newPremise{s-promoForget-prem}}
 \UnaryInfC{$M;\refConcept{vpContext}(vp_0) \vdash_T \value\baseterm;: \void$}
     \AxiomC{\newPremise{s-promoForget-ts0}}
     \UnaryInfC{$M;\emptyset \vdash_C (lms_0, vp_0, ts_0) : \type{\void \rightarrow \tau}$}
     \RightLabel{\small \typeRule{TC-StatClo}}
   \BinaryInfC{$M;\emptyset \vdash_C (lms_0, vp_0, \si{\value\baseterm;}\ ts_0): \type{\void \rightarrow \tau}$}
   \RightLabel{\small\typeRule{T-Config}}
 \UnaryInfC{$M;\emptyset\vdash[gs_0\,\vert\,(lms_0, vp_0, \si{\value\baseterm;}\ ts_0)] : \tau$}
\end{prooftree}

We want to prove $C_1 : \tau$:
\begin{prooftree}\small
  \AxiomC{\newPremise{s-promoForget-ts0-exp}}
  \UnaryInfC{$M;\emptyset \vdash_C(lms_0, vp_0, ts_0): \type{void \rightarrow \tau}$}
   \RightLabel{\small \typeRule{T-Config}}
  \UnaryInfC{$M;\emptyset\vdash[gs_0\,\vert\,(lms_0, vp_0, ts_0)] : \tau$}
\end{prooftree}

Realizing that \premise{s-promoForget-ts0-exp} = \premise{s-promoForget-ts0} finishes the proof.


\item[\semRule{OP-ReturnVoidImpl}] $TE_0 = \circM$. From $C_0:\tau$ we know:
\begin{prooftree}
\small
 \AxiomC{\newPremise{s-retImpl-ts0}}
 \UnaryInfC{$M;\emptyset \vdash_C (lms_0, vp_G, ts_0) : \type{\mathit{\refConcept{typeOf}(@retVal,[vp_G \circ_G vp])} \rightarrow \tau}$}
 \RightLabel{\small \typeRule{TC-RetImpl}}
 \UnaryInfC{$M;\emptyset \vdash_C (lms_0, [vp_G \circ_G vp], \,\si{\circM}\ ts_0): \type{\void \rightarrow \tau}$}
   \RightLabel{\small\typeRule{T-Config}}
 \UnaryInfC{$M;\emptyset\vdash[gs_0\,\vert\,(lms_0, [vp_G \circ_G vp], \,\si{\circM}\ ts_0)] : \tau$}
\end{prooftree}

We want to prove $C_1 : \tau$:

\noindent{\small
   \AxiomC{}
   \UnaryInfC{$M;\refConcept{vpContext}(vp_G) \vdash_T \intVal{\none}{\bot}{void}: \void$}
     \AxiomC{\newPremise{s-retImpl-StatClo-Right}}
     \UnaryInfC{$M;\emptyset \vdash_C (lms_0,vp_G, ts_0) : \type{\void \rightarrow \tau}$}
    \RightLabel{\small \typeRule{TC-ExprClo}}
    \BinaryInfC{$M;\emptyset \vdash_C(lms_0, vp_G, \si{\intVal{\none}{\bot}{void}}\ ts_0): \type{void \rightarrow \tau}$}
      \RightLabel{\small \typeRule{T-Config}}
    \UnaryInfC{$M;\emptyset\vdash[gs_0\,\vert\,(lms_0, vp_G, \si{\intVal{\none}{\bot}{void}}\ ts_0)] : \tau$}
\hspace*{-1.1cm}\centerline{\DisplayProof}}

We must prove that $\refConcept{typeOf}(@retVal,[vp_G \circ_G vp]) = \void$.

The only way to get $\circM$ to the stack is by invoking method $m$ using rule \semRule{OP-DoMethodCallCl} which also sets $\refConcept{typeOf}(@retVal,[vp_G \circ_G vp])$ to the \refConcept{return type} \type{T} of the method $m$. The method is \refConcept[well-typed method]{well-typed}.

Let \type{T} be not \void. From the premises of typing rule \typeRule{T-Method}, this means that $\refConcept{RetOk}(\refConcept[method typing context]{M_B}(m))$ is satisfied. However, by Lemma \ref{lemma-eval-retok-to-return} we know that there must be a \baseterm{return;} or \baseterm{return }$\value$\baseterm{;} statement evaluated during evaluation of method $m$'s body. This would replace $\circM$. But this contradicts our assumption that $TE = \circM$. Therefore $\refConcept{typeOf}(@retVal,[vp_G \circ_G vp]) = \void$.

The proof of this case is now finished as \premise{s-retImpl-StatClo-Right} is equal to \premise{s-retImpl-ts0}.


\item[\semRule{OP-Skip}] $TE_0 = \baseterm;$. From $C_0:\tau$ we know:
\begin{prooftree}
\small
 \AxiomC{}
 \RightLabel{\small \typeRule{T-Skip}}
 \UnaryInfC{$M;\refConcept{vpContext}(vp_0) \vdash_T \baseterm;: \void$}
     \AxiomC{\newPremise{s-skip-ts0}}
     \UnaryInfC{$M;\emptyset \vdash_C (lms_0, vp_0, ts_0) : \type{\void \rightarrow \tau}$}
     \RightLabel{\small \typeRule{TC-StatClo}}
   \BinaryInfC{$M;\emptyset \vdash_C (lms_0, vp_0, \si{\baseterm;}\ ts_0): \type{\void \rightarrow \tau}$}
   \RightLabel{\small\typeRule{T-Config}}
 \UnaryInfC{$M;\emptyset\vdash[gs_0\,\vert\,(lms_0, vp_0, \si{\baseterm;}\ ts_0)] : \tau$}
\end{prooftree}

We want to prove $C_1 : \tau$:
\begin{prooftree}\small
  \AxiomC{\newPremise{s-skip-ts0-exp}}
  \UnaryInfC{$M;\emptyset \vdash_C(lms_0, vp_0, ts_0): \type{void \rightarrow \tau}$}
   \RightLabel{\small \typeRule{T-Config}}
  \UnaryInfC{$M;\emptyset\vdash[gs_0\,\vert\,(lms_0, vp_0, ts_0)] : \tau$}
\end{prooftree}

Realizing that \premise{s-skip-ts0-exp} = \premise{s-skip-ts0} finishes the proof.


\item[\semRule{OP-VarDecl}] $TE_0 = T\ x\baseterm;$. From $C_0:\tau$ we know:
\begin{prooftree}
\small
   \AxiomC{\newPremise{s-varDecl-ts0}}
   \UnaryInfC{$M; \vdash_C (lms_0, [vp_G \circ_G [vp \circ_L vp_L']], ts_0) : \type{\void \rightarrow \tau}$}
   \RightLabel{\small \typeRule{TC-VarDeclOne}}
   \UnaryInfC{$M;\emptyset \vdash_C (lms_0, [vp_G \circ_G [vp \circ_L vp_L]], \si{T\ x\baseterm;}\ ts_0): \type{\void \rightarrow \tau}$}
   \RightLabel{\small\typeRule{T-Config}}
 \UnaryInfC{$M;\emptyset\vdash[gs_0\,\vert\,(lms_0, [vp_G \circ_G [vp \circ_L vp_L]], \si{T\ x\baseterm;}\ ts_0)] : \tau$}
\end{prooftree}

We want to prove $C_1 : \tau$:
\begin{prooftree}\small
  \AxiomC{\newPremise{s-varDecl-ts0-exp}}
  \UnaryInfC{$M;\emptyset \vdash_C(lms_0, [vp_G \circ_G [vp \circ_L vp_L'']], ts_0): \type{void \rightarrow \tau}$}
   \RightLabel{\small \typeRule{T-Config}}
  \UnaryInfC{$M;\emptyset\vdash[gs_0\,\vert\,(lms_0, [vp_G \circ_G [vp \circ_L vp_L'']], ts_0)] : \tau$}
\end{prooftree}

where $vp_L'' = vp_L[x \mapsto \none]_{+,var}[x \mapsto T]_{+,type}$.

By \refConcept[variable properties]{definition}, $vp_L' = (f'_{var},f'_{ch},f'_{qa},f'_{type})$ and $vp_L'' = (f''_{var},f''_{ch},f''_{qa},f''_{type})$. From definitions of \semRule{OP-VarDecl} and \typeRule{TC-VarDeclOne}, we get $f'_{type} = f''_{type}$. This is enough to see that \premise{s-varDecl-ts0-exp} is equal to \premise{s-varDecl-ts0} as the other parts of the variable properties tuples ({\it ie.} $f'_{var},f'_{ch},f'_{qa}, f''_{var},f''_{ch},f''_{qa}$) are never used in typing.


\item[\semRule{OP-VarDeclAlF}, \semRule{OP-VarDeclChE}] The proof is a straightforward alteration of the proof of \semRule{OP-VarDecl}.

\end{description}
\end{proof}

{\lemma[Type preservation for $\longrightarrow_{ret}$] Let $C_0 = [gs_0\,\vert\,(lms_0, [vp_G \circ_G vp], TE_0\ \dots\circM ts_0)]$, $C_1 = [gs_1\,\vert\,(lms_1, vp_G, ts_1)]$ be two configurations such that $C_0 \longrightarrow_{ret} C_1$. If $C_0:\tau$ then $C_1:\tau$.}

\begin{proof}
We prove this lemma for each rule of relation $\longrightarrow_{ret}$:

\begin{description}
\item[\semRule{OP-ReturnValue}] $TE_0 = \return\ \value\baseterm;$. From $C_0:\tau$ we know (here we denote $vpc = \refConcept{vpContext}([vp_G \circ_G vp])$ and $\sigma = \refConcept{typeOf}(@retVal,[vp_G \circ_G vp])$):
\begin{prooftree}
\small
 \AxiomC{}
 \RightLabel{\typeRule{T-Value}}
 \UnaryInfC{$M;vpc \vdash_T \value : \sigma$}
             \AxiomC{\newPremise{ret-rval-ts0}}
             \UnaryInfC{$M;\emptyset \vdash_C (lms_0,vp_G, ts_0) : \type{\sigma \rightarrow \tau}$}
       \RightLabel{\small \typeRule{TC-RetExpr}}
    \BinaryInfC{$M;\emptyset \vdash_C (lms_0, [vp_G \circ_G vp], \si{\return\ \value\baseterm;}\ \dots \circM ts_0): \type{\void \rightarrow \tau}$}
      \RightLabel{\small\typeRule{T-Config}}
    \UnaryInfC{$M;\emptyset\vdash[gs_0\,\vert\,(lms_0, [vp_G \circ_G vp], \si{\return\ \value\baseterm;}\ \dots\circM ts_0)] : \tau$}
\end{prooftree}

We want to prove $C_1 : \tau$:
\begin{prooftree}\small
 \AxiomC{}
 \RightLabel{\typeRule{T-Value}}
 \UnaryInfC{$M;\refConcept{vpContext}(vp_G) \vdash_T \value : \sigma$}
          \AxiomC{\newPremise{ret-rval-ExprClo-Right}}
          \UnaryInfC{$M;\emptyset \vdash_C(lms_0, vp_G, ts_0): \type{\sigma \rightarrow \tau}$}
    \RightLabel{\small \typeRule{TC-ExprClo}}
    \BinaryInfC{$M;\emptyset \vdash_C(lms_0, vp_G, \si{\value}\ ts_0): \type{void \rightarrow \tau}$}
      \RightLabel{\small \typeRule{T-Config}}
    \UnaryInfC{$M;\emptyset\vdash[gs_0\,\vert\,(lms_0, vp_G, \si{\value}\ ts_0)] : \tau$}
\end{prooftree}

The type of return value $\value$ does not depend on the context, therefore it is always $\sigma$. Realizing that \premise{ret-rval-ExprClo-Right} = \premise{ret-rval-ts0} finishes the proof.


\item[\semRule{OP-ReturnVoid}] $TE_0 = \return\baseterm;$. From $C_0:\tau$ we know:
\begin{prooftree}
\small
    \AxiomC{\newPremise{ret-rvoid-ts0}}
    \UnaryInfC{$M;\emptyset \vdash_C (lms_0,vp_G, ts_0) : \type{\void \rightarrow \tau}$}
       \RightLabel{\small \typeRule{TC-RetVoid}}
    \UnaryInfC{$M;\emptyset \vdash_C (lms_0, [vp_G \circ_G vp], \si{\return\baseterm;}\ \dots \circM ts_0): \type{\void \rightarrow \tau}$}
      \RightLabel{\small\typeRule{T-Config}}
    \UnaryInfC{$M;\emptyset\vdash[gs_0\,\vert\,(lms_0, [vp_G \circ_G vp], \si{\return\baseterm;}\ \dots\circM ts_0)] : \tau$}
\end{prooftree}

We want to prove $C_1 : \tau$ (here we denote $vpc = \refConcept{vpContext}(vp_G)$):

\noindent{\small
 \AxiomC{}
 \RightLabel{\typeRule{T-Value}}
 \UnaryInfC{$M;vpc \vdash_T \intVal{\none}{\bot}{void} : \void$}
          \AxiomC{\newPremise{ret-rvoid-ExprClo-Right}}
          \UnaryInfC{$M;\emptyset \vdash_C(lms_0, vp_G, ts_0): \type{\void \rightarrow \tau}$}
    \RightLabel{\small \typeRule{TC-ExprClo}}
    \BinaryInfC{$M;\emptyset \vdash_C(lms_0, vp_G, \si{\intVal{\none}{\bot}{void}}\ ts_0): \type{void \rightarrow \tau}$}
      \RightLabel{\small \typeRule{T-Config}}
    \UnaryInfC{$M;\emptyset\vdash[gs_0\,\vert\,(lms_0, vp_G, \si{\intVal{\none}{\bot}{void}}\ ts_0)] : \tau$}
\hspace*{-1.1cm}\centerline{\DisplayProof}}

Realizing that \premise{ret-rvoid-ExprClo-Right} = \premise{ret-rvoid-ts0} finishes the proof.

\end{description}

\end{proof}

\subsection{Type soundness}

\begin{theorem}[Type soundness for single-process programs]
 For any program which does not contain \baseterm{fork}, \baseterm{send} and \baseterm{recv} constructs, if a \refConcept{configuration} $C_0 = [gs \,\vert\, ls]: \tau$ evolves to a \refConcept[terminal configuration]{terminal} \refConcept{configuration} $C_n$, then either $C_n$ is an \refConcept{errorneous configuration}, or $C_n: \tau$.
\end{theorem}

\begin{proof}
 This is the corollary of progress and type preservation lemmata.
\end{proof}

\begin{theorem}[Type soundness for noncommunicating part of the language]
 For any program which does not contain {\bf send} and {\bf recv} constructs, if a \refConcept{configuration} $C_0 = [gs_0\,\vert\,ls_{0,1}\parallel\dots\parallel ls_{0,m}]:\tau$ evolves to a \refConcept[terminal configuration]{terminal} \refConcept{configuration} $C_n$, then either $C_n$ is an \refConcept{errorneous configuration}, or $C_n: \tau \times \tau'$ for some $\tau'$.
\end{theorem}

\begin{proof}
 This is the corollary of progress and type preservation lemmata.
\end{proof}

\section{Conclusion and future work}

We have described the LanQ imperative quantum programming language. This language can be used for implementation of both quantum algorithms and quantum protocols. We have formalized its syntax, both concrete and internal, typing, operational semantics, and proved type soundness of the non-communicating part of the language using proofs of standard lemmata in style of Wright and Felleisen \cite{WriFel94} and \cite{BieParPit2004}.

The language can be used for proving correctness of implemented quantum algorithms. If a correct pairing of sending and receiving processes is assured, it is also possible to prove correctness of quantum protocols.

By-reference usage of variables causes sometimes unwanted behaviour: it is possible to declare a program where a process sends a qubit and then it attempts to measure it. However, this is impossible as the qubit is then not owned by the process. Such situations are handled by \refConcept[runtime error]{runtime errors} what also helps in debugging programs and protocols written in LanQ.

The simulator of LanQ is also being developed and it is publicly available from address  \makebox{\url{http://lanq.sourceforge.net/}}.

An example of a truly random number generator and its evaluation sequence can be found in Appendix \ref{sec:example}.

\section*{Acknowledgements}

I would like to express my thanks to Jan Bouda, Simon Gay, Philippe Jorrand, Rajagopal Nagarajan, Nick Papanikolaou, Igor Peterl\'\i{}k, and Libor \v{S}karvada for their invaluable comments on LanQ and related discussions. I also wish to thank my supervisor, Jozef Gruska, for his guidance.

\bibliography{bib}

\newcommand{\etalchar}[1]{$^{#1}$}
\begin{thebibliography}{BBC{\etalchar{+}}93}

\bibitem[AG04]{AltGra04}
Thorsten Altenkirch and Jonathan Grattage.
\newblock {A functional quantum programming language}.
\newblock {\em quant-ph/0409065}, 2004.

\bibitem[BB84]{BB84}
C.H. Bennett and G.~Brassard.
\newblock {Quantum Cryptography: Public Key Distribution and Coin Tossing}.
\newblock In {\em {Proceedings of IEEE International Conference on Computers
  Systems and Signal Processing}}, pages 175--179, India, December 1984.
  Bangalore.

\bibitem[BBC{\etalchar{+}}93]{Ben_93}
Charles~H. Bennett, Gilles Brassard, Claude Cr\'epeau, Richard Jozsa, Asher
  Peres, and William~K. Wootters.
\newblock Teleporting an unknown quantum state via dual classical and
  einstein-podolsky-rosen channels.
\newblock {\em Phys. Rev. Lett.}, 70(13):1895--1899, Mar 1993.

\bibitem[BCS01]{BetCalSer01}
S.~Bettelli, T.~Calarco, and L.~Serafini.
\newblock Toward an architecture for quantum programming, 2001.

\bibitem[Ben92]{B92}
C.~H. Bennett.
\newblock Quantum cryptography using any two nonorthogonal states.
\newblock {\em Phys. Rev. Lett.}, 68(21):3121--3124, 1992.

\bibitem[BPP03]{BieParPit2004}
Gavin Bierman, Matthew Parkinson, and Andrew Pitts.
\newblock {MJ}: {A}n imperative core calculus for {J}ava and {J}ava with
  effects.
\newblock Technical Report 563, Cambridge University Computer Laboratory, April
  2003.

\bibitem[Eke91]{E91}
Artur~K. Ekert.
\newblock Quantum cryptography based on bell's theorem.
\newblock {\em Phys. Rev. Lett.}, 67(6):661--663, 1991.

\bibitem[GH05]{GayHol05}
Simon Gay and Malcolm Hole.
\newblock Subtyping for session types in the pi calculus.
\newblock {\em Acta Informatica}, 42(2):191--225, 2005.

\bibitem[GN04]{GayNag04}
Simon~J. Gay and Rajagopal Nagarajan.
\newblock {Communicating Quantum Processes}.
\newblock {\em quant-ph/0409052}, 2004.

\bibitem[GN05]{GayNag05}
Simon~J. Gay and Rajagopal Nagarajan.
\newblock Communicating quantum processes.
\newblock In {\em POPL '05: Proceedings of the 32nd ACM Symposium on Principles
  of Programming Languages}, pages 145--157, 2005.

\bibitem[GN06]{GayNag06}
Simon~J. Gay and Rajagopal Nagarajan.
\newblock {Types and typechecking for Communicating Quantum Processes}.
\newblock {\em Mathematical. Structures in Comp. Sci.}, 16:375--406, 2006.

\bibitem[JL04]{JorLal04b}
Philippe Jorrand and Marie Lalire.
\newblock Toward a quantum process algebra.
\newblock In {\em CF '04: Proceedings of the 1st conference on Computing
  frontiers}, pages 111--119, New York, NY, USA, 2004. ACM Press.

\bibitem[KPT96]{KobPieTur96}
Naoki Kobayashi, Benjamin~C. Pierce, and David~N. Turner.
\newblock Linearity and the pi-calculus.
\newblock In {\em POPL '96: Proceedings of the 23rd ACM SIGPLAN-SIGACT
  symposium on Principles of programming languages}, pages 358--371, New York,
  NY, USA, 1996. ACM Press.

\bibitem[Lal05]{Lal05}
Marie Lalire.
\newblock {A Probabilistic Branching Bisimulation for Quantum Processes}.
\newblock {\em quant-ph/0508116}, 2005.

\bibitem[Lal06]{Lal06}
Marie Lalire.
\newblock {\em D\'eveloppement d'une notation alorithmique pour le calcul
  quantique}.
\newblock PhD thesis, 2006.

\bibitem[LGP06]{LamGinPap06}
M.~Lampis, K.~G. Ginis, and N.~S. Papaspyrou.
\newblock Quantum data and control made easier.
\newblock In Peter Selinger, editor, {\em Preliminary Proceedings of the 4th
  International Workshop on Quantum Programming Languages}, pages 73--86, 2006.
\newblock The final version will be published in Electronic Notes in
  Theoretical Computer Science.

\bibitem[LJ04]{JorLal04}
Marie Lalire and Philipe Jorrand.
\newblock {A process-algebraic approach to concurrent and distributed quantum
  computation: operational semantics}.
\newblock {\em quant-ph/0407005}, 2004.

\bibitem[{\"O}me00]{Oem00}
B.~{\"O}mer.
\newblock Quantum programming in {QCL}.
\newblock Master's thesis, TU Vienna, 2000.

\bibitem[Sel04]{Sel04b}
Peter Selinger.
\newblock {Towards a quantum programming language}.
\newblock {\em Mathematical. Structures in Comp. Sci.}, 14(4):527--586, 2004.

\bibitem[Sho94]{Shor94}
P.~W. Shor.
\newblock {Algorithms for quantum computation: Discrete logarithms and
  factoring}.
\newblock In {\em {Proceedings of the 35th Annual Symposium on Foundations of
  Computer Science}}. IEEE Computer Society Press, 1994.

\bibitem[SV05]{SelVal05}
Peter Selinger and Beno\^{\i}t Valiron.
\newblock {A Lambda Calculus for Quantum Computation with Classical Control}.
\newblock {\em Lecture Notes in Computer Science}, 3461 / 2005:354--368, 2005.

\bibitem[vT03]{Ton03}
Andr\'e{} van Tonder.
\newblock {A Lambda Calculus for Quantum Computation}.
\newblock {\em quant-ph/0307150}, 2003.

\bibitem[WF94]{WriFel94}
Andrew~K. Wright and Matthias Felleisen.
\newblock A syntactic approach to type soundness.
\newblock {\em Information and Computation}, 115(1):38--94, 1994.

\bibitem[WZ82]{WooZur82}
W.K. Wootters and W.H. Zurek.
\newblock A single quantum cannot be cloned.
\newblock {\em Nature}, 299:802--803, 1982.

\bibitem[Zul01]{Zul01}
Paolo Zuliani.
\newblock {\em {Quantum Programming}}.
\newblock PhD thesis, University of Oxford, 2001.

\end{thebibliography}

\newpage
\appendix
\section{Program execution example}
\label{sec:example}

The probabilistic nature of measurement of quantum particles allows us to create generator of truly random numbers: Let us have a quantum particle in the state $\ket\psi = \frac{1}{2}(\ket{0} + \ket{1})$. Now we apply a measurement of this particle in the basis \{\ket0, \ket1\} (so called {\em standard basis}). The result of the measurement is 0 or 1 with equal probability.

The random number generator can be implemented as shown in Figure \ref{example:rand}.

{\renewcommand{\baselinestretch}{1}
\begin{figure}[h]
 \method{int}{main}{}{%
 \varQ{q}; \\%
 $q$ = \new \type{qbit}(); \\
 \return \Measure$(StdBasis, q)$;
 }
 \caption{Program example: Random number generator}
 \label{example:rand}
\end{figure}}

Before the execution, we must specify the \refConcept{method typing context} $M = (M_T, M_H, M_B)$. We have a program containing only a method $main$, hence the domain of the functions in \refConcept[method typing context]{$M$} is $\{main\}$. $\refConcept[method typing context]{M}$ is specified as: 
\begin{eqnarray*}
\refConcept[method typing context]{M_T}(main) &= & \type{\void \longrightarrow int} \\
\refConcept[method typing context]{M_H}(main) &= & \type{int}\ main() \\
\refConcept[method typing context]{M_B}(main) &= &\text{\{ \varQ{q}; $q$ = \new \type{qbit}(); }\\
	& & \qquad \text{\return \Measure($\intVal{\none}{StdBasis}{MeasurementBasis}, q$); \}}
\end{eqnarray*}

The execution of the program is shown in Figure \ref{example:rand:execution}. For typographical reasons we have simplified \refConcept{variable properties tuple} -- we do not show states of $f_{ch}, f_{qa}$ and $f_{type}$ as only the $f_{var}$ element of the tuple is needed in the example. 

\begin{sidewaysfigure}
\setlength{\baselineskip}{1.5\baselineskip}
\begin{center}
$ \bf{start} = [(((1), []), [])\,\vert\,(([],[],[],[]), \blacksquare, \si{main()})]$\\

$\downarrow_e$ \makebox[0pt][l]{\semRule{OP-DoMethodCallCl}}

$[(((1), []), [])\,\vert\,(([],[],[],[]), [\blacksquare \circ_G [\refConcept[square]{\square} \circ_L \refConcept[empty variable properties tuple]{\lozenge}]], \si{\text{\{ \varQ{q}; $q$ = \new \type{qbit}(); \return \Measure($\intVal{\none}{StdBasis}{MeasurementBasis}, q$); \}}}\ \si{\circM})]$

$\downarrow_s$ \makebox[0pt][l]{\semRule{OP-Block}}

$[(((1), []), [])\,\vert\,(([],[],[],[]), [\blacksquare \circ_G [[\refConcept[square]{\square} \circ_L \refConcept[empty variable properties tuple]{\lozenge}] \circ_L \refConcept[empty variable properties tuple]{\lozenge}]], \text{ \si{\varQ{q}; $q$ = \new \type{qbit}(); \return \Measure($\intVal{\none}{StdBasis}{MeasurementBasis}, q$);}}\ \si{\circL}\ \si{\circM})]$

$\downarrow_r$ \makebox[0pt][l]{\semRule{OP-BlockHead}}

$[(((1), []), [])\,\vert\,(([],[],[],[]), [\blacksquare \circ_G [[\refConcept[square]{\square} \circ_L \refConcept[empty variable properties tuple]{\lozenge}] \circ_L \refConcept[empty variable properties tuple]{\lozenge}]], \text{ \si{\varQ{q};} \si{$q$ = \new \type{qbit}(); \return \Measure($\intVal{\none}{StdBasis}{MeasurementBasis}, q$);}}\ \si{\circL}\ \si{\circM})]$

$\downarrow_s$ \makebox[0pt][l]{\semRule{OP-VarDecl}}

$[(((1), []), [])\,\vert\,(([],[],[],[]), [\blacksquare \circ_G [[\refConcept[square]{\square} \circ_L \refConcept[empty variable properties tuple]{\lozenge}] \circ_L [q \mapsto \none]]], \text{\si{$q$ = \new \type{qbit}(); \return \Measure($\intVal{\none}{StdBasis}{MeasurementBasis}, q$);}}\ \si{\circL}\ \si{\circM})]$

$\downarrow_r$ \makebox[0pt][l]{\semRule{OP-BlockHead}}

$[(((1), []), [])\,\vert\,(([],[],[],[]), [\blacksquare \circ_G [[\refConcept[square]{\square} \circ_L \refConcept[empty variable properties tuple]{\lozenge}] \circ_L [q \mapsto \none]]], \text{\si{$q$ = \new \type{qbit}();} \si{\return \Measure($\intVal{\none}{StdBasis}{MeasurementBasis}, q$);}}\ \si{\circL}\ \si{\circM})]$

$\downarrow_e$ \makebox[0pt][l]{\semRule{OP-PromoExpr}}

$[(((1), []), [])\,\vert\,(([],[],[],[]), [\blacksquare \circ_G [[\refConcept[square]{\square} \circ_L \refConcept[empty variable properties tuple]{\lozenge}] \circ_L [q \mapsto \none]]], \text{\si{$q$ = \new \type{qbit}()} \si{$\bullet$ ;} \si{\return \Measure($\intVal{\none}{StdBasis}{MeasurementBasis}, q$);}}\ \si{\circL}\ \si{\circM})]$

$\downarrow_e$ \makebox[0pt][l]{\semRule{OP-AssignExpr}}

\end{center}
 \captcont{Program example: Random number generator execution (to be continued)}
\end{sidewaysfigure}

\begin{sidewaysfigure}
\setlength{\baselineskip}{1.5\baselineskip}
\begin{center}
$\downarrow_e$ \makebox[0pt][l]{\semRule{OP-AssignExpr}}

$[(((1), []), [])\,\vert\,(([],[],[],[]), [\blacksquare \circ_G [[\refConcept[square]{\square} \circ_L \refConcept[empty variable properties tuple]{\lozenge}] \circ_L [q \mapsto \none]]], \text{\si{\new \type{qbit}()} \si{$q$ = $\bullet$} \si{$\bullet$ ;} \si{\return \Measure($\intVal{\none}{StdBasis}{MeasurementBasis}, q$);}}\ \si{\circL}\ \si{\circM})]$

$\downarrow_v$ \makebox[0pt][l]{\semRule{OP-AllocQ}}

$[((\frac{1}{2}\begin{pmatrix}1 & 0 \\ 0 & 1\end{pmatrix}, [2]),
 [])\,\vert\,(([],[(Quantum,[1])\mapsto(GQuantum,[1])],[],[]), [\blacksquare \circ_G [[\refConcept[square]{\square} \circ_L \refConcept[empty variable properties tuple]{\lozenge}] \circ_L [q \mapsto \none]]],$ \\[-0.33\baselineskip]
\hfill $\si{\intVal{(Quantum,[1])}{(GQuantum,[1])}{Q_2}}\ \text{\si{$q$ = $\bullet$} \si{$\bullet$ ;} \si{\return \Measure($\intVal{\none}{StdBasis}{MeasurementBasis}, q$);}}\ \si{\circL}\ \si{\circM})]$

$\downarrow_e$ \makebox[0pt][l]{\semRule{OP-SubstE}}

$[((\frac{1}{2}\begin{pmatrix}1 & 0 \\ 0 & 1\end{pmatrix}, [2]),
 [])\,\vert\,(([],[(Quantum,[1])\mapsto(GQuantum,[1])],[],[]), [\blacksquare \circ_G [[\refConcept[square]{\square} \circ_L \refConcept[empty variable properties tuple]{\lozenge}] \circ_L [q \mapsto \none]]],$ \\[-0.33\baselineskip]
\hfill $\text{\si{$q = \intVal{(Quantum,[1])}{(GQuantum,[1])}{Q_2}$}  \si{$\bullet$ ;} \si{\return \Measure($\intVal{\none}{StdBasis}{MeasurementBasis}, q$);}}\ \si{\circL}\ \si{\circM})]$

 $\downarrow_v$ \makebox[0pt][l]{\semRule{OP-AssignQValue}}

$[((\frac{1}{2}\begin{pmatrix}1 & 0 \\ 0 & 1\end{pmatrix}, [2]),
 [])\,\vert\,(([],[(Quantum,[1])\mapsto(GQuantum,[1])],[],[]), [\blacksquare \circ_G [[\refConcept[square]{\square} \circ_L \refConcept[empty variable properties tuple]{\lozenge}] \circ_L [q \mapsto (Quantum,[1])]]],$ \\[-0.33\baselineskip]
\hfill $\text{\si{$\intVal{(Quantum,[1])}{(GQuantum,[1])}{Q_2}$}  \si{$\bullet$ ;} \si{\return \Measure($\intVal{\none}{StdBasis}{MeasurementBasis}, q$);}}\ \si{\circL}\ \si{\circM})]$

$\downarrow_e$ \makebox[0pt][l]{\semRule{OP-SubstS}}

$[((\frac{1}{2}\begin{pmatrix}1 & 0 \\ 0 & 1\end{pmatrix}, [2]),
 [])\,\vert\,(([],[(Quantum,[1])\mapsto(GQuantum,[1])],[],[]), [\blacksquare \circ_G [[\refConcept[square]{\square} \circ_L \refConcept[empty variable properties tuple]{\lozenge}] \circ_L [q \mapsto (Quantum,[1])]]],$ \\[-0.33\baselineskip]
\hfill $\text{\si{$\intVal{(Quantum,[1])}{(GQuantum,[1])}{Q_2}$ ;} \si{\return \Measure($\intVal{\none}{StdBasis}{MeasurementBasis}, q$);}}\ \si{\circL}\ \si{\circM})]$

$\downarrow_s$ \makebox[0pt][l]{\semRule{OP-PromoForget}}
\end{center}
 \captcont{Program example: Random number generator execution (continued) (to be continued)}
\end{sidewaysfigure}

\begin{sidewaysfigure}
\setlength{\baselineskip}{1.5\baselineskip}
\begin{center}
$\downarrow_s$ \makebox[0pt][l]{\semRule{OP-PromoForget}}

$[((\frac{1}{2}\begin{pmatrix}1 & 0 \\ 0 & 1\end{pmatrix}, [2]),
 [])\,\vert\,(([],[(Quantum,[1])\mapsto(GQuantum,[1])],[],[]), [\blacksquare \circ_G [[\refConcept[square]{\square} \circ_L \refConcept[empty variable properties tuple]{\lozenge}] \circ_L [q \mapsto (Quantum,[1])]]],$ \\[-0.33\baselineskip]
\hfill $\text{\si{\return \Measure($\intVal{\none}{StdBasis}{MeasurementBasis}, q$);}}\ \si{\circL}\ \si{\circM})]$

$\downarrow_e$ \makebox[0pt][l]{\semRule{OP-ReturnExpr}}

$[((\frac{1}{2}\begin{pmatrix}1 & 0 \\ 0 & 1\end{pmatrix}, [2]),
 [])\,\vert\,(([],[(Quantum,[1])\mapsto(GQuantum,[1])],[],[]), [\blacksquare \circ_G [[\refConcept[square]{\square} \circ_L \refConcept[empty variable properties tuple]{\lozenge}] \circ_L [q \mapsto (Quantum,[1])]]],$ \\[-0.33\baselineskip]
\hfill $\si{\Measure(\intVal{\none}{StdBasis}{MeasurementBasis}, q)}\ \si{\return\ \bullet;}\ \si{\circL}\ \si{\circM})]$

$\downarrow_e$ \makebox[0pt][l]{\semRule{OP-MeasureExpr}}

$[((\frac{1}{2}\begin{pmatrix}1 & 0 \\ 0 & 1\end{pmatrix}, [2]),
 [])\,\vert\,(([],[(Quantum,[1])\mapsto(GQuantum,[1])],[],[]), [\blacksquare \circ_G [[\refConcept[square]{\square} \circ_L \refConcept[empty variable properties tuple]{\lozenge}] \circ_L [q \mapsto (Quantum,[1])]]],$ \\[-0.33\baselineskip]
\hfill $\si{q}\ \si{\Measure(\intVal{\none}{StdBasis}{MeasurementBasis}, \bullet)}\ \si{\return\ \bullet;}\ \si{\circL}\ \si{\circM})]$

$\downarrow_v$ \makebox[0pt][l]{\semRule{OP-Var}}

$[((\frac{1}{2}\begin{pmatrix}1 & 0 \\ 0 & 1\end{pmatrix}, [2]),
 [])\,\vert\,(([],[(Quantum,[1])\mapsto(GQuantum,[1])],[],[]), [\blacksquare \circ_G [[\refConcept[square]{\square} \circ_L \refConcept[empty variable properties tuple]{\lozenge}] \circ_L [q \mapsto (Quantum,[1])]]],$ \\[-0.33\baselineskip]
\hfill $\si{\intVal{(Quantum,[1])}{(GQuantum,[1])}{Q_2}}\ \si{\Measure(\intVal{\none}{StdBasis}{MeasurementBasis}, \bullet)}\ \si{\return\ \bullet;}\ \si{\circL}\ \si{\circM})]$

$\downarrow_e$ \makebox[0pt][l]{\semRule{OP-SubstE}}

$[((\frac{1}{2}\begin{pmatrix}1 & 0 \\ 0 & 1\end{pmatrix}, [2]),
 [])\,\vert\,(([],[(Quantum,[1])\mapsto(GQuantum,[1])],[],[]), [\blacksquare \circ_G [[\refConcept[square]{\square} \circ_L \refConcept[empty variable properties tuple]{\lozenge}] \circ_L [q \mapsto (Quantum,[1])]]],$ \\[-0.33\baselineskip]
\hfill $\si{\Measure(\intVal{\none}{StdBasis}{MeasurementBasis},\intVal{(Quantum,[1])}{(GQuantum,[1])}{Q_2})}\ \si{\return\ \bullet;}\ \si{\circL}\ \si{\circM})]$

$\downarrow_v$ \makebox[0pt][l]{\semRule{OP-DoMeasure}}
\end{center}
 \captcont{Program example: Random number generator execution (continued) (to be continued)}
\end{sidewaysfigure}

\begin{sidewaysfigure}
\setlength{\baselineskip}{1.5\baselineskip}
\begin{center}
$\downarrow_v$ \makebox[0pt][l]{\semRule{OP-DoMeasure}}

\quad$0.5 \bullet [((\begin{pmatrix}1 & 0 \\ 0 & 0\end{pmatrix}, [2]),
 [])\,\vert\,(([],[(Quantum,[1])\mapsto(GQuantum,[1])],[],[]), [\blacksquare \circ_G [[\refConcept[square]{\square} \circ_L \refConcept[empty variable properties tuple]{\lozenge}] \circ_L [q \mapsto (Quantum,[1])]]],$ \\[-0.33\baselineskip]
\hfill $\si{\intVal{\none}{0}{int}}\ \si{\return\ \bullet;}\ \si{\circL}\ \si{\circM})]$\\[-0.33\baselineskip]
$\boxplus\ 0.5 \bullet [((\begin{pmatrix}0 & 0 \\ 0 & 1\end{pmatrix}, [2]),
 [])\,\vert\,(([],[(Quantum,[1])\mapsto(GQuantum,[1])],[],[]), [\blacksquare \circ_G [[\refConcept[square]{\square} \circ_L \refConcept[empty variable properties tuple]{\lozenge}] \circ_L [q \mapsto (Quantum,[1])]]],$ \\[-0.33\baselineskip]
\hfill $\si{\intVal{\none}{1}{int}}\ \si{\return\ \bullet;}\ \si{\circL}\ \si{\circM})]$

$^{0.5}\!\!\!\swarrow$ \quad \semRule{NP-ProbEvol} \quad $\searrow\!\!\!^{0.5}$

\hfill Evolution of measurement branch 0 \hfill \hfill Evolution of measurement branch 1 \hfill\hfill \\[1.5\baselineskip]

\vfill
\end{center}

\setlength{\baselineskip}{0.6666\baselineskip}
We continue showing the program evolution of the branch where the measurement returned the value 0 only. The other measurement branch evolves obviously the same way, the only difference is in the measured value and the global quantum state.\\[1.5\baselineskip]

\begin{center}
$\downarrow^{0.5}$ \makebox[0pt][l]{\semRule{NP-ProbEvol}}

$[((\begin{pmatrix}1 & 0 \\ 0 & 0\end{pmatrix}, [2]),
 [])\,\vert\,(([],[(Quantum,[1])\mapsto(GQuantum,[1])],[],[]), [\blacksquare \circ_G [[\refConcept[square]{\square} \circ_L \refConcept[empty variable properties tuple]{\lozenge}] \circ_L [q \mapsto (Quantum,[1])]]],$ \\[-0.33\baselineskip]
\hfill $\si{\intVal{\none}{0}{int}}\ \si{\return\ \bullet;}\ \si{\circL}\ \si{\circM})]$\\[-0.33\baselineskip]

$\downarrow_e$ \makebox[0pt][l]{\semRule{OP-SubstS}}

$[((\begin{pmatrix}1 & 0 \\ 0 & 0\end{pmatrix}, [2]),
 [])\,\vert\,(([],[(Quantum,[1])\mapsto(GQuantum,[1])],[],[]), [\blacksquare \circ_G [[\refConcept[square]{\square} \circ_L \refConcept[empty variable properties tuple]{\lozenge}] \circ_L [q \mapsto (Quantum,[1])]]],$ \\[-0.33\baselineskip]
\hfill $\si{\return\ \intVal{\none}{0}{int};}\ \si{\circL}\ \si{\circM})]$\\[-0.33\baselineskip]

$\downarrow_v$ \makebox[0pt][l]{\semRule{OP-ReturnValue}}

$[((\begin{pmatrix}1 & 0 \\ 0 & 0\end{pmatrix}, [2]),
 [])\,\vert\,(([],[(Quantum,[1])\mapsto(GQuantum,[1])],[],[]), \blacksquare,$ \si{$\intVal{\none}{0}{int}$})]\\[-0.33\baselineskip]

\end{center}
 \label{example:rand:execution}
 \caption{Program example: Random number generator execution (continued)}
\end{sidewaysfigure}

\end{document}